\documentclass[11pt]{article}

\makeatletter
\normalsize
\usepackage{comment}
\usepackage{draft}
\usepackage{putex}
 
 \usepackage[table]{xcolor}
 \usepackage{dsfont}
\usepackage[all]{xy}

\usepackage{stackrel,amssymb}
\usepackage{graphicx}
\usepackage{bm}
\usepackage{array}
\usepackage{makecell}
\usepackage{subcaption}
\usepackage{epstopdf}
\usepackage{tensor}
\usepackage{slashed}
\usepackage[utf8]{inputenc}
\usepackage{rotating}
\usepackage{ytableau}
\usepackage{tikz-cd}
\usetikzlibrary{arrows.meta}
\usetikzlibrary{bending}

\usepackage{adjustbox}
\usepackage{multirow}

\usetikzlibrary{intersections}

\def\le{\left(}
\def\ri{\right)} 

\def\Hom{\operatorname{Hom}}

\def\XXint#1#2#3{{\setbox0=\hbox{$#1{#2#3}{\int}$ }
		\vcenter{\hbox{$#2#3$ }}\kern-.6\wd0}}

\numberwithin{equation}{section}

\newtheorem{theorem}{Theorem}

\def\<{\langle}
\def\>{\rangle}

\def\id{\mathds{1}}

\divide\@tempdima\baselineskip
\@tempcnta=\@tempdima
\skip\@mpfootins = \skip\footins
\renewcommand\section{\@startsection{section}{1}{\z@}%
                                   {-3.5ex \@plus -1.3ex \@minus -.7ex}%
                                   {2.3ex \@plus.4ex \@minus .4ex}%
                                   {\normalfont\large\bfseries}}
\renewcommand\subsection{\@startsection{subsection}{2}{\z@}%
                                   {-2.3ex\@plus -1ex \@minus -.5ex}%
                                   {1.2ex \@plus .3ex \@minus .3ex}%
                                   {\normalfont\normalsize\bfseries}}
\renewcommand\subsubsection{\@startsection{subsubsection}{3}{\z@}%
                                   {-2.3ex\@plus -1ex \@minus -.5ex}%
                                   {1ex \@plus .2ex \@minus .2ex}%
                                   {\normalfont\normalsize\bfseries}}
\renewcommand\paragraph{\@startsection{paragraph}{4}{\z@}%
                                   {1.75ex \@plus1ex \@minus.2ex}%
                                   {-1em}%
                                   {\normalfont\normalsize\bfseries}}
\renewcommand\subparagraph{\@startsection{subparagraph}{5}{\z@}%
                                   {1.75ex \@plus1ex \@minus .2ex}%
                                   {-1em}%
                                   {\normalfont\normalsize\itshape}}

\renewcommand{\@dotsep}{10000}

\def\fnum@figure{\textbf{\figurename\nobreakspace\thefigure}}
\def\fnum@table{\textbf{\tablename\nobreakspace\thetable}}

\long\def\@makecaption#1#2{%
  \vskip\abovecaptionskip
  \sbox\@tempboxa{\small #1. #2}%
  \ifdim \wd\@tempboxa >\hsize
    \small #1. #2\par
  \else
    \global \@minipagefalse
    \hb@xt@\hsize{\hfil\box\@tempboxa\hfil}%
  \fi
  \vskip\belowcaptionskip}

\renewenvironment{thebibliography}[1]{%
\begin{oldthebibliography}{#1}%
\small%
\raggedright%
\setlength{\itemsep}{5pt plus 0.2ex minus 0.05ex}%
}%
{%
\end{oldthebibliography}%
}
\makeatother

\usepackage{scalerel}
\usepackage{graphicx}
\usepackage{amsmath,amssymb}
\usepackage{amsfonts}
\usepackage{physics}
\usepackage{bbold}
\usepackage{mathtools}
\usepackage{verbatim}
\usepackage{cite} 
\usepackage{mdframed}
\usepackage{enumitem}
\usepackage[normalem]{ulem}

\usepackage[font=footnotesize,labelfont=bf,width=.9\textwidth]{caption}

\usepackage[colorlinks=true, linkcolor=blue, filecolor=blue, urlcolor=blue, citecolor=blue, anchorcolor=blue, menucolor=blue , linktocpage=true]{hyperref}

\allowdisplaybreaks
\numberwithin{equation}{section}

\makeatletter
\newcommand{\subalign}[1]{%
  \vcenter{%
    \Let@ \restore@math@cr \default@tag
    \baselineskip\fontdimen10 \scriptfont\tw@
    \advance\baselineskip\fontdimen12 \scriptfont\tw@
    \lineskip\thr@@\fontdimen8 \scriptfont\thr@@
    \lineskiplimit\lineskip
    \ialign{\hfil$\m@th\scriptstyle##$&$\m@th\scriptstyle{}##$\hfil\crcr
      #1\crcr
    }%
  }%
}
\makeatother

\usepackage[textsize=scriptsize,textwidth=2.45cm]{todonotes}

\newcommand{\eq}[1]{\begin{equation}#1\end{equation}}

\newcommand{\eqsp}[1]{\begin{equation}\begin{split}#1\end{split}\end{equation}}

\newcommand{\overbar}[1]{\mkern 1.5mu\overline{\mkern-1.5mu#1\mkern-1.5mu}\mkern 1.5mu}

\renewcommand{\title}[1]{\vbox{\center\LARGE{#1}}\vspace{5mm}}
\renewcommand{\author}[1]{\vbox{\center#1}\vspace{5mm}}

\newcommand{\email}[1]{\vspace{5mm}\vbox{\center\footnotesize\tt#1}\vspace{5mm}}

\begin{document}

\thispagestyle{empty}

\phantom{a}
\vskip20mm

\begin{center} 

\title{
Generalized Symmetries and \\
Deformations of Symmetric Product Orbifolds
}

\author{Nathan Benjamin$^{\,a}$, Suzanne Bintanja$^{\,b,\,c}$, Yu-Jui Chen$^{\,a}$, \\ Michael Gutperle$^{\,b}$, Conghuan Luo$^{\,a}$, Dikshant Rathore$^{\,b}$
}

\vskip4.1mm

\begin{minipage}[c]{0.89\textwidth}\centering \footnotesize \em {\it $^{a}$Department of Physics and Astronomy \\
University of Southern California,
Los Angeles, CA 90089, USA}
\\ \vspace{0.5em} 
{\it $^{b}$Mani L. Bhaumik Institute for Theoretical Physics, Department of Physics and Astronomy,
University of California Los Angeles, Los Angeles, CA 90095, USA}\\ 
\vspace{0.5em} 
{\it $^{c}$CERN, Theory Division,
Geneva 23, CH-1211, Switzerland} 
\end{minipage}

\email{nathanbe@usc.edu, sbintanja@physics.ucla.edu, yujuiche@usc.edu, \\ gutperle@ucla.edu, luocongh@usc.edu, drathore@physics.ucla.edu}
\end{center}

\vskip6mm

\begin{abstract}
\noindent We construct generalized symmetries in two-dimensional symmetric product orbifold CFTs $\text{Sym}^N(\mathcal{T}),$ for a generic seed CFT $\mathcal{T}$. 
These symmetries are more general than the universal and maximally symmetric ones previously constructed. We show that, up to one fine-tuned example when the number of copies $N$ equals four, the only symmetries that can be preserved under twisted sector marginal deformations are invertible and maximally symmetric. The results are obtained in two ways. First, using the mathematical machinery of $G$-equivariantization of fusion categories, and second, via the projector construction of topological defect lines. As an application, we classify all preserved symmetries in symmetric product orbifold CFTs with the seed CFT given by any $A$-series $\mathcal{N}=(2,2)$ minimal model. We comment on the implications of our results for holography.

\end{abstract}

\eject

\phantom{a}
\vspace{-5em}
{
\tableofcontents
}

\newpage

\section{Introduction}

The orbifold construction in two-dimensional conformal field theory (CFT) provides a way to obtain new CFTs from old ones \cite{Dixon:1985jw,Dixon:1986qv}. It is used in the construction of perturbative string theory vacua, the replica method, AdS/CFT, and many other areas.
In general, taking an orbifold involves the gauging of a global symmetry $G$ acting on the CFT, which constitutes projecting out states that are not invariant under $G$ and adding twisted sectors to restore modular invariance  \cite{Dijkgraaf:1989hb,Vafa:1989ih}. In this work, we are interested in symmetric product orbifolds, where one takes the tensor product of $N$ copies of a seed CFT and gauges the permutation symmetry, i.e., the symmetric group  $G=S_N$ that permutes the $N$ copies. At finite $N$, symmetric product orbifolds display several notable features, including a second quantized form of the partition function given by the DMVV formula \cite{Dijkgraaf:1996xw}, as well as the covering space method for calculating correlation functions of operators in the twisted sector \cite{Lunin:2000yv,Lunin:2001pw}. In the large-$N$ limit, symmetric product orbifold theories display many of the characteristic features of three-dimensional AdS gravity through the AdS$_3$/CFT$_2$ correspondence. 
For example, these theories exhibit a sparse spectrum \cite{Keller:2011xi,Hartman:2014oaa,Belin:2016yll}, a universal form of the thermal partition function and thermal correlation functions \cite{Keller:2011xi,Belin:2025nqd}, and a form of large-$N$ factorization  \cite{Jevicki:1998bm,Lunin:2000yv,Pakman:2009zz,Belin:2015hwa,Roumpedakis:2018tdb}, where the role of single-trace/single-particle operators in gauge theory is played by operators in twisted sectors labelled by a single-cycle conjugacy class. However, symmetric product orbifold CFTs also display features that are incompatible with a holographic interpretation in terms of a local AdS supergravity dual, such as a universal Hagedorn growth of light states \cite{Hartman:2014oaa,Keller:2011xi,Belin:2014fna}, the existence of an infinite tower of higher spin currents \cite{Baggio:2015jxa,Gaberdiel:2015uca,Gaberdiel:2015mra},
and the associated non-chaotic behavior of out-of-time-ordered correlators \cite{Belin:2017jli}. These properties render symmetric product orbifolds unsuitable as prospective duals to a semiclassical AdS supergravity theory. In some sense, symmetric product orbifolds are as far removed as possible from holographic CFTs --- at least within the class of CFTs that have a well-defined large-$N$ limit, satisfy large-$N$ factorization, and satisfy the sparseness condition of \cite{Hartman:2014oaa}.
Consequently, the dual has to be stringy and involve nonstandard features such as a tensionless limit \cite{Seiberg:1999xz}. A precise duality was established between the symmetric product orbifold of a seed $\mathcal{N}=(4,4)$ superconformal theory with target space ${\cal M}_4$ given by $\mathbb{T}^4$ or K3, and type IIB string theory on $AdS_3\times S^3\times {\cal M}_4$ with one unit of NS-NS three-form flux \cite{Gaberdiel:2017oqg,Gaberdiel:2018rqv,Eberhardt:2018ouy,Giribet:2018ada,Eberhardt:2019ywk}. One possible way to obtain a CFT which is dual to a semiclassical supergravity theory from the symmetric product orbifold CFT is to deform it away from the symmetric product orbifold point via a marginal deformation in the twisted sector (which exists in many symmetric product orbifolds of superconformal seed theories) \cite{Avery:2010er,David:1999ec,David:2002wn}.  This deformation 
lifts higher spin currents and drives the CFT to a strongly coupled point in its moduli space. From the string theory perspective, turning on the deformation increases the string tension, thereby lifting the stringy states from the spectrum.

Generalized and non-invertible symmetries in quantum field theories have become an active area of research in recent years.\footnote{See \cite{Shao:2023gho,Bhardwaj:2023kri,Schafer-Nameki:2023jdn,Kong:2022cpy} for some reviews and lectures with additional references.} They extend the notion of ordinary symmetries by interpreting symmetries as topological (extended) operators in spacetime (in Euclidean signature) that commute with the stress tensor, and provide a much richer structure of selection rules in QFTs. In addition, they also generalize the form of 't Hooft anomaly, which brings forth strong constraints on the IR phases along the RG flow. While the first examples of non-invertible symmetries were constructed in two-dimensional CFTs \cite{Verlinde:1988sn,Petkova:2000ip}, these ideas have since been generalized to higher-dimensional QFTs, higher form symmetries\cite{Gaiotto:2014kfa}, and have been formalized using category theory\cite{Fuchs:2002cm,Bhardwaj:2017xup,Chang:2018iay,Moore:1988qv,Moore:1989vd}. In this paper, we focus on two-dimensional QFT, where the 0-form global symmetries are described by fusion categories\cite{Etingof:2002vpd}.

Global symmetries of QFTs can be described by a topological field theory in one spacetime dimension higher, also known as the Symmetry Topological Field Theory (SymTFT)\cite{Apruzzi:2021nmk,Freed:2012bs,Freed:2022qnc}. In the context of holography, in some cases, the SymTFT can be obtained from supergravity by direct construction, see e.g. \cite{GarciaEtxebarria:2022vzq,Apruzzi:2022rei}. 
Physically, the topological operators in the boundary CFT should become dynamical branes when pulled into the bulk according to the ``no global symmetry conjecture'', which link nontrivially with Wilson surfaces that carry nontrivial representations and end on the boundary \cite{Harlow:2018tng,Banks:2010zn,Heckman:2024oot}. In the setup of ${\rm AdS}_3/{\rm CFT}_2$, the topological sector in the bulk is mathematically described by a modular tensor category known as the Drinfeld center (or its continuous generalization\cite{Brennan:2024fgj}). For a physical presentation, see e.g. \cite{Lin:2022dhv} and references therein. As a result, we expect the existence of global symmetries on the boundary to impose constraints on bulk dynamics. For instance, they provide nontrivial selection rules for correlation functions of (quasi-local) operator insertions with nontrivial gauge charges. Since symmetric product orbifold CFTs exhibit many desirable properties in the context of holography and are useful in the search for holographic CFTs, it is essential to analyze their generalized symmetry structure. Furthermore, one should understand what happens to this symmetry structure under deformations that potentially give rise to holographic CFTs, as well as its consequences for the putative gravitational duals. This work aims to leverage the power of the mathematical description of generalized symmetries to study these questions.

We want to emphasize that the results of this work apply to any symmetric product orbifold theory, as well as permutation orbifold theories, whether or not it is related to a holographic CFT. Moreover, the results are valid not just in the large-$N$ limit, but also at any finite value of $N$.

\subsection*{Summary of results}

The first goal of this paper is to construct the generalized symmetries of symmetric product orbifold theories. More precisely, we provide a systematic description of all the global symmetries in the symmetric product orbifolds that originate from the seed theory via the discrete gauging procedure. In Sec.~\ref{sec:review} we review generalized symmetries in two-dimensional QFTs. Zero-form symmetries are viewed as topological defect lines (TDLs), which are described mathematically by a fusion category. Since the orbifold procedure can be understood as gauging by a discrete group $G$, we review this gauging procedure using the language of algebra objects, and describe TDLs in the gauged theory in terms of $A$-bimodules. The action of TDLs on local operators can be systematically derived, given the information from the seed theory. The categorical realization of global symmetries in a symmetric product orbifold goes under the name of $G$-equivariantization \cite{Drinfeld:2009nez}.  This mathematical framework, in terms of fusion categories, allows us to efficiently and cleanly describe the generalized symmetries of symmetric product orbifolds.
  
 In Sec.~\ref{sec:symn}, we apply the machinery of the previous section to construct the generalized symmetries in symmetric product orbifolds explicitly. After a brief review of symmetric product orbifolds, including the spectrum of local operators and partition functions, we construct the TDLs in Sec.~\ref{sec:symNtdls}. The first class of TDLs contains the so-called universal defects \cite{Gutperle:2024vyp}, which depend only on the representations of the symmetric group $S_N$ and are independent of the details of the seed CFT. We then construct non-universal defects built using nontrivial TDLs of the seed theory, generalizing the construction of \cite{Gutperle:2024vyp}  to an arbitrary set of seed TDLs. In addition, we provide formulas for the action of these TDLs on local operators that will be important when we consider deformations of symmetric product orbifolds in the following section. Throughout this analysis, we employ both abstract mathematical language and provide explicit examples, particularly when the seed theory is chosen to be rational, for which the actions of TDLs can be written in terms of the modular $S$-matrix of the seed RCFT. As a caveat, we note that our construction should not be viewed as a complete description of all global symmetries in symmetric product orbifolds, even if we consider as input all global symmetries of the seed theory. Generically, we cannot rule out the possibility of (additional) emergent symmetries, which we discuss briefly in Sec.\,\ref{sec:symNtdls}.
 
 The second goal of the paper is to determine which, if any, of the generalized symmetries survive under exactly marginal deformations of the symmetric product orbifold CFT.\footnote{More precisely, we do not consider exactly marginal deformations induced purely by the seed theory, i.e., marginal deformations in the untwisted sector, which are less interesting in the context of holography.} The condition for a TDL to be preserved is that the local deformation operator commutes with the TDL associated with the symmetry. In Sec.~\ref{sec:4}, we check this condition using general arguments. Again, we derive the statements using the abstract formulation and illustrate them with the concrete construction using RCFTs. Our main result is given in Theorem \ref{thm:4}, which states that the only TDLs that commute with a local operator in the twisted sector are of a very special form: they are built from a single invertible TDL in the seed CFT and are maximally symmetric.\footnote{In \cite{Gaberdiel:2021kkp,Knighton:2024noc,Gutperle:2024vyp} the term maximally fractional was used, in the following we will use maximally symmetric instead.} Consequently, we prove that all non-invertible symmetries are broken by an exactly marginal deformation.\footnote{We note that there is an exception when $N=4$. A maximally symmetric non-invertible TDL associated to the representation $R=(2,2)$ of $S_4$ can survive. We discuss this in more detail in Sec.~\ref{sec:4}.} In addition, we classify all possible preserved TDLs in the case where the seed CFT is an $A$-series $\cN=(2,2)$ minimal model. For such seed CFTs, the existence of exactness of certain marginal deformations is guaranteed by supersymmetry, and due to their rationality, the marginal operators as well as the TDLs in the seed theory are known. 
 We close this section with a brief discussion on the implications of our results for holographic theories related to large-$N$ limits of symmetric product orbifolds. 
 
 We conclude with a discussion of our results as well as some directions for future research in Sec.~\ref{sec:discussion} and relegate some technical details, calculations, and conventions to the appendices.

\section{Non-invertible symmetries and their gauging}
\label{sec:review}

In this section, we review salient properties of non-invertible global symmetries in QFT. We 
adopt the modern perspective of generalized symmetries, in which a global symmetry is associated with topological defects\cite{Gaiotto:2014kfa}. Specifically, a $p$-form symmetry in $d$-dimensional spacetime corresponds to a codimension-$(p+1)$ topological operator. This leads to (at minimum) two new perspectives on symmetries. First, in a Lorentzian QFT, this framework unifies the notions of symmetry operators (which commute with the Hamiltonian) acting on states in a Hilbert space supported on a spatial slice with topological defects that can be inserted along the time direction and construct twisted Hilbert spaces. These two notions are encoded by the same set of topological data.\footnote{There are exotic exceptions to this interpretation. One example is a $(d-1)$-form symmetry that corresponds to a topological local operator, and it makes no sense to ``insert'' it along the time direction and create a twisted Hilbert space. Another example is a $(-1)$-form symmetry that should probably be interpreted as a generalization of the $\theta$ angle.} Second, in Euclidean QFT, it allows for the exploitation of locality and study of (the analogue of) OPEs for extended operators. This gives constraints for bootstrapping extended operators, which in turn can constrain the allowed sets of topological data.

From now on, we specialize to two-dimensional QFTs with $p=0$, where global symmetries are described by topological defect lines (TDLs). A well-known class of topological defects is one that appears in the context of rational conformal field theories (RCFT), originally studied in \cite{Petkova:2000ip,Frohlich:2006ch,Petkova:2009pe,Petkova:2013yoa}. These defects are examples of global symmetries that are not necessarily equipped with a group multiplication law, hence the name \emph{non-invertible} (see e.g. \cite{Bhardwaj:2017xup,Chang:2018iay,Shao:2023gho,Kitaev:2005hzj} for reviews). The corresponding mathematical structure that describes this family of global symmetries is known as a tensor category; for a comprehensive mathematical review, see \cite{etingof2015tensor}.

\subsection{Two-dimensional global symmetries as fusion categories}
\label{subsec:fusion_category}

Mathematically, TDLs in two-dimensional QFTs are described by fusion categories. 
Instead of directly stating the definition of a fusion category, we will summarize its salient properties and provide some physical interpretation. Throughout the rest of this work, we consider unitary compact QFTs with a unique vacuum. 

\paragraph{Simple objects and fusion rules.}

A TDL $\cL$ naturally induces a twisted Hilbert space $\cH_{\cL}$ on $S^1$ by inserting it perpendicular to the circle. The direct sum of two TDLs $\cL_1 \oplus \cL_2$ is defined in terms of the direct sum of their twisted Hilbert spaces $\cH_{\cL_1} \oplus \cH_{\cL_2}$, or equivalently, the sum of partition functions with corresponding TDL insertions $Z(\cL_1 \oplus \cL_2) = Z(\cL_1) + Z(\cL_2)$. We call a TDL \emph{simple} if $\cH_{\cL}$ is indecomposable, i.e., a simple TDL cannot be written as a direct sum of multiple TDLs. In what follows, we denote the set of simple objects of a category $\cC$ as ${\rm Irr}(\cC)$. With this interpretation, objects in $\cC$ can only be constructed by a linear combination of simple objects with non-negative integer coefficients. 

Furthermore, given two parallel TDLs, we can fuse them and obtain a new TDL $\cL_1 \otimes \cL_2$. The resulting TDL is, in general, not simple. However, we will work under the assumption that all TDLs can always be written as a direct sum of a finite number of simple TDLs, which we refer to as semi-simplicity.\footnote{Semi-simplicity is true for all Verlinde lines in RCFTs, which are important examples of TDLs we use in this work. However, note that semisimplicity does not always hold, even for TDLs in RCFT. Examples of TDLs that are not semisimple include the non-compact lines of \cite{Chang:2020imq}, and TDLs in logarithmic CFTs \cite{Gaberdiel:2001tr}.} 
Due to the interpretation in terms of a (twisted) Hilbert space, the fusion coefficients must be non-negative integers. In other words, if we consider a set of TDLs that is closed under fusion, denoted with $\cC$, the fusion rules must form a $\mZ^+$-ring, called the fusion ring, with fusion rules
\ie \label{eq:tdlfusion}
\cL_i \otimes \cL_j = \oplus_{k\in {\rm Irr}(\cC)} N_{ij}^k \cL_k, \quad N_{ij}^k \in \mZ_{\geq 0} \,.
\fe
The term non-invertible comes from the fact that, given an arbitrary TDL $\cL$, generically there is no other TDL $\cL'$ such that $\cL \otimes \cL' = \id$. Hence, $\cL$ has no inverse.

\paragraph{Morphisms.}

\begin{figure}[t!]
    \centering
    \begin{subfigure}[t]{0.5\textwidth}
        \centering
        \includegraphics[height=45mm]{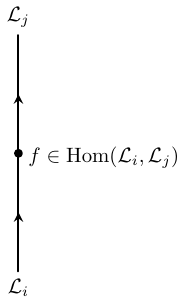}
        \caption{\quad\quad\quad}
        \label{subfig:2pt_junction}
    \end{subfigure}
    \begin{subfigure}[t]{0.5\textwidth}
        \centering
        \includegraphics[height=38mm]{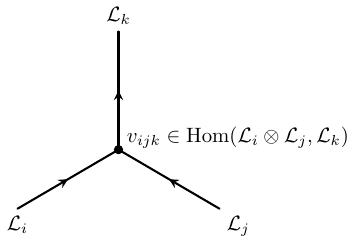}
    \caption{\quad\quad\quad\quad\quad\quad}
    \label{subfig:3pt_junction}
    \end{subfigure}
    \caption{Two- and three-point junctions.}
    \label{fig:2&3pt_junction}
\end{figure}

In the framework of category theory, we can package the data of TDLs efficiently. In the language of categories, TDLs, and their associated defect Hilbert spaces, should be considered as the  ``objects'' of a category $\cC$. Given two TDLs $\cL_i$, $\cL_j$, one can consider isomorphisms between their associated twisted Hilbert spaces 
$\cH_{\cL_i}$, $\cH_{\cL_j}$ (these do not necessarily exist). These isomorphisms are called the morphisms of the category $\cC$, denoted as $\Hom_{\cC}(\cL_i, \cL_j)$. The dimension of the Hom space, as a linear vector space on $\mC$, counts the number of linearly independent topological point operators one can insert that interpolate between the two TDLs $\cL_i$ and $\cL_j$, as shown in Figure~\ref{subfig:2pt_junction}.
For simple objects denoted by $\cL_i$, we have $\Hom_{\cC}(\cL_i, \cL_i) \cong \mC$. This represents the freedom of a total rescaling when defining an indecomposable Hilbert space. On the other hand, for simple objects $\cL_i\neq \cL_j$, we have $\Hom_{\cC}(\cL_i, \cL_j) = \varnothing$. Moreover, for a generic fusion channel we have $\Hom_{\cC}(\cL_i \otimes \cL_j, \cL_k) \cong \mC^{N_{ij}^k}$. In Euclidean signature, the fusion coefficients count the number of linearly independent topological point operators one can insert on the three-way junctions of TDLs, as shown in Figure~\ref{subfig:3pt_junction}.\footnote{For comparison of notation, \cite{Chang:2018iay} defined a completely analogous notion called a junction vector, as the weight-$(0,0)$ state on the defect Hilbert space $\cH_{\cL_i \otimes \cL_j \otimes \bar\cL_k}$. This notion is isomorphic to the morphism structures $\Hom_{\cC}(\cL_i \otimes \cL_j, \cL_k)$ discussed here. }

\paragraph{Identity line and dual lines.}

In the context of a QFT, there are a few more physical properties that the set of simple TDLs ${\rm Irr}(\cC)$ has to satisfy. First, there exists a trivial (simple) TDL $\id$, that fuses trivially with all TDLs: $\id \otimes \cL = \cL \otimes \id = \cL$. Furthermore, given a TDL $\cL$, there exists the orientation reversal of $\cL$, which is a TDL denoted by $\bar\cL$. The fusion of $\bar\cL$ satisfies $\cL \otimes \bar\cL = \id \oplus \cdots$. $\bar\cL$ is called the dual object to $\cL$. 
A category $\cC$ that satisfies all the above assumptions is called a fusion category. Note that mathematically, a fusion category satisfies an extra assumption, stating that ${\rm Irr}(\cC)$ is a finite set, so that every object in $\cC$ can be decomposed into a finite basis. We abuse the notion ``fusion category'' to denote a semisimple tensor category, so that our results generalize to continuous symmetries that satisfy the semisimplicity assumption. 

\paragraph{Hermitian structure.}

Because the QFTs we consider are unitary, we focus on unitary fusion categories, which can be endowed with a Hermitian structure
\ie
\dagger: \Hom_{\cC}(\cL_i, \cL_j) \to \Hom_{\cC}(\cL_j, \cL_i) \,,
\fe
such that the composition $f^{\dagger} \circ f,$ with $ f \in \Hom_{\cC}(\cL_i, \cL_j),$ is positive semi-definite in $\Hom_{\cC}(\cL_i, \cL_i)$. In other words, we can choose a unitary gauge, such that an arbitrary morphism $f \in \Hom_{\cC}(\cL, \cL')$ can be represented by a unitary matrix $U$ when expanded in the basis of simple objects, and $f^{\dagger}$ becomes $U^{\dagger}$.

\paragraph{F symbols and the generalized 't Hooft anomaly.}

The topological data discussed so far is not enough to fully specify a fusion category $\cC$. In particular, there are two ways of fusing three TDLs $\cL_i \otimes \cL_j \otimes \cL_k$. These two distinct ways to fuse three TDLs can be related to one another via a basis transformation known as an F move. Such basis transformations are encoded by what are known as F symbols. 
They can be understood at the level of correlation functions as follows. The most general form of a correlation function on a two-dimensional manifold involves an insertion of a network of TDLs. This correlation function cannot change as the TDLs are deformed (up to possible isotopy anomalies \cite{Chang:2018iay}), as long as they do not cross each other or any local operators. F moves, however, are allowed, and constitute a local operation illustrated by the following equation
 \begin{equation}\label{eq:Fmove}
    \vcenter{\hbox{\includegraphics[width=10cm]{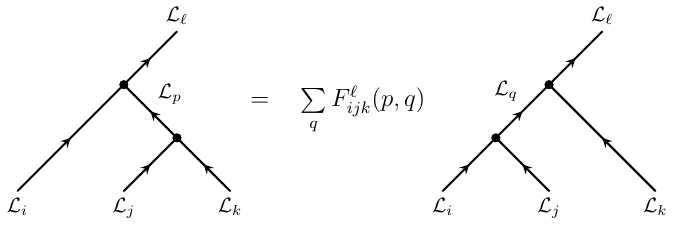}}}\,.
  \end{equation}
Here, the coefficients $F^\ell_{ijk}(p,q)$ are the F symbols associated with the F move, and we have simplified the expression by omitting the indices labeling three-way junction vectors. 

The F symbols have to satisfy additional consistency conditions called the pentagon equations (see e.g. \cite{Chang:2018iay,Choi:2023xjw} for details).
In general, given a fusion ring, there can be multiple or no solutions to the pentagon equations. Hence, to fully specify a fusion category $\cC$, one has to pick a set of consistent F symbols. F symbols are the generalization of 't Hooft anomalies of invertible global symmetries. This is guaranteed by a mathematical theorem named Ocneanu rigidity\cite{Etingof:2002vpd}, stating that the solutions to pentagon equations are discrete and admit no smooth deformations. Therefore, they must be preserved under RG flows. One can also check that the definition of F symbols coincides with the $H^3(G,U(1))$ classification of the 't Hooft anomaly of a symmetry group $G$.

\subsection{Gauging via an algebra object}
\label{subsec:gauging}

Next, we consider orbifolding by a discrete group $G$, i.e., gauging a finite group, in the language of fusion categories. Of course, orbifolding by discrete groups has been studied in great detail (see e.g. \cite{DiFrancesco:1997nk,Polchinski:1998rq}) previously. The interpretation in terms of states on $S^1$ is as follows. We include all the twisted sectors obtained by adding a $g\in G$ holonomy along the circle, and project out states that are charged under the symmetry $G$. Both holonomy and symmetry operators can be implemented by inserting TDLs; they are simplicial cohomology classes valued on $G$. Hence, we can write the gauging procedure as follows
\ie\label{eq:suma}
Z_{\rm orb} = \sum_{a \in H^1(\cM, G)} Z(a) \,,
\fe
where $a$ is the background gauge connection of $G$, and $\cM$ is the spacetime manifold on which we consider the theory. As an example, consider  $\cM = T^2$ and $G = \mathbb{Z}_2$ generated by an invertible TDL $\eta$ such that $\eta^2 = 1.$ In this example, the sum over $a$ in \eqref{eq:suma} amounts to summing over $\mathbb{Z}_2$ connections and projecting onto $\mathbb{Z}_2$-invariant states. This is done by inserting $\eta$ defect lines along the two non-contractible cycles of the torus, giving us the partition function for the gauged theory
\ie
Z_{\rm orb}(\tau) = \frac{1}{2}\Big(\;
\vcenter{\hbox{\begin{tikzpicture}[scale=0.85]
  \draw[line width=0.45pt] (0,0) rectangle (1,1);
\end{tikzpicture}}}
\;+\;
\vcenter{\hbox{\begin{tikzpicture}[scale=0.85]
  \draw[line width=0.45pt] (0,0) rectangle (1,1);
  \draw[line width=0.8pt] (0.5,0) -- (0.5,1);
  \node[anchor=west] at (0.5,0.5) {$\eta$};
\end{tikzpicture}}}
\;+\;
\vcenter{\hbox{\begin{tikzpicture}[scale=0.85]
  \draw[line width=0.45pt] (0,0) rectangle (1,1);
  \draw[line width=0.8pt] (0,0.5) -- (1,0.5);
  \node[anchor=south] at (0.5,0.47) {$\eta$};
\end{tikzpicture}}}
\;+\;
\vcenter{\hbox{\begin{tikzpicture}[scale=0.85]
  \draw[line width=0.45pt] (0,0) rectangle (1,1);
  \draw[line width=0.8pt] (0.5,0) -- (0.5,1);
  \draw[line width=0.8pt] (0,0.5) -- (1,0.5);
\end{tikzpicture}}}
\; \Big).
\fe
If $G$ is Abelian, it is straightforward to add the dependence of the background gauge connection of the quantum symmetry on the left-hand side, such that the gauging process looks like a Fourier transform of cohomology groups\footnote{We have omitted a normalization factor in the phase that depends on $G$. } 
\ie
Z_{\rm orb}(b) = \sum_{a \in H^1(\cM, G)} Z(a) e^{i\int a \cup b} \,,
\label{gauging}
\fe
where we denote the background gauge connection of the quantum symmetry $\hat{G}\cong H^1(G,U(1))$ as $b$. It follows that the defect partition functions of the gauged theory (partition functions with topological line insertions) can be computed from the defect partition functions of the original theory, as well as the action of dual TDLs on local operators.\footnote{Here, and in what follows, we use the term ``dual TDLs'' to denote TDLs in the gauged theory, which are not to be confused with the notion ``dual objects'' in the fusion category that we introduced in the previous subsection. }

The language of cohomology groups works nicely in the context of gauging an Abelian group. However, cohomology groups are not directly applicable in the context of non-invertible symmetries. In fact, this already happens when one considers gauging a non-Abelian group $G$. When $G$ is non-Abelian, its corresponding Wilson lines are not invertible. To systematically determine the defect partition functions with non-invertible TDL insertions, we thus need a new framework to describe the gauged theory. This is the framework of fusion categories we introduced in Sec.\,\ref{subsec:fusion_category}.

\begin{figure}[htbp]
    \centering
     \begin{subfigure}[t]{0.3\textwidth}
        \centering
        \includegraphics[height=50mm]{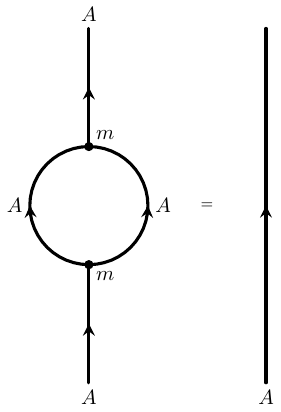}
    \caption{}
    \label{subfig:separability}
    \end{subfigure}
    \quad\quad\quad\quad\quad\quad
    \begin{subfigure}[t]{0.3\textwidth}
        \centering
        \includegraphics[height=22mm]{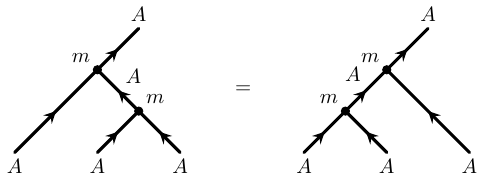}
        \caption{}
        \label{subfig:A_associativity}
    \end{subfigure}
    
    \begin{subfigure}[t]{0.3\textwidth}
        \centering
        \includegraphics[height=25mm]{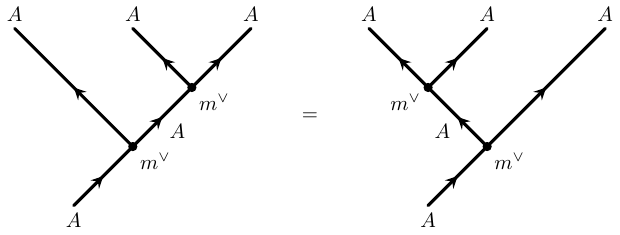}
    \caption{}
    \label{subfig:A_associativity_dual}
    \end{subfigure}
    \quad\quad\quad\quad\quad\quad\quad\quad\quad
    \begin{subfigure}[t]{0.3\textwidth}
        \centering
        \includegraphics[height=30mm]{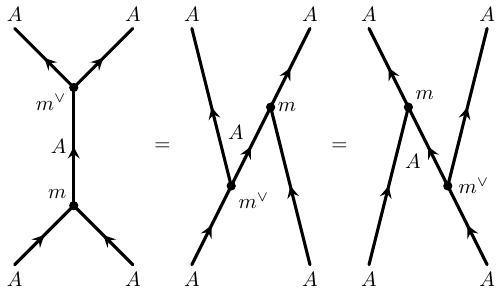}
    \caption{}
    \label{subfig:Frobenius}
    \end{subfigure}
   
    \caption{The separability (a), associativity (b), coassociativity (c), and Frobenius condition (d) of an algebra object $(A, m, m^{\vee})$. At every junction in the diagrams, we simply apply the map $m$ or $m^\vee$ as appropriate.}
    \label{fig:A_consistent}
\end{figure}

In the framework of generalized symmetries in terms of fusion categories, the notion of discrete gauging has been generalized accordingly \cite{Bhardwaj:2017xup}. Discrete gauging is described by a symmetric special separable Frobenius algebra object, which we call an algebra object for short. For a precise mathematical definition, see also \cite{Fuchs:2002cm,Kong:2013aya,Diatlyk:2023fwf}. Roughly speaking, an algebra object is defined by a triplet $(A,m,m^{\vee})$, where $A \in\cC$, and $m\in \Hom(A\otimes A, A)$, $m^{\vee} \in \Hom(A, A \otimes A)$ are called a multiplication map and a comultiplication map respectively. Furthermore, $m$ and $m^\vee$ satisfy the consistency conditions illustrated in Figure \ref{fig:A_consistent}.\footnote{Technically speaking, one also needs to introduce a unit $u\in \Hom(\id, A)$ and a counit $u^{\vee}\in \Hom(A, \id)$ that satisfy certain consistency conditions. They are important for fixing the overall normalization of the multiplication and comultiplication maps. However, we omit them in the main text discussions for brevity.}

\begin{figure}[htbp]
    \centering
        \includegraphics[height=40mm]{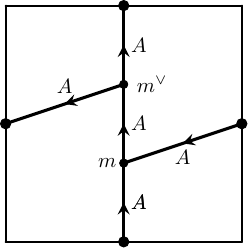}
    \caption{Gauging procedure from an algebra object.}
    \label{fig:gauging}
\end{figure}

The procedure of gauging an algebra object is given by stacking a dual triangulation network of $A$ on the spacetime manifold $M$, as depicted in Figure~\ref{fig:gauging}.
Intuitively, the consistency condition for an algebra object states that the gauging does not depend on the shape of dual triangulations, and bubbles of $A$ on the network can be freely removed. Furthermore, we note that for unitary theories, one can always choose a gauge for $A$ such that $m^{\vee} = m^{\dagger}$. Therefore, we will only consider multiplication maps $m$ in the rest of this work. 

As a simple example, which will be important in this paper, we consider gauging a finite (possibly non-Abelian) group $G$. In this example, we have $A = \bigoplus_{g\in G} \cL_g$, where $\cL_g$ denotes the TDL labeled by 
group element $g$. To specify the multiplication map $m$, one needs to specify linear maps $m(g,h):\mathbb{C}\rightarrow \mC$, one for each  $\mathcal{L}_g\otimes\mathcal{L}_h \to \mathcal{L}_{gh}$.
In this case, the maps are thus labeled by $G \times G$. It is further restricted to be a 2-cocycle in $H^2(G, U(1))$ by imposing the consistency conditions in Figure~\ref{fig:A_consistent}, as well as modding out gauge degrees of freedom. 
The multiplication map $m$ characterizes a choice of discrete torsion, and when $G$ is Abelian, it is equivalent to a choice of SPT phases $i\int a \cup a$ that can be added to the action in~\eqref{gauging}. 
In the QFT context, it can also be viewed as a counter-term that describes an invertible TQFT.
Since in this work our main focus is on gauging a finite group, which is a drastic simplification, we will not describe the most general form of algebra objects relevant to non-invertible gauging.

\subsection[Dual TDLs as \texorpdfstring{$A$}{A}-bimodules]{Dual TDLs as \texorpdfstring{$\boldsymbol{A}$}{A}-bimodules}
\label{subsec:A-bimodules}

\begin{figure}[htbp]
    \centering
     \begin{subfigure}[t]{0.3\textwidth}
        \centering
        \includegraphics[height=30mm]{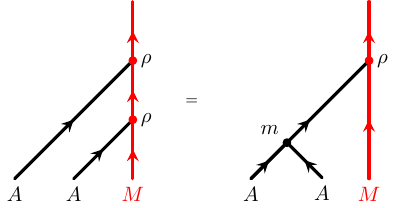}
    \caption{}
    \label{subfig:L_A_module}
    \end{subfigure}
    \quad\quad\quad\quad\quad
    \begin{subfigure}[t]{0.3\textwidth}
        \centering
        \includegraphics[height=30mm]{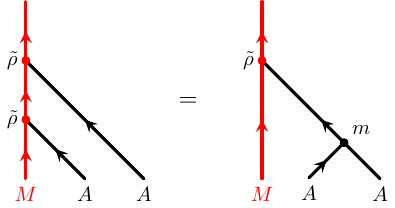}
        \caption{}
        \label{subfig:R_A_module}
    \end{subfigure}
    
    \begin{subfigure}[t]{0.3\textwidth}
        \centering
        \includegraphics[height=30mm]{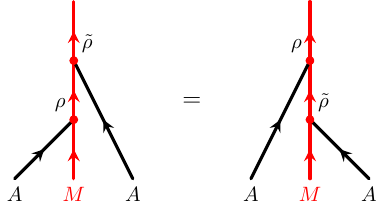}
    \caption{}
    \label{subfig:AA_module}
    \end{subfigure}

    \caption{Compatibility conditions of $A$-bimodules. The red line denoted by $M$ is the $A$-bimodule, and the black lines denoted by $A$ are the algebra objects. }
    \label{fig:bimodule_compatibility}
\end{figure}

The TDLs in the gauged theory, which we call dual TDLs, are described by $A$-bimodules, which are reviewed in, e.g., \cite{Fuchs:2002cm, Bhardwaj:2017xup}. $A$-bimodules are defined by a triplet $(M, \rho, \tilde\rho)$ which satisfies the compatibility conditions with the algebra object $A$ illustrated in Figure\,\ref{fig:bimodule_compatibility}. Note that $M \in \cC$, and we call $\rho \in \Hom(A\otimes M, M)$, and $\tilde\rho \in \Hom(M \otimes A, M)$ the left- resp. right action of $A$. Physically, the $A$-bimodules correspond to topological defects originating from $\cC$, which can move freely within the mesh of $A$.

As a simple example, when we gauge a discrete group $G$, we have $A = \bigoplus_{g\in G} \cL_g$ and $M = n\bigoplus_{g\in G} \cL_g$, where $n$ is a positive integer left to be determined. We can gauge fix the left action $\rho$ to be $n \times n$ identity matrices, and the solutions (to the equations illustrated in Figure~\ref{fig:bimodule_compatibility}) of the right action $\tilde\rho$ must be in one-to-one correspondence with representations of $G$, according to the right associativity condition. More concretely, $\tilde\rho_{h,g} = R(g)$, where $R$ is an $n$-dimensional representation of $G$. One can further prove that the fusion rules and F symbols are given by the tensor product and 6j symbols of representations of $G$, so the dual fusion category is Rep$(G)$.

\paragraph{Actions of dual TDLs on local operators}

\begin{figure}[htbp]
    \centering
        \includegraphics[height=20mm]{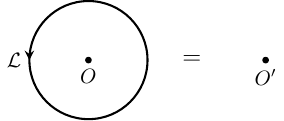}
    \caption{Linking action on a plane. }
    \label{fig:circle_L}
\end{figure}

\begin{figure}[htbp]
    \centering
        \includegraphics[width = 0.9\textwidth]{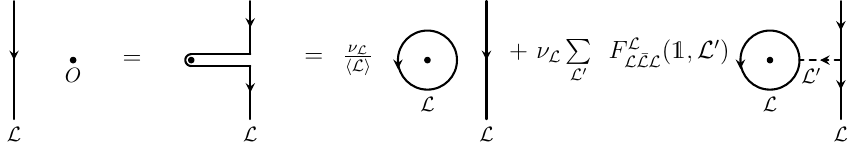}
    \caption{Passing action on a plane. The coefficients in the sum on the right-hand side are given by the F symbols introduced in \eqref{eq:Fmove}. }
    \label{fig:lasso}
\end{figure}

In this paper, we mainly focus on the linking action of TDLs on a local operator, depicted in Figure~\ref{fig:circle_L}. The TDL can be shrunk towards the local operator $O$, thereby acting on $O$ to produce another local operator---or annihilating it. 
Schematically, one can write
\ie
\cL \cdot O = O' \,.
\fe
Generally, a symmetry operator can map a local operator to a non-local operator, but the non-local action is not captured by the linking action. Instead, it is captured by the passing action shown in Figure~\ref{fig:lasso}, which can also be organized using the tube algebra\cite{ocneanu1994chirality,Lin:2022dhv}.

\begin{figure}[htbp]
    \centering
        \includegraphics[height=40mm]{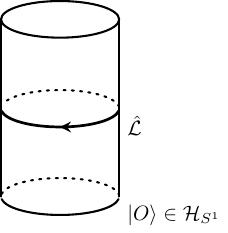}
    \caption{TDLs as linear operators on a cylinder.}
    \label{fig:cynlinder_L}
\end{figure}

In a CFT, there is an equivalent description of the linking action through the state-operator correspondence. This description is in terms of the action of TDLs winding around a cylinder, acting on the states living in the Hilbert space $\cH_{S^1}$. One can treat $\hat\cL$ 
as a linear operator acting on the Hilbert space (this is what we mean when we add a hat to a TDL $\cL$), as depicted in Figure\,\ref{fig:cynlinder_L}. More generally, information of the tube algebra is captured by the cylinder partition function with the most general TDL insertions around the $S^1$. 

Depending on the gauge choice of isotopy anomaly (see e.g. \cite{Chang:2018iay}), the linking action on a plane and on a cylinder can differ by a phase. Here we choose the convention where for a non-self-dual TDL $\cL$, they are identical, while for a self-dual TDL $\cL$, they differ by the Frobenius-Schur indicator $\nu_{\cL} = \pm 1$. One can define their vacuum expectation value on a plane $\la\cL\ra_{\mR^2}$ and on a cylinder 
\begin{equation} \la \cL \ra \coloneqq \la 0 | \hat{\cL} | 0 \ra\,.
\label{eq:quantum_dim}
\end{equation}
We thus have
\ie
\la\cL\ra_{\mR^2} = \nu_{\cL} \la \cL \ra \,.
\fe
The cylinder vacuum expectation value $\la \cL \ra$ is also known as the quantum dimension of $\cL$. In the following, we will, by default, refer to the linking action on a cylinder. For example, we have $\cL \cdot 1 = \la \cL \ra 1$. From Figure~\ref{fig:lasso} we also note that the local part of the passing action on $O$ differs from the linking action by a factor of $\la \cL \ra$. 

Notably, the information of the local linking action on local operators is also included in the torus partition function with a TDL inserted along the spatial cycle
\ie
Z^\cL \coloneqq \Tr \hat\cL \; q^{L_0 - \frac{c}{24}} \bar{q}^{\bar{L}_0 - \frac{c}{24}} = \sum_{O} \la O | \hat\cL | O \ra \chi_{O} \bar\chi_{O} \,,
\fe
where the summation is over all the Virasoro primaries, and $\chi_O$ is the Virasoro character of $O$. 
A local operator $O$ will not contribute if $\hat\cL|O\ra$ is a state belonging to some defect Hilbert space.  

\begin{figure}[htbp]
    \centering
        \includegraphics[height=30mm]{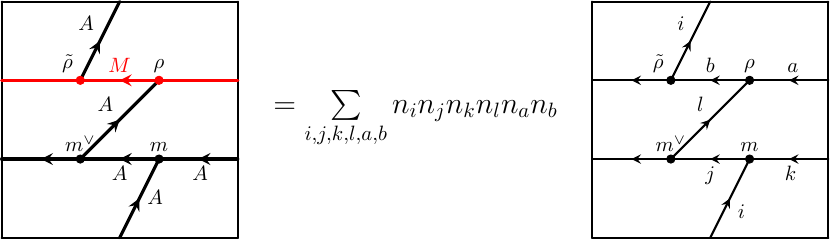}
    \caption{The action of dual TDLs in the gauged theory in terms of actions of TDLs in the original theory. On the right-hand side, the coefficients $n_i,n_j,n_k,n_l$ are due to the expansion of $A$ into simple objects of $\cC$, while $n_a,n_b$ are the coefficients of expanding $M$. }
    \label{fig:bimodule_insertion}
\end{figure}

Analogously, one can write the cylinder partition function with TDL insertions in the gauged theory. Moreover, using the language of $A$-bimodules, one can write it explicitly in terms of the partition functions of the ungauged theory, as shown in Figure~\ref{fig:bimodule_insertion}. This relation implies that the TDL actions on local operators in the gauged theory are completely determined by the TDL actions in the ungauged theory. To explicitly work out the dictionary, we only need to solve the corresponding algebra object and bimodule structures.\footnote{For the most general form of defect partition functions, we also need to solve the three-way junction of $A$-bimodules. In this paper, we do not consider this.} We provide a more detailed derivation of the action from this point of view in App.~\ref{app:action}. 

We once more return to the example of gauging a discrete group $G$. The dual TDLs are labeled by representations of $G$. On the other hand, local operators in the gauged theory are labeled by conjugacy classes of $G$. Colloquially, operators labeled by a non-trivial conjugacy class are called twisted operators. A local operator labeled by $[g]$ is represented by the part of the partition function where we have $\cL_g \subset A$ going around the temporal cycle, along with an $A$ projector along the spatial cycle. The loop action of a dual TDL labeled by $R$ can be obtained by inserting an $A$-bimodule labeled by $R$ along the spatial cycle. After fixing the normalization where $\cL_R |0\ra = \la \cL_R \ra |0\ra$, we have
\begin{equation}
    \cL_R \cdot \phi_{[g]} = \Tr [\rho_R(g) ] \ \phi_{[g]}= \chi_R(g) \ \phi_{[g]} \,.
    \label{repG_action}
\end{equation}
In this case, because we have gauged the entire $G$ symmetry, the original $G$ action completely drops out.\footnote{According to the stabilizer-orbit theorem, we can choose a unified normalization for all the actions. Note that this needs to be true in order for the bimodule conditions to be satisfied.}

One should note that in general, we are mostly interested in gauging a subsymmetry, represented by an algebra object $A$, that is embedded (not necessarily trivially) in a larger symmetry. The $A$-bimodule actions only capture the action on local operators of the dual subsymmetry. To fully describe the actions of all TDLs, we need to incorporate the information of the tube algebra of the original theory. We will see explicit examples of this when we discuss TDLs in symmetric product orbifolds. 

\subsection[\texorpdfstring{$G$}{G}-equivariantization of fusion categories]{\texorpdfstring{$\boldsymbol{G}$}{G}-equivariantization of fusion categories}
\label{subsec:equivariantization}

We consider symmetric product orbifolds, as well as their subgroup cousins called permutation orbifolds. The most generic form of a global symmetry in these theories is described by a special class of fusion categories, known as the $G$-equivariantization of fusion categories, where we take $G$ to be the symmetric group $S_N$ or one of its subgroups.\footnote{Roughly speaking, this is the analogue of central group extensions except that objects in the exact sequence are replaced by fusion categories. More specifically, it is defined by the following central exact sequence of fusion categories:
\ie
{\rm Rep}(G) \to \cC^G \to \cC \,.
\fe
} 
The use of $G$-equivariantization turns out to be very convenient, and provides a simple way to understand and manipulate the symmetries in symmetric product orbifolds. 
Therefore, in this section, we give a minimal presentation of all the ingredients necessary, including the simple objects, fusion rules, and bimodule structures, which are used to describe TDL actions on local operators. More technical details are deferred to App.~\ref{app:equivariantization}. 

$G$-equivariantization has been studied extensively in the math literature \cite{Drinfeld:2009nez,burciu2013fusion} and is important to classify braided group-theoretical fusion categories\cite{naidu2009fusion,naidu2010crossed}. One famous example is the Drinfeld center of ${\rm Vec}_G^{\omega}$ where $G$ is a finite group, which describes anyons in the 3D Dijkgraaf-Witten gauge theory\cite{Dijkgraaf:1989pz}. 
It is also used in the description of gauging the global symmetry in a symmetry-enriched topological phase\cite{Barkeshli:2014cna}, which is mathematically characterized by the $G$-crossed braided tensor categories\cite{Etingof:2009yvg}.

Intuitively, we start with a non-invertible symmetry along with an automorphism group that preserves its fusion structure, i.e., $\cC \rtimes G$. That is to say, we can define an action of $G$ on $\cC$ denoted by $\alpha_g$ which acts on TDLs by fusion conjugation $\cL \mapsto g \cL g^{-1}$. In this subsection, we simplify the notation by writing $\cL_g$ as $g$ and dropping the tensor product symbol ``$\otimes$''. For a more precise definition of its fusion categorical structure, see App.~\ref{app:equivariantization}. We can then obtain the $G$-equivariantization of $\cC$ by gauging the subgroup $G$, which is, by definition, non-anomalous. We denote the resulting non-invertible symmetry as $\cC^G$.

\paragraph{Simple Objects in $\cC^{G}$.}

First, we describe the TDLs of the gauged theory. When we gauge a finite group $G$, the corresponding invertible TDLs should be trivialized in the gauged theory, so no line operators in the gauged theory can feel their existence. However, the group $G$ defines an action by conjugation on the TDLs $X \in\cC$. Therefore, given a simple TDL $X \in\cC$, we expect $X$ to combine with all TDLs in the $G$-orbit to form a new TDL in the dual theory
\begin{equation}
\bigoplus_{g\in G/G_X} \alpha_g(X) \,,
    \label{eqn:dual_TDL}
\end{equation}
where $G_X$ is the subgroup of $G$ that contains the elements whose actions on $X$ are trivial. We call $G_X$ the stabilizer subgroup of $X$. 

In other words, all simple TDLs in the same $G$-orbit as $X$ are composed together to become a TDL in the gauged theory, such that the resulting TDL commutes with $\cL_g$ for all $g\in G$.\footnote{One may consider a more stringent requirement that all $\cL_g$ should be able to end topologically on dual TDLs. This requirement is too strong to be implemented literally, given that $\cL_g$ does not even exist in the gauged theory. Such TDLs only exist in the form of a (trivial) condensation defect \cite{Roumpedakis:2022aik}, or mathematically, what we called an algebra object before. } On the other hand, we should also be able to see the outset of gauging the stabilizer subgroup $G_X$ in the gauged theory, so we expect the TDL in the dual theory that originates from $X$ to carry an extra label of the representations of $G_X$, namely the same labels as the Wilson lines of $G_X$.
As a result, the dual TDL should be viewed as a summation over the coset $G/G_X$ as written in \eqref{eqn:dual_TDL}, with an extra label denoting a representation of $G_X$. 
In the following, we, in general, label the simple objects in $\cC^G$ as a pair $(X,R)$, where $X\in {\rm Irr}(\cC)$ is a representative of a $G$-orbit in $\cC$, and $R$ is an irreducible representation of $G_{X}$. We denote the set of isomorphism classes of simple objects in $\cC$ as ${\rm Irr}(\cC)$, and the set of its $G$ orbits by ${\rm Irr}(\cC)/G$. Furthermore, we denote the orbit that $X$ belongs to as $\cO(X)$. See App.~\ref{app:equivariantization} for more details. Note that two simple objects $(X,R_1)$, $(Y,R_2)$ are isomorphic if and only if $X$ and $Y$ belong to the same $G$-orbit, and $R_1 \simeq R_2$.

To illustrate the method and notation, we work out one of the simplest examples of equivariantization explicitly. Consider a theory with two copies of a non-anomalous $\mZ_2$ symmetry, along with a nontrivial $\mZ_2$ action that acts by interchanging the $\mZ_2$'s. The full symmetry is thus given by $G =\mZ_2^2 \rtimes \mZ_2 \simeq D_8$. Even though it is well-known that gauging the non-normal $\mZ_2$ subgroup of $D_8$ results in a ${\rm Rep}(D_8)$ dual symmetry, we presume not to know this, and derive the dual symmetry using $\mZ_2$-equivariantization.

We denote the $\mZ_2$ lines in $\cC$ as $\eta_1$ and $\eta_2$, and the permuting $\mZ_2$ line as $\theta$. The lines thus satisfy the fusion rule $\theta \eta_1 \theta = \eta_2$. After gauging the non-normal $\mZ_2$, the totally symmetric TDLs of $\mZ_2^2$ ($\id$ and $\eta_1 \eta_2$) survive and acquire a label from the quantum symmetry ${\rm Rep}(\mZ_2)$.\footnote{Formally, the totally symmetric lines, $\id$ and $\eta_1 \eta_2$, should be written as $\id \boxtimes \id'$ and $\eta_1 \boxtimes \eta_2$. For simplicity, we omit the Deligne tensor product signs here.}
In total, we thus have four invertible TDLs forming a $\mZ_2^2$ subgroup. Moreover, the nontrivial $\mZ_2$ orbit $\cN := \cO(\eta_1) \simeq \eta_1 \oplus \eta_2$ becomes a non-invertible TDL with $\la \cN \ra = 2$ after the gauging. We denote the elements of ${\rm Rep}(\mZ_2) \cong \mZ_2$ as 1 and $\theta'$. The simple objects in the gauged theory are thus
\ie
(\id,1), (\id, \theta'), (\eta_1\eta_2, 1), (\eta_1\eta_2, \theta'), (\eta_1 \oplus \eta_2, 1) \,.
\fe
This dual fusion category has the same set of simple objects as ${\rm Rep}(D_8)$. 

In App.~\ref{app:bimodule_equivariantization}, we provide a more precise derivation of the simple objects using the language of $A$-bimodules.

\paragraph{Fusion Rules in $\cC^{G}$.}

The most general form of the fusion rules is stated and proven in \cite{burciu2013fusion}. For a more physical presentation, see \cite{Barkeshli:2014cna}. We will quote the statement here, and refer to Theorem 3.9 of \cite{burciu2013fusion} for the proof. To provide some intuition, we work out several examples in subsequent sections. 

\begin{theorem}
    The fusion coefficient of three simple objects labeled by $(X_i,R_i)$, $(X_j,R_j)$, $(X_k,R_k)$ is given by the following expression:
    \ie
    N_{ij}^k = \sum_{D \in G_{X_i} \backslash G / G_{X_j}} \sum_{\substack{1 \leq l \leq n, g_l^{-1} h_l \in D \\ {\rm Hom}_{\cC}(X_k,\alpha_{g_l}(X_i) \otimes \alpha_{h_l}(X_j) ) \neq 0} } m_{T_l}(R_k|_{T_l}, R_{\alpha_{g_l}(i)}|_{T_l} \otimes R_{\alpha_{h_l}(j)}|_{T_l} ) \,.
    \label{fusion_rule}
    \fe    

A double coset $H_1 \backslash G / H_2$ is defined by the equivalence relations $g \sim g'$ if there exists $h_1 \in H_1$, $h_2 \in H_2$ such that $h_1 g h_2 = g'$. In the second summation, $(g_1,h_1), \dots, (g_n,h_n)$ are representatives of $n$ $G_{X_k}$-orbits (actions defined as diagonal left multiplication, to be explained later) in $G/G_{X_i} \times G/G_{X_j}$, and in the summand, $T_l = G_{X_k} \cap G_{\alpha_{g_l}(X_i)} \cap G_{\alpha_{h_l}(X_j)}$. Here, $R_{\alpha_{g_l}(i)}$ is the same irreducible representation in $G_{\alpha_{g_l}(X_i)}$ as $R_i$, which is a representation of $G_{X_i}$. Finally
\ie
m_{T_l}(R,R') := {\rm dim\, Hom}_{{\rm Rep}(T_l)}(R, R') \,.
\fe
\label{thm:fusion_rule}
\end{theorem}

We give an interpretation of this result by separating the information of bimodule structures into two parts: first, there is the information of the fusion properties of $\cO(X)$ originating from $\cC$, and second, there is information regarding the representation theory related to the stabilizer subgroups. 
For the first part, the fusion $\cO(X_i) \otimes \cO(X_j)$, we take a naive point of view by fusing all the objects inside the orbits, namely, we obtain a set of $G/G_{X_i} \times G/G_{X_j}$:
\ie
\{ gX_ig^{-1} hX_j h^{-1} | g\in G/G_{X_i}, h\in G/G_{X_j} \} \,.
\fe
The resulting objects lead to the same dual TDL if they fall into the same orbit, namely, we have
\ie
g'gX_ig^{-1} hX_j h^{-1} g'^{-1} \sim gX_ig^{-1} hX_j h^{-1} \,, \quad g'\in G \,.
\fe
Note that the orbits of such a diagonal left action of $G$ on $G/G_{X_i} \times G/G_{X_j}$ have one-to-one correspondence with elements in a double coset $G_{X_i} \backslash G / G_{X_j}$. Thus, at the end of the day, we should sum over all possible double cosets in the fusion channel. Given a choice of final TDL $(X_k,R_k)$, the diagonal left action is defined faithfully only on the coset $G/G_{X_k}$. 
Note that fusion channels from $\cO(X_i) \otimes \cO(X_j)$ to $\cO(X_k)$ that can be admitted are generically not unique, and they all contribute to the fusion coefficient. 

Next, we turn to the second part of the information coming from the bimodule structures. 
For each fusion channel, we should look at the fusion channels of the tensor products of the common stabilizer group $T_l = G_{X_k} \cap G_{\alpha_{g_l}(X_i)} \cap G_{\alpha_{h_l}(X_j)}$. The number of admitted fusion channels is then given by $m_{T_l}(R, R')$, which counts the time of appearance of an irreducible representation $R$ inside a representation $R'$.

To provide some intuition, we return to the simple example $({\mZ_2^2})^{\mZ_2} \simeq {\rm Rep}(D_8)$ and rederive the fusion rules using Theorem~\ref{thm:fusion_rule}. 
In this example, we can work out the fusion rules for $\cN \otimes \cN$ explicitly. Let $X_i = X_j = \eta_1$, then $G_{X_i}$ and $G_{X_j}$ are trivial, and hence $T_l$ is also trivial. Therefore, $m_{T_l}=1$ for every permitted fusion channel. Furthermore, we note that the fusion channel for two $\cN$ is always given by $\alpha_{g_l}(\eta_1) \otimes \alpha_{h_l}(\eta_1)$, which is equal to either $\eta_1\eta_2$ or $\id$, hence we have $G_{X_k} = \mZ_2$. We thus have $G_{X_i} \backslash G / G_{X_j}=\mZ_2$, and there is only one $G_{X_k}$-orbit in $G/G_{X_i} \times G/G_{X_j}$ for each $D \in G_{X_i} \backslash G / G_{X_j}$. As a result, we get $N_{ij}^k = 1$ for $X_k \in\{ 1, \eta_1\eta_2\}$ and $R_k$ arbitrary, while $N_{ij}^k=0$ for $X_k \in\{ \eta_1, \eta_2\}$. In other words, the fusion of two non-invertible objects results in a sum over invertible objects with coefficients equal to one; non-invertible objects do not enter the fusion channel. We thus arrive at the following fusion rule
\ie
\cN^2 = (\id, 1) \oplus (\id, \theta') \oplus (\eta_1 \eta_2, 1) \oplus (\eta_1 \eta_2, \theta') \,.
\fe
This is identical to the fusion rule of two duality defects in ${\rm Rep}(D_8)$.\footnote{The fusion rules are almost manifest in this simple example, because we have only one choice for the $\cN \otimes \cN$ fusion. }

In our example, we have now shown that the simple objects and fusion rules are consistent with a Rep$(D_8)$ dual fusion category. One might be tempted to conclude indeed that the $\mZ_2$ equivariantization of $\mZ_2^2\rtimes\mZ_2$ results in Rep$(D_8)$. This, however, is too quick. To complete the proof, one has to show that the F symbols also agree. We show that this is indeed the case in App.~\ref{app:F_symbols}.
For a review of the discrete gauging related to Rep$(D_8)$, see e.g. \cite{Diatlyk:2023fwf}.

\section{Global symmetries in symmetric product orbifolds}\label{sec:symn}

The class of theories that forms the main interest of this work goes under the name of symmetric product orbifolds. In particular, we will study deformations of symmetric product orbifolds and the fate of their TDLs under such deformations. Symmetric product orbifolds form a class of two-dimensional CFTs with a well-defined large-$N$ limit, making them particularly suitable for studying holography. We will start this section by discussing their construction in Sec. \ref{sec:symNdef}, after which we will describe their TDLs in detail in Sec. \ref{sec:symNtdls}. 

\subsection{Definition of the theory}\label{sec:symNdef}

The starting point of the construction of a symmetric product orbifold is a compact unitary two-dimensional CFT $\mathcal{T}$ that we call the seed theory. Usually, the seed theory is chosen such that it is under good mathematical control, but for the construction of the symmetric product orbifold, the details of the seed do not matter.

From the starting point of the seed theory $\mathcal{T}$, the symmetric product orbifold $\text{Sym}^N(\mathcal{T})$ can be constructed in two steps. The first step consists of taking the $N$-fold tensor power of $\mathcal{T}$
\eq{\label{eq:tensorprod}
\underbrace{\mathcal{T}\otimes \mathcal{T}\otimes\cdots\otimes \mathcal{T}\otimes \mathcal{T}}_{N\text{ factors}}\,.
}
Taking this tensor product enhances the central charge by a factor $N$, and introduces a global discrete $S_N$ symmetry that acts by permuting the different seed copies \eqref{eq:tensorprod}. The product theory thus contains TDLs labeled by permutations $g\in S_N$. These TDLs act on local operators inserted on seed copy $j$ by sending it to copy $g(j)$. Using these TDLs, one can define defect Hilbert spaces in the twist-$g$ sector by inserting the associated TDL along the time direction. 

The second step in the construction is then to gauge the $S_N$ symmetry. In practice, this has two effects: the local operators attached to TDLs associated with the $S_N$ symmetry are promoted to local operators in twisted sectors of the gauged theory, and the Hilbert space is projected onto the $S_N$ invariant subspace, i.e., we only keep the operators that are uncharged under $S_N$. This projection ensures that all operators in the symmetric product orbifold are gauge invariant. The resulting theory thus contains local operators labeled by the twisted sector they belong to. Due to the symmetrization over $S_N$, or equivalently, due to gauge invariance of local operators, the twisted sectors are labeled not by elements of $S_N$, but by its conjugacy classes $[g]$, which are in one-to-one correspondence to integer partitions of $N$. This can be seen by the fact that a conjugacy class in $S_N$ is uniquely and completely fixed by the cycle structure of the permutations of its constituents.

The resulting theory is called the symmetric product orbifold of $\mathcal{T}$ and is denoted by
\eq{
\text{Sym}^N(\mathcal{T})\coloneqq\frac{\mathcal{T}^{\otimes N}}{S_N}\,.
}
Note that we also could have chosen not to gauge the entire $S_N$ symmetry, but to gauge a subgroup $G_N\subset S_N$ \cite{Klemm:1990df}. In this case, the gauging procedure is the same, and the twisted sector operators are labeled by the conjugacy classes of $G_N$.  The resulting theories go under the name of permutation orbifolds, and their large-$N$ limits have been studied in the context of holography in \cite{Haehl:2014yla,Belin:2014fna,Belin:2015hwa,Belin:2017jli,Keller:2017rtk}.

One of the main reasons for studying symmetric product orbifolds and permutation orbifolds comes from the fact that any of their observables can be computed in terms of the CFT data of the seed theory. Hence, these constructions give rise to families of CFTs with a tractable large-$N$ limit. Moreover, the group structure of the orbifold gives rise to many universal properties of these theories in the $N\rightarrow \infty$ limit, meaning that these properties are independent of the choice of $\mathcal{T}$. These have been extensively studied in the case of symmetric product orbifolds. Some universal properties of symmetric product orbifolds at large $N$ are: their phase structure \cite{Keller:2011xi,Belin:2014fna}, large-$N$ factorization of the light spectrum\cite{Belin:2015hwa}, the presence of an infinite tower of higher spin currents in the spectrum\cite{Baggio:2015jxa,Gaberdiel:2015uca,Gaberdiel:2015mra,Apolo:2022fya}, the behaviour of the out-of-time ordered correlators\cite{Belin:2017jli}, and thermal correlation functions \cite{Belin:2025nqd}. Of particular interest to this work is the presence of universal TDLs \cite{Gutperle:2024vyp}, whose construction we will review in Sec.~\ref{sec:unitdls}. The fact that these properties are universal implies that they can be understood purely from the group structure of $S_N$ at large $N$ and universal properties of two-dimensional CFT. It would be interesting to find an interpretation for some of these universal properties using the $S_N$ TDLs. We provide a first step towards achieving this goal by studying the torus partition function of permutation orbifolds using the $G_N$ TDLs.

\subsubsection*{Torus partition function}

As mentioned, all observables in symmetric product orbifolds (and permutation orbifolds) can be computed in terms of seed theory quantities. Of particular interest is the torus partition function, which we review next. We will use the techniques first introduced by \cite{Bantay:1997ek,Bantay:1999us}, and recently reviewed in \cite{Belin:2025nqd}. For now, we will keep the group $G_N$ we orbifold by general, even though our main interest is the symmetric group, and we will restrict ourselves to $S_N$ when we discuss the explicit construction of TDLs. 

To find the partition function of the orbifold theory we can follow the gauging procedure. We start with the seed torus partition function, which we denote by $Z_\mathcal{T}(\tau)$.\footnote{For notational convenience, we suppress the anti-holomorphic dependence here and below.} To get the orbifold partition function we should add the twisted sector contributions, by summing over the product partition function $Z_{\mathcal{T}^{\otimes N}}(\tau)= Z_{\mathcal{T}}(\tau)^N$ with $G_N$ TDLs inserted along the temporal cycle of the torus. Furthermore, we can implement the symmetrization over $G_N$ by also summing over $G_N$ TDLs inserted along the spatial cycle, since this will project onto the part of the spectrum that is uncharged under the $G_N$ symmetry. To understand the contributions to the orbifold torus partition function, we thus have to study the torus partition function in the product theory with $G_N$ TDLs inserted along both cycles of the torus.

As we have seen before, the $G_N$ TDLs act on the product theory by changing the boundary conditions of local operators: they are multivalued with their nontrivial monodromy determined by the permutation $g\in G_N$ of the TDL. To use the seed data, we have to map these multivalued objects to a cover space where they are single-valued. In order to understand the torus partition functions, we thus should study cover surfaces of the torus (with modular parameter $\tau$). Such cover surfaces of the torus are characterized by group homomorphisms $\mathbb{Z}\oplus\mathbb{Z}\rightarrow G_N$, where we recognize $\mathbb{Z}\oplus\mathbb{Z}$ as the fundamental group of the torus. These group homomorphisms are in one-to-one correspondence to all pairs of commuting elements $g,h\in G_N$ such that $gh=hg$.
There is a neat interpretation of this characterization of cover tori in terms of TDLs inserted along the different cycles of the torus. The group elements $g$ and $h$ encode the permutation of the seed copies as one moves around the two independent cycles of the torus. 

The cover surface labeled by commuting group elements $g$ and $h$ is not necessarily connected. It can have multiple connected components that correspond to the different nontrivial cycles that make up $g$ and $h$. However, each of the connected components is again a torus by the Riemann-Hurwitz formula. Each connected component of the cover is in one-to-one correspondence to an orbit of the abelian subgroup $\langle g,h\rangle$ acting in the natural way on the set $\{1,\dots,N\}$. We will denote such orbits by $\xi\in O(g,h)$. An orbit $\xi$ in turn is characterized by three integers:
\begin{enumerate}
    \item[$\lambda_\xi$:] the length of $g$-orbits inside $\xi$
    \item[$\mu_\xi$:] the number of $g$-orbits inside $\xi$
    \item[$\kappa_\xi$:] the smallest nonnegative integer such that $g^{\kappa_\xi}=h^{\mu_\xi}$ on every point of $\xi$
\end{enumerate}
The modular parameter of the cover tori is then equal to
\eq{\label{eq:tauxi}
\tau_\xi=\frac{\mu_\xi\tau+\kappa_\xi}{\lambda_\xi}\,,
}
reflecting the inclusion of the nontrivial permutations while going around the different cycles of the torus.

Adding the above ingredients together, we can now give an explicit expression for the product theory partition function with a $g$-line inserted along the temporal cycle and a $h$-line inserted along the spatial cycle (where $gh=hg$)
\eq{\label{eq:Zgh}
Z_{g,h}(\tau)=\prod_{\xi\in O(g,h)}Z_{\mathcal{T}}(\tau_{\xi})\,.
}
Note that this expression is constant on the conjugacy classes of $G_N$, that is $Z_{g_1,h_1}(\tau)=Z_{g_2,h_2}(\tau)$ if and only if $[g_1]=[g_2]$ and $[h_1]=[h_2]$. To obtain the partition function of the orbifold theory, all we have to do is sum over all possible insertions of $G_N$ TDLs inserted along the two different cycles, and normalize appropriately. The resulting partition function is thus
\eq{\label{eq:bantay}
Z_{G_N}(\tau)=\frac{1}{\abs{G_N}}\sum_{\substack{g,h\in G_N\\gh=hg}}Z_{g,h}(\tau)\,.
}
For any permutation orbifold, this formula expresses the torus partition function as a function of the seed theory partition function \cite{Bantay:1997ek,Bantay:1999us}, which is the realization of the gauging by an algebra object $A = \sum_{g \in G_N} \cL_g$ introduced in Section~\ref{subsec:gauging} with 
a trivial discrete torsion. Via the interpretation of $g$ and $h$ as inserting TDLs along the temporal and spatial cycles in the product theory, it should come as no surprise that the terms in the sum with a fixed choice of $g$ give rise to the twist-$g$ sector. In other words, the sum over $g$ includes the twisted sectors, while the sum over $h$ takes care of the symmetrization over $G_N$ \cite{Belin:2014fna}. We note again that the twisted sectors only depend on the conjugacy classes of $G_N$, since \eqref{eq:Zgh} depends only on the conjugacy classes.

In the special case $G_N=S_N$, there is another expression for the torus partition function. In fact, in this case, it is convenient to consider a so-called grand canonical partition function, where we introduce a chemical potential $\rho$ conjugate to the central charge parameter $N$ (recall that the central charge is equal to $c=c_\mathcal{T}N$)
\eq{\label{eq:grandZ}
\mathcal{Z}(\tau,\bar{\tau},\rho)\coloneqq \sum_{N=0}^\infty e^{2\pi i\rho N} Z_{S_N}(\tau,\bar{\tau})=\text{exp}\left(\sum_{L=0}^\infty \frac{e^{2\pi i\rho L}}{L}T_{L} Z_\mathcal{T}(\tau,\bar{\tau})\right)\,.
}
Here $T_L$ denotes the $L$-th Hecke operator, which acts as
\eq{
T_L f(\tau,\bar{\tau})=\sum_{\substack{a,d=1\\ ad=L}}^L\sum_{b=0}^{d-1} f\left(\frac{a\tau+b}{d},\frac{a\bar{\tau}+b}{d}\right)\,.
}
From the grand canonical partition function \eqref{eq:grandZ}, the partition function of the $N$-th symmetric product orbifold can be found by isolating the coefficient of $e^{2\pi i \rho N}$. Ref. \cite{Bantay:2000eq} explicitly shows that the coefficient agrees with \eqref{eq:bantay}. We will not prove this statement here, but let us remark on some of the links between the two expressions. First, the parameters in the Hecke operator $a$, $b$, and $d$ are related one-to-one to the integers $\mu$, $\kappa$, and $\lambda$ characterizing the orbits $\xi$. Second, the $L$-th Hecke operator encodes the (modular invariant) contribution to the partition function of a single cycle permutation. In the language of Bantay, this corresponds to all the orbits $\xi$ with $\mu_\xi\lambda_\xi=L$. The exponent then generates all the different possible configurations of cycles into all permutations of $S_N$. This shows why \eqref{eq:grandZ} is only valid for $S_N$, and not in the more general permutation orbifolds. The grand canonical partition function \eqref{eq:grandZ} can be rewritten elegantly as an infinite product using the so-called DMVV product formula \cite{Dijkgraaf:1996xw,Dijkgraaf:1998zd}
\eq{\label{eq:dmvv}
\mathcal{Z}(\tau,\bar{\tau},\rho)=\prod_{m>0}\prod_{\substack{n,\bar{n}\\n-\bar{n}=0\text{\ mod\ }m}}\left(1-e^{2\pi i(\rho m+ (\tau  n+  \bar{\tau} \bar{n})/m)}\right)^{-d(n,\bar{n})}\,,
}
where the powers $d(n,\bar{n})$ are the seed theory degeneracies, i.e.,
\eq{
Z_{\mathcal{T}}(\tau,\bar{\tau})=\sum_{n,\bar{n}}d(n,\bar{n})e^{2\pi i (n\tau+\bar{n}\bar{\tau})}\,,
}
and the restriction to $n-\bar{n}=0\text{\ mod\ }m$ ensures that operators have integer spin. Through the arguments of \cite{Bantay:2000eq,Dijkgraaf:1996xw}, and the above considerations, we can thus understand the DMVV formula through TDLs!

\subsection{TDLs in symmetric product orbifolds}\label{sec:symNtdls}

In this section, we describe the TDLs in symmetric product orbifolds. We provide two complementary descriptions; one comes from the language of fusion category, in which the focus is on $G$-equivariantization, which packages all the topological data in the gauged theory systematically. The other is the projector method construction of Petkova and Zuber \cite{Petkova:2000ip}, which is physically intuitive and easily worked out for commutative fusion categories. Some immediate TDLs after gauging are constructed from the TDLs associated with $S_N$, which were first described by \cite{Gutperle:2024vyp}. There are, however, many more TDLs in the theory than the universal ones studied in \cite{Gutperle:2024vyp}, which are generally labeled by $S_N$ orbits and representations of the corresponding stabilizer groups. 

In the language of $G$-equivariantization, reviewed in Sec.~\ref{subsec:equivariantization}, the TDLs in the symmetric product orbifolds can be described as follows. Given a fusion category $\mathcal{C}$ that describes the global symmetries of the seed theory, the most generic form of the global symmetry in the symmetric product orbifold theory is given by the $S_N$-equivariantization $(\cC^{\boxtimes N})^{S_N}$. The discussion can be easily generalized to any subgroup $G \subset S_N$, so called permutation orbifolds \cite{Bantay:1997ek, Bantay:1999us, Bantay:2000eq, Keller:2017rtk,Belin:2014fna,Belin:2015hwa,Haehl:2014yla}.

Symmetric product orbifolds form a special realization of $S_N$-equivariantization. For example, its defect torus partition function can be written as a generalization of the (DMVV) formula \cite{Dijkgraaf:1996xw}. The twisted sectors of $S_N$ can be reorganized as tensor products of the seed Hilbert spaces on $S^1$, where the size of the circle depends on the twist. Moreover, as is true for any observable in the symmetric product orbifold, the action of TDLs in the orbifold theory is completely determined by the action of TDLs in the seed theory. We describe this in detail in Sec.~\ref{subsubsec:non-universal_defects}.

We should note that it can be the case that in symmetric product orbifolds, there exist additional emergent global symmetries that are not included in $(\cC^{\boxtimes N})^{S_N}$. One instance of this occurs when there exist spin-$1$ (non-local) currents in the tensor product theory $\cT^{\otimes N}$. Such a spin-1 current naturally leads to a continuous family of (generically non-invertible) symmetries in the symmetric product orbifold \cite{Delmastro:2025ksn}.\footnote{We thank Seolhwa Kim for bringing this idea to our attention.} 
For example, the theory given by the symmetric product orbifold $\text{Sym}^2(\text{Ising})$ is a point on the $c=1,$ $S^1/\mathbb Z_2$ orbifold branch (at radius $2\sqrt2 r_{\text{self-dual}}$). This theory has two continuous families of non-invertible symmetries generated by two non-local spin-$1$ currents $\psi_1\psi_2$, $\tilde\psi_1\tilde\psi_2$ \cite{Chang:2020imq,Thorngren:2021yso}. 
In this work, we do not consider the possible additional emergent symmetries in our analysis; we plan to generalize our results and include them in the future.

\subsubsection[Universal defects from Rep\texorpdfstring{$(S_N)$}{(SN)}]{Universal defects from Rep\texorpdfstring{$\boldsymbol{(S_N)}$}{(SN)}}
\label{sec:unitdls}

We first review the universal TDLs that are present in every symmetric product orbifold theory, irrespective of the choice of the seed CFT \cite{Gutperle:2024vyp}. These arise purely from the ${\rm Rep}(S_N)$ quantum symmetry that emerges from gauging $S_N$ and are therefore built on the trivial defect of the seed theory $\mathcal{T}$ \cite{Thorngren:2021yso, Vafa:1989ih, Tachikawa:2017gyf}. We construct the universal defects using the projector method of \cite{Petkova:2000ip}. For every irrep $R$ of $S_N,$ we define a defect operator as
\eq{\label{eq:unvrslDef}
\cL_R = \sum_{[g]} \chi_R([g]) \; P_{[g]}\,,
}
where $[g]$ denotes the twisted sector in the $ \cT^{\otimes N}/S_N$ orbifold labeled by the corresponding conjugacy class of $S_N$ and $P_{[g]}$ is the identity projector in the $[g]$-twisted sector. By construction, this defect operator satisfies the commutation relations
\eq{
[L_n, \cL_R] = [\bar{L}_n, \cL_R] = 0,
}
for all Virasoro modes $L_n.$
The expression \eqref{eq:unvrslDef} captures the action of a TDL on a local operator, with the action derived in~\eqref{repG_action}.   Using the orthogonality of the characters $\chi_R([g]) $, the defect operators can be shown to satisfy the fusion algebra
\eq{
\cL_{R_1} \; \cL_{R_2} = \sum_{R_3} N^{R_3}_{R_1 R_2} \; \cL_{R_3}\,,
}
where $N_{R_1 R_2}^{R_3} $ are the Kronecker coefficients of the decomposition of the product $R_1\otimes R_2$ into irreducible representations $R_3$. For any state $\ket{\Phi_{[g]}}$ in the $[g]$-twisted sector, the action of the defect operator is given by
\eq{
\cL_R \ket{\Phi_{[g]}} = \chi_R([g]) \ket{\Phi_{[g]}}\,.
}
An important quantity that we will make use of is the quantum dimension $\la \cL \ra$, which is defined as the vacuum expectation value of $\cL$ on a cylinder, see \eqref{eq:quantum_dim}. For the universal defects, the quantum dimension is simply given by
\eq{\label{eq:TDLdim}
\la 0 | \hat{\cL} | 0 \ra = \chi_R([e]) = {\rm dim}(R)\,.
}
For $S_N$ with $N>2$, the only invertible defects are therefore the ones characterized by the trivial and the alternating representation, for which ${\rm dim}(R) =1.$

\subsubsection*{Defect partition function and gauging}

Now that we understand how the universal defects act on local operators, we can derive the defect partition function. The defect partition functions have appeared before in \cite{Gutperle:2024vyp,Knighton:2024noc}, and from the perspective of boundary states, they have been discussed in \cite{Belin:2021nck,Gaberdiel:2021kkp}. To find the defect partition function of a universal defect labeled by the irrep $R$ inserted along the spatial cycle of the torus, we just need to understand
\eq{
\Tr\left[e^{2\pi i \tau (L_0-c/24)}e^{2\pi i \bar{\tau} (\bar{L}_0-c/24)}\mathcal{L}_R\right]\,,
}
where the trace is over the orbifolded Hilbert space. We can now use what we know about the defect line acting on local operators. The contribution from the twist-$[g]$ is multiplied by $\chi_R(g)$, and hence, using \eqref{eq:bantay} we find
\eq{
Z_{R,\rm{Id}}(\tau)=\frac{1}{\abs{G_N}}\sum_{\substack{g,h\in G_N\\ gh=hg}}\chi_R([g])Z_{g,h}(\tau)\,.
}
Here we denote with $Z_{R,R'}$ the partition function with a universal defect labeled by $R$ inserted along the spatial cycle, and with a universal defect labeled by $R'$ inserted along the temporal cycle. We can use modular invariance to find the defect partition function with a defect inserted along the temporal cycle, since an $S$-transformation $\tau\mapsto-\frac{1}{\tau}$ exchanges the cycles of the torus. We thus find
\eq{
Z_{\rm{Id},R}(\tau)=\frac{1}{\abs{G_N}}\sum_{\substack{g,h\in G_N\\ gh=hg}}\chi_R([h])Z_{g,h}(\tau)\,.
}
Ref. \cite{Knighton:2024noc} generalized this expression in the case of invertible TDLs and considered multiple windings around the different cycles of the torus.

\begin{figure}[htbp]
     \centering
     \begin{subfigure}[b]{0.2\textwidth}
         \centering
         \includegraphics[width=\textwidth]{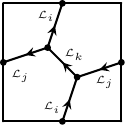}
         \caption{}
         \label{fig:4pt_resol1}
     \end{subfigure}
     \quad\quad\quad\quad
     \begin{subfigure}[b]{0.2\textwidth}
         \centering
         \includegraphics[width=\textwidth]{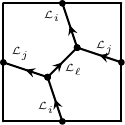}
         \caption{}
         \label{fig:4pt_resol2}
     \end{subfigure}
     \caption{The most general insertions of TDLs on a torus. A four-point junction can be resolved into two three-point junctions in two ways, illustrated here in (a) and (b).}
     \label{fig:4pt_resol}
\end{figure}

A general expression of partition functions with defect insertions is harder to write down. One needs to specify it for each resolution of the form shown in Figure \ref{fig:4pt_resol}.
We refrain from giving the complete answer here, but only to mention, as a consistency check, that the defect partition function with nontrivial defects inserted along both cycles of the torus that contribute to gauging ${\rm Rep}(G_N)$ is of the form\footnote{For simplicity of notation we assume that the representations are real; the generalization to complex representations is straightforward.}
\eq{
Z_{R,R'}(\tau)=\frac{1}{\abs{G_N}}\sum_{\substack{g,h\in G_N\\ gh=hg}}\chi_{R}([g])\chi_{R'}([h])Z_{g,h}(\tau)\,.
\label{modu_inv_partition}
}
Recall that in general the procedure of gauging ${\rm Rep}(G_N)$ can be described by an algebra object given by the regular representation $A = \bigoplus_{R} \cL_R$, where the gauged partition function is obtained by inserting a dual triangulation of $A$ with corresponding multiplication maps. To work out every partition function of the form of Figure \ref{fig:4pt_resol} explicitly is intricate. Instead,~\eqref{modu_inv_partition} is the total contribution from all the partition functions that have holonomies labeled by $R$ and $R'$ along the temporal and spatial cycles, respectively, which can be derived with the help of modular invariance. With this expression in hand, we can now explicitly see how gauging the Rep$(G_N)$ symmetry gives back the tensor product theory, at least from the perspective of the torus partition function. 
We posit
\eq{\label{eq:rrp}
Z_{\text{gauged}}(\tau)=\frac{1}{\abs{G_N}}\sum_{R,R'\in\text{Rep}(G_N)}\chi_R(e)\chi_{R'}(e)Z_{R,R'}(\tau)\,,
}
where the sum is understood to be only over irreducible representations of $G_N$. This is the appropriate normalization since
\eq{
\sum_{R\in\text{Rep}(G_N)}\chi_R(e)^2=\abs{G_N}\,.
}
We can use the orthogonality of irreducible characters to simplify \eqref{eq:rrp}. In particular,
\eq{
\sum_{R\in\text{Rep}(G_N)}\chi_R(e)\chi_R([g])=
\begin{cases}
\abs{G_N}\qquad & g=e\\
0 &\text{otherwise}
\end{cases}\,.
}
Hence, the sums over $R$ and $R'$ collapse the sums over $g$ and $h$ to $g=h=e$ in \eqref{eq:rrp}. We thus end up with
\eq{
Z_{\text{gauged}}(\tau)=Z_{e,e}(\tau)=Z_{\mathcal{T}}(\tau)^N\,,
}
where, to obtain the final expression, we use that for $g=h=e$, all $N$ orbits $\xi\in O(e,e)$ are trivial. Thus, the product over orbits simply reduces to a product over seed partition functions with $\tau_\xi=\tau$. As advertised, we recover the partition function we started with before gauging $G_N$, that is, the partition function of $\mathcal{T}^{\otimes N}$!\footnote{Note that our arguments generalize to any theory gauged by a group-like symmetry $G$, since $Z_{e,e}(\tau)$ is just the torus partition function of the theory before gauging $G$.}

\subsubsection{Non-universal defects}
\label{subsubsec:non-universal_defects}

Next, we describe the non-universal TDLs that appear in symmetric product orbifolds. 
These TDLs depend on the TDLs of the seed theory and are characterized as follows.

In this subsection, to simplify notation, we assume that the set of isomorphism classes of simple TDLs in a seed theory is given by a finite set $\cL_i, i \in\{ 1, \dots, r\}$, with $r$ a finite integer denoting the rank of the fusion category $\mathcal{C}$ (or the number of simple Verlinde lines in a RCFT with a diagonal partition function). Note that it is straightforward to generalize the subsequent discussion to incorporate continuous symmetry. 
The data characterizing the non-universal TDLs contains two parts. First, we pick $N$ simple TDLs in the tensor product theory $\mathcal{T}^{\otimes N}$, which we organize into an ordered list of the ${\mathcal L}_i$

\begin{align}\label{prod_statesbb}
 X= \Big(\underbrace{
    {\mathcal L_1}, {\mathcal L_1},\cdots,{\mathcal L_1} }_{n_1}, \underbrace{
    {\mathcal L_2}, {\mathcal L_2},\cdots,{\mathcal L_2} }_{n_2}, \cdots, \underbrace{
    {\mathcal L_r}, {\mathcal L_r},\cdots,{\mathcal L_r} }_{n_r}\Big)\,.
\end{align}
Second, we choose a set of irreducible  representations  $R_i$  of $S_{n_i}$ for  the stabilizer group $G_X= S_{n_1}\times S_{n_2}\times \cdots \times S_{n_r}$
\begin{align}\label{rep-nonun}
    R= (R_1,R_2, \cdots, R_r)\,.
\end{align}
In App.\,\ref{app:c} we give an explicit construction of the TDL for RCFT seed theories using the folded boundary state formalism. A maximally symmetric defect is a defect with $n_i=N$ for one $i\in\{1,\dots , r\}$, and $n_{j\neq i}=0$, such that the stabilizer group satisfies $G_X=S_N$.

The fusion rules can be found using Theorem~\ref{thm:fusion_rule}, by enumerating all the inequivalent fusing channels of the fusion of $S_N$ orbits. To illustrate how the fusion works, let us consider $N=2$. The fusion coefficient of $(\cL_{i_1}, \cL_{i_2}; R_1)$, $(\cL_{j_1}, \cL_{j_2}; R_2)$ and $(\cL_{k_1}, \cL_{k_2}; R_3)$ is nonzero if either 
\[
\Hom(\cL_{i_1} \otimes \cL_{j_1}, \cL_{k_1})\neq\emptyset\neq \Hom(\cL_{i_2} \otimes \cL_{j_2}, \cL_{k_2})\,,
\]
or 
\[
\Hom(\cL_{i_1} \otimes \cL_{j_2}, \cL_{k_1})\neq\emptyset\neq\Hom(\cL_{i_2} \otimes \cL_{j_1}, \cL_{k_2})\,.
\]
If this is the case, the fusion coefficient can be computed using equation~\eqref{fusion_rule}. 

Let us give an example of such a fusion coefficient. For instance, we take $i_1 \neq i_2$ and $j_1 \neq j_2$, which implies that $R_1 = R_2 = I$. Then we have 
\eqsp{
&\la (\cL_{i_1}, \cL_{i_2}; I) \otimes (\cL_{j_1}, \cL_{j_2}; I), (\cL_{k_1}, \cL_{k_2}; R)\ra = \\
&\hspace{10pt}\begin{cases}
\la \cL_{i_1} \otimes \cL_{j_1}, \cL_{k_1} \ra \la \cL_{i_2} \otimes \cL_{j_2}, \cL_{k_2} \ra  & k_1 = k_2 \\
\la \cL_{i_1} \otimes \cL_{j_1}, \cL_{k_1} \ra \la \cL_{i_2} \otimes \cL_{j_2}, \cL_{k_2} \ra + \la \cL_{i_1} \otimes \cL_{j_2}, \cL_{k_1} \ra \la\cL_{i_2} \otimes \cL_{j_1}, \cL_{k_2} \ra  & k_1 \neq k_2
\end{cases}\,.
}
Here, we denote $\la \cL_i, \cL_j \ra \equiv \dim \Hom(\cL_i, \cL_j)$.  Generalizations to larger values of $N$, and to the cases with $i_1=i_2$ or $j_1=j_2,$ are straightforward but more cumbersome to write down explicitly.

The quantum dimension of the non-universal TDL $\mathcal{L}_{X,R}$ is derived in App.~\ref{app:action} as~\eqref{eqn:qdim_L}
\begin{align}\label{q-dimlxr}
   \langle {\mathcal L}_{X,R}\rangle= |\cO(X)|\dim(R)\la X \ra = \frac{|S_N|}{|G_X|} \prod_{i=1}^r  {\rm dim}(R_i) \langle\mathcal L_i\rangle^{n_i}\,,
\end{align}
where we recall that $\langle\mathcal L_i\rangle$ is the quantum dimension of the $i$-th TDL in the seed theory and $n_i$ is the number of times it appears in $X$ defined in (\ref{prod_statesbb}). Note that from the stabilizer-orbit theorem, we have $|\cO(X)| = |S_N/G_X|$. As an illustration, we explicitly calculate (\ref{q-dimlxr}) 
for TDLs of rational seed CFTs. in App.\,\ref{app:c}.

\paragraph{Action on local operators}

A systematic way of packaging the information of linking actions is via the description of TDLs in the gauged theory via $A$-bimodules, as reviewed in Sec.~\ref{subsec:A-bimodules}. We have derived the general structures of $A$-bimodules for all $G$-equivariantizations in App.~\ref{app:equivariantization}. Here, we discuss what it means physically, especially in the scenario of symmetric product orbifolds. 

Recall that the dual TDL as an object in the original fusion category, can be written as~\eqref{bimodule_object}, written here for convenience (for our conventions on notation see App.\,\ref{app:equivariantization})
\ie
M = n \sum_{Y \in \cO(X), g\in G} Y^g \,,
\fe
and its action on local operators is captured in Figure~\ref{fig:bimodule_insertion}. 
The left and right actions are computed from~\eqref{left_associativity} to~\eqref{eqn:right_action}. 

We immediately notice that the linking action vanishes on local operators that are labeled by conjugacy classes $[g]\in G$ that do not include any elements that stabilize $X$, because no fusion channels are admitted on the cylinder. As an example, in the symmetric product orbifold, the conjugacy class characterized by a single cycle of length $N$ 
clearly does not stabilize any TDLs except the maximally symmetric ones. So we have
\ie
(X,R) \cdot O_{(N)} = 0 \,,
\label{action_vanish_N}
\fe
for any non-maximally symmetric defects $(X,R)$ on any local operators labeled by $(N)$. More generally, we have 
\ie
(X,R) \cdot O_{[g]} = 0 \,,
\fe
if ${}^gX \neq X$ for all $g \in [g]$. Equivalently, the action vanishes if ${}^gY \neq Y$ for all $Y \in \cO(X)$ given a $g \in [g]$. 

As a corollary, we mention that the large-$N$ limit of any general TDL that is not maximally symmetric is nontrivial, because in the leading large-$N$ limit, it maps $O_{(N)}$ to non-local operators.

On the other hand, when $g \in G_X$ for some $g \in [g]$, we derive the general formula for the action of $\cL_{X,R}$ on local operators from this twisted sector in App.~\ref{app:action}, the final result being
\ie
\cL_{X,R} \cdot O_{[g]} = \sum_{ \{h^i \in H | {}^{h_i}X \in \cO_g(X)\} } \la {}^{h_i}XO_g \ra \chi_R(h_i g h_i^{-1}) O_{[g]} \,,
\label{action_generic_symmetric_orbifold}
\fe
where we denote $\cO_g(X) = \{Y \in \cO(X) | {}^gY = Y \}$ and $\la {}^{h_i}XO_g \ra = \la O_g| {}^{h_i}X | O_g \ra$ denotes the action in the ungauged theory of ${}^{h_i}X$ on $O_g$, which is a local operator attached to a TDL labelled by $g$. This action satisfies $\la {}^{h_i}XO_g \ra \leq \la X \ra$. Moreover, $H$ is a representative set of $G_X \backslash G$, which we use to implement gauge fixing of the bimodule structures in App.~\ref{app:bimodule_equivariantization}. By deriving \eqref{action_generic_symmetric_orbifold}, we have separated the dynamics of the ungauged theory from the kinetics of gauging $S_N$.

Up to this point, all discussions apply generically to all $G$-equivariantizations. For symmetric product orbifolds, the missing part of the action comes from the action of TDLs of the seed theory, which can be neatly packaged in a DMVV-like formula, which we describe next. 

Following the spirit of Figure~\ref{fig:bimodule_insertion}, we need to specify the linking actions on the twisted sectors in the seed theory. Consider a conjugacy class $[g]$ of $S_N$ described by the cycle structure $\prod_{j=1}^N (j)^{n_j} $, where $\sum_{j=1}^N j n_j = N$. The state belonging to $\cH_{g}$, where $g \in [g]$, can be viewed as $\otimes_{j=1}^N (\cH_{2\pi j}^{\mZ_j})^{n_j}$, where $\cH_{2\pi j}^{\mZ_j}$ denotes the untwisted Hilbert space living on a circle with radius $2\pi j$ with a projection on spin-$j\mZ$ states. Then, the action on $\cH_g$ can be obtained by acting on the tensor product Hilbert space and using~\eqref{action_vanish_N} inductively. 

Let us give an example. In the seed theory, we denote the local Virasoro primary operators as $\phi_a$, and we denote the action of $\cL_i$ on them as
\ie
\cL_i |\phi_a \ra = \sigma^i_{ab} |\phi_b \ra \,.
\fe
Then, we can denote a local operator in $\cH_g$ as $\vec{\phi} = (\phi_1, \dots, \phi_M)$, where $M = \sum_{i=1}^N n_i$. The action of the totally symmetric TDL $\tilde\cL_{i;R} \coloneqq (\cL_i, \cL_i, \dots, \cL_i; R)$ on $\vec{\phi}$ is given by
\ie
\tilde\cL_{i;R} |\vec{\phi}\ra = \chi_R(g) \le \prod_{j=1}^M \sigma^i_{j b_j} \ri |\phi_{b_1}, \dots, \phi_{b_M} \ra \,.
\fe

The torus partition function with an insertion of the totally symmetric TDL $\tilde\cL_{i; I}$ can be written down in terms of the following generating function (cf. \eqref{eq:dmvv})
\ie
\sum_{N=0}^{\infty} p^N Z^{\cL}(\text{Sym}^N(\mathcal{T}); \tau, \bar\tau, \lambda_i) = \prod_{m>0}\prod_{\substack{n,\bar{n}\\n-\bar{n}=0\text{\ mod\ }m}}\left(1-\lambda_i p^m e^{2\pi i (\tau  n+  \bar{\tau} \bar{n})/m}\right)^{-c(n,\bar{n}, i)}  \,,
\fe
where the coefficients $c(n, \bar{n}, i)$ are defined by the expansion
\ie
Z^{\cL}(\cT; \tau, \bar\tau, \lambda_i) = \sum_{n, \bar{n} \geq 0, i} c(n, \bar{n}, i) \lambda_i e^{2\pi i (n\tau + \bar{n} \bar\tau)}  \,,
\fe
and $\{\lambda_i\}$ is the set of eigenvalues of local operators, namely $\la \phi | \hat{\cL} | \phi \ra$. In the special case where the fusion rules are commutative, the $\lambda_i$'s must be the solutions of the algebraic equations described by the fusion rules. 

We end this section by illustrating the action on local operators more explicitly in the case of a single cycle (of length $w$) twisted sector 
 primary operator, or in short, a twist-$w$ primary operator $\phi_{[w]}^l$ in the symmetric product orbifold, where we can find a basis of local operators such that $\phi_{[w]}^l$ is a one-dimensional representation of the fusion category.\footnote{Generically this is not possible. In the examples in Sec.~\ref{sec:examples}, we focus on Verlinde lines, which are by definition commutative, so that we do not have to worry about this.} 
The examples presented here will be useful for the analysis of preserved TDLs under marginal deformations in Sec.\,\ref{sec:preserveddefects} and are of particular interest in the context of holography. The simplest case is a maximally symmetric TDL which has $n_i=N$ for one $i\in \{1,2,\cdots, r\}$, while all the other $n_{j\neq i}=0$. That is, $X= ({\mathcal L}_i, {\mathcal L}_i,\cdots,{\mathcal L}_i)$. The stabilizer satisfies $G_X=S_N$, and hence, the maximally symmetric TDL can have a non-vanishing action on twisted sector states in any conjugacy class $[g]$. The action is explicitly computed in App.\,\ref{app:c}, resulting in
\begin{align}\label{phiL-max-bulk}
\mathcal{L}_{X,R} \cdot  \phi^l_{[w]}  &= \Big(\chi_R([w]) \langle {\mathcal L}_i \phi^l \rangle \langle {\mathcal L}_i\rangle^{N-w} \Big) \;  \phi^l_{[w]} \,.
 \end{align}
 A second example is the action of a general non-universal TDL with $X$ as in (\ref{prod_statesbb}), with generic values of $n_i$, $i\in\{1,\dots ,r\}$ on a twist-$w$ state of the $l$-th primary  $\phi^l_{[w]}$. The action is found to statisfy (see App.\,\ref{app:c} for a derivation)
 \begin{align}
\hspace{-5pt}\mathcal{L}_{X,R} \cdot \phi^l_{[w]}  &=  \Big(
 \sum_{i=1}^r\delta(n_i\ge w)\frac{\chi_{R_i}(w) (N-w)!}{(n_i-w)!}  \langle {\mathcal L}_i \phi^l \rangle \langle {\mathcal L}_i\rangle^{n_i-w}\prod_{\substack{j=1\\j\neq i}}^r\frac{{\rm dim}(R^j)}{n_j!} \langle {\mathcal L}_j\rangle^{n_j}\Big)  \phi^l_{[w]}\,.
\end{align}
The fact that the right-hand side vanishes for twisted sector states with $w>n_i$, $i\in\{1,\cdots,r\}$ illustrates the general principle laid out above, since $[g]=[w]$ does not stabilize $G_X$. The explicit form of the action of TDL on multi-cycle twist states can, in principle, be worked out using similar methods as those used in App.\,\ref{app:c}. However, in most examples we consider in Sec.\,\ref{sec:4}, the marginal deformations are related to states in the single-cycle twisted sector, and multi-cycle twisted sectors will not be needed. Furthermore, in the large-$N$ limit, relevant for holography, deformations by multi-cycle twisted operators do not give rise to large anomalous dimensions, and hence, cannot lead to a supergravity point on moduli space.

\section{Deformations of symmetric product orbifolds}\label{sec:4}

One of the main results of this work regards the fate of TDLs under deformations of the underlying CFT. We start this section by reviewing deformations of CFTs in Sec.\,\ref{sec:defs}. Then we continue to discuss the fate of TDLs under deformations of symmetric product orbifolds in Sec.\,\ref{sec:preserveddefects}, and discuss explicit examples of preserved TDLs in Sec.\,\ref{sec:examples}. We end the section with a short discussion on applications for holography in Sec.\,\ref{sec:hol}.  The fate of TDLs under marginal deformation has been discussed for other CFTs, such as Gepner models \cite{Cordova:2023qei}, K3 sigma models \cite{Angius:2024evd}, and toroidal $c=2$ CFTs \cite{Damia:2024xju}.

\subsection{Deformations of CFTs}\label{sec:defs}

A deformation of a CFT can be thought of as adding a term to the action of the form 
\eq{\label{eq:deltas}
\Delta S \coloneqq \lambda \int d^d x \ \Phi(x)\,,
}
where $\Phi$ has scaling dimension satisfying $\Delta_\Phi \leq d$ and is called the deformation operator. We will be mostly interested in (exactly) marginal deformations $\Delta_\Phi=d$, but one could just as easily consider relevant deformations. When $\Phi$ is exactly marginal, the coupling constant $\lambda$ corresponds to a modulus of the CFT.

Exactly marginal operators or moduli preserve conformal symmetry, and thus give rise to a conformal manifold of CFTs, also called a moduli space. On the other hand, in the case of a relevant deformation, $\lambda$ is interpreted as a mass or coupling constant that introduces a new dimensionful scale to the theory once it is turned on. 

Even when the CFT has no known description in terms of an action, deformations of the form \eqref{eq:deltas} can be made sense of. Normalized observables in the deformed theory can be computed in terms of the undeformed theory using the intuition coming from the partition function\footnote{For simplicity, we write the explicit expression for correlation functions of local operators; it is a straightforward generalization to also include non-local operators.}
\eq{\label{eq:defcor}
\langle O_1(z_1)\cdot O_n(z_n)\rangle_\lambda\coloneqq\frac{\left\langle O_1(z_1)\cdots O_n(z_n) \ e^{\lambda\int d^d x \ \Phi(x)}\right\rangle}{\left\langle e^{\lambda\int d^d x \ \Phi(x)}\right\rangle}\,,
}
where the right-hand side of this equation is evaluated in the undeformed theory. Properties of the $\lambda\neq 0$ theory can be inferred in two ways. First, when $\lambda$ is small, one can compute observables in the deformed theory using conformal perturbation theory. The idea is to expand the exponentials in \eqref{eq:defcor} to some fixed order in $\lambda$. In the context of symmetric product orbifolds, this method has been used to compute anomalous dimensions and changes in certain OPE coefficients under exactly marginal deformations \cite{Avery:2010er,Gaberdiel:2015uca,Keller:2019yrr,Keller:2019suk,Guo:2020gxm,Benjamin:2021zkn,Apolo:2022fya,Benjamin:2022jin,Apolo:2022pbq,Apolo:2024hoa}.

The second way one can infer properties of the deformed theory is to study properties that are independent of $\lambda$. This forms an incredibly powerful tool, since one can infer properties of continuous families of theories by performing a computation in the $\lambda=0$ theory, which usually is a much simpler theory than the deformed theory. Perhaps the most well-known property that does not depend on $\lambda$ is the elliptic genus, which is an index (and thus constant) on moduli space. The elliptic genus has been studied in the context of holography in e.g. \cite{deBoer:1998us, deBoer:1998kjm, Benjamin:2015hsa, Benjamin:2015vkc,Belin:2016knb,Belin:2018oza,Belin:2019jqz,Belin:2019rba,Belin:2020nmp,Benjamin:2022jin,Apolo:2024kxn}. The preservation of topological data is another such property that does not depend on the value of $\lambda$. 

For example, an extended topological operator (associated to the symmetry) is preserved under a deformation if it commutes with the associated deformation operator $\Phi$. Whether an operator commutes with such a symmetry operator is independent of the value of $\lambda$, since any correlation function evaluated at $\lambda\neq 0$ can be written in terms of a correlation function in the undeformed theory using \eqref{eq:defcor}. Thus, if the symmetry operator commutes with $\Phi$ at $\lambda=0$, it also commutes at any nonzero value of $\lambda$. When the deformation is exactly marginal, a preserved symmetry operator provides insight into the global symmetries of the entire conformal manifold. 

In the case of a relevant deformation, we have to be more careful, and the more conservative statement that holds is that at the IR fixed point, the information concerning the 't Hooft anomaly is identical to that of the undeformed theory and puts strong constraints on the IR theory \cite{tHooft:1979rat,Chang:2018iay,Thorngren:2019iar}.

In the following, we will restrict ourselves to $d=2$ and exactly marginal deformations, though we pause to note that the arguments in this section that are based solely on group theory apply to any deformation (such as RG flows of relevant deformations). In order for an operator $\Phi$ to be exactly marginal in a two-dimensional CFT, it has to have scaling dimension $(h,\bar{h})_{\Phi}=(1,1)$, and this scaling dimension cannot receive corrections at any order in perturbation theory. 

There are two main ways that have been used to construct exactly marginal deformations.\footnote{It would be interesting to understand if there are other constructions, or prove that there cannot be.} Both rely on the fact that one uses symmetry to protect the marginal operator from receiving corrections to its scaling dimension. The first type of exactly marginal deformations goes under the name of current-current deformations. The deformation is constructed using a set of conserved holomorphic $U(1)$  currents $J^i$, $(h_{J^i},\bar{h}_{J^i})=(1,0)$, and conserved anti-holomorphic $U(1)$ currents $\bar{J^j}$ with $(h_{\bar{J}^j},\bar{h}_{\bar{J}^j})=(0,1)$. It was shown in \cite{Chaudhuri:1988qb} that the operator  $\Phi =\sum_{ij} c_{ij} J^i\bar{J}^j$ is exactly marginal for arbitrary real coefficients $c_{ij}$. See \cite{Borsato:2023dis} for a recent review of current-current deformations.

The second type of exactly marginal deformations we discuss, and the type that will be relevant for the examples in Sec.\,\ref{sec:examples}, appear in supersymmetric theories. In a theory with at least $\mathcal{N}=2$ supersymmetry (see App.\,\ref{app:N=2} for our conventions), the bottom component of a short multiplet with the correct scaling dimensions is exactly marginal \cite{Dixon:1987bg,Seiberg:1988pf,Lerche:1989uy,Baggio:2012rr}. More explicitly, the exactly marginal operator is given by the $G_{-1/2}^-$ ($G_{-1/2}^+)$ descendant of an (anti-) chiral primary with conformal weight and $U(1)$ R-charge given by
\eq{
h=\frac{1}{2}\,,\qquad Q=1(-1)\,.
}
The resulting modulus is then any (Hermitian) combination of chiral/anti-chiral primary in the spectrum with the appropriate $G_{-1/2}$ and $\overbar{G}_{-1/2}$ descendant.
In addition to preserving conformal invariance, exactly marginal deformations constructed in this way will preserve the $\mathcal{N}=2$ superconformal algebra.

For the theories under consideration, symmetric product orbifolds, there exist various kinds of exactly marginal deformations, with different effects on the CFT data. Within the context of holography, single-trace twisted sector deformations are the most interesting. This is the only type of deformation that can give rise to large anomalous dimensions for generic operators \cite{Belin:2020nmp,Apolo:2022pbq}. Multi-trace deformations, on the other hand, can turn on certain OPE coefficients by a large amount, and can break large-$N$ factorization in a controlled manner \cite{Apolo:2022pbq,Apolo:2024hoa}.

In this work, we will only consider deformations that interact non-trivially with the Rep$(S_N)$ symmetry structure of the symmetric product orbifold. We will thus restrict ourselves to deformations coming from the twisted sectors. The twisted sector ground state has a nontrivial weight that increases monotonically with the length of the twist and the central charge of the seed.
\footnote{For example, the ground state of the single-cycle length $n$ twisted sector has weight $h=\bar{h}=\frac{c_\mathcal{T}}{24}\left(n-\frac{1}{n}\right)$.} This means that given a twisted sector, there is an upper bound on the seed central charge for moduli to exist in that twisted sector, see \cite{Apolo:2022fya}. In particular, no twisted sector deformation exists whenever $c_\mathcal{T}>6$ \cite{Belin:2020nmp}. In Sec.\,\ref{sec:examples}, we explicitly construct the twisted sector exactly marginal deformations under consideration for the symmetric product orbifold of the $A$-series $\mathcal{N}=2$ minimal models.

\subsection{Preservation of TDLs under twisted deformations}\label{sec:preserveddefects}

Now that we understand the deformation operators, we move our attention to the fate of TDLs under those deformations. As mentioned before, a TDL is preserved under a deformation if and only if the TDL commutes with the deformation operator. In other words, the TDL acts trivially on the deformation operator $\Phi$
\ie\label{eq:comm}
\cL \cdot \Phi = \la \cL \ra \Phi \,.
\fe
Here $\la \cL \ra$ is the quantum dimension of $\cL$, defined as its vacuum expectation value on a cylinder, see \eqref{eq:quantum_dim}. Moreover, TDLs that commute with $\Phi$ form a subcategory within the fusion category. This can be seen as follows. First, we note that quantum dimensions satisfy the fusion rules
\ie
\la \cL_i \ra \la \cL_j \ra = \sum_k N_{ij}^k \la \cL_k \ra \,.
\label{eqn:fusion_vev}
\fe
This equation can be obtained by taking the expectation value of \eqref{eq:tdlfusion}.
The second ingredient we need is the fact that for a unitary theory, the action of $\cL$ on any local operator cannot exceed its quantum dimension, or more precisely:
\ie
|\la \phi | \hat\cL | \phi \ra|  \leq \la \cL \ra \,,
\label{eqn:lemma_qdim}
\fe
for all local operators $\phi$. This inequality is saturated if and only if $\phi$ commutes or anti-commutes with $\cL$. 

\begin{figure}[htbp]
    \centering
        \includegraphics[width=\textwidth]{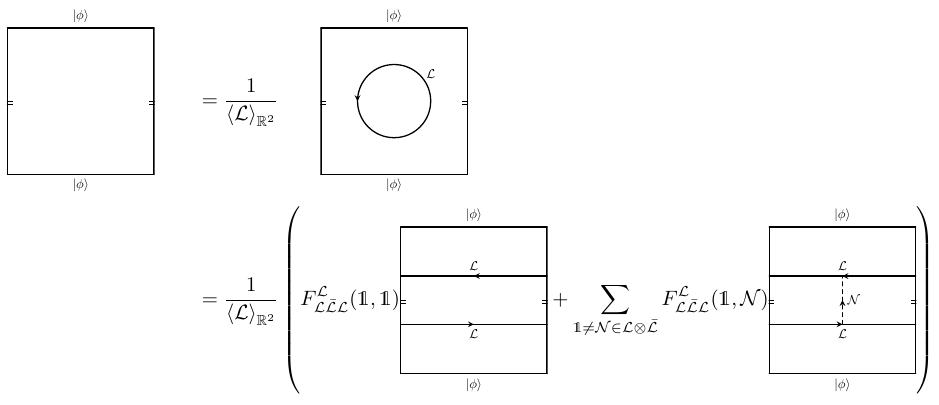}
    \caption{Pictorial proof for the inequality \eqref{eqn:lemma_qdim} making use of F moves.}
    \label{fig:cyn}
\end{figure}
We provide a physical proof of this statement in Figure\,\ref{fig:cyn}. We consider a cylinder with some state $|\phi\ra$ on the boundaries, and insert a bubble TDL $\cL$, at the cost of a factor $\frac{1}{\la\cL\ra_{\mR^2}}$. We can then implement an F move that leads to the equation as shown on the second line of Figure~\ref{fig:cyn}. Note that one can always choose a gauge for F symbols such that for a chosen TDL $\cL$, we have\footnote{See e.g. \cite{Chang:2018iay,Kitaev:2005hzj} for a discussion on gauge fixing. }
\ie
F_{\cL\bar\cL\cL}^{\cL}(\id, \id) = \frac{1}{\la\cL\ra_{\mR^2}} \,,\quad \la\cL\ra_{\mR^2} F_{\cL\bar\cL\cL}^{\cL}(\id, \cN) \geq 0 \,, \quad \forall\ \cN \in \cL \otimes \bar\cL \,.
\fe
Now~\eqref{eqn:lemma_qdim} follows from the following inequalities
\ie
\la \cL \ra^2 = \la\cL\ra_{\mR^2}^2 = \sum_{|\psi\ra \in \cH} |\la \phi | \hat{\cL} | \psi \ra|^2 + \la\cL\ra_{\mR^2} \sum_{\id \neq \cN \in \cL \otimes \bar\cL} F_{\cL\bar\cL\cL}^{\cL}(\id, \cN) |\hat{\cL}_{\cN}|\phi\ra|^2  \geq |\la \phi | \hat{\cL} | \phi \ra|^2 \,,
\fe
where we denote the Lasso operator related to $\cL$ that maps a state in $\cH$ to $\cH_{\cN}$ as $\cL_{\cN}$. In the second equality, we have inserted a complete basis of $\cH$ between two $\hat\cL$ operators. Moreover, we make use of the unitarity of the theory. 

Hence, if $\cL_i$ and $\cL_j$ commute with $\Phi$, it follows from~\eqref{eqn:fusion_vev} that all the TDLs $\cL_k$ that appear in the fusion $\cL_i \otimes \cL_j$ satisfy
\ie
\sum_k N_{ij}^k \la \hat\cL_k \ra = \la \hat\cL_i \ra \la \hat\cL_j\ra = \la \phi | \hat\cL_i | \phi \ra \la \phi | \hat\cL_j | \phi \ra  = \la \phi | \hat\cL_i \hat\cL_j |\phi \ra = \sum_k N_{ij}^k \la \phi | \hat\cL_k | \phi \ra \,.
\fe
Thus, all $\cL_k$ saturate the inequality~\eqref{eqn:lemma_qdim} without a relative sign. We have thus shown that the commuting TDLs with some local operator $\phi$ form a subcategory.\footnote{We assume that all TDLs discussed in this paper are compact. Note the existence of non-compact TDLs in a compact theory, as discussed in \cite{Lin:2022dhv}. If there are non-compact TDLs, the proof of Theorem~\ref{thm_commute} does not hold because the quantum dimension is infinite in the conventional normalization scheme. } 

\begin{theorem}
    For a unitary theory with global symmetry described by $\cC$, TDLs that commute with a local operator must form a subcategory of $\cC$. 
    \label{thm_commute}
\end{theorem}

Next, we discuss the preservation of TDLs in the theories of interest: symmetric product orbifolds. As in Sec.\,\ref{sec:symn}, we start by discussing the arguments that depend solely on the Rep$(S_N)$ structure, after which, we include the dependence on the seed theory.

\subsubsection[Universal defects from Rep\texorpdfstring{$(S_N)$}{(SN)}]{Universal defects from Rep\texorpdfstring{$\boldsymbol{(S_N)}$}{(SN)}}

In this section, we study the commutation of local (twisted sector) operators and universal TDLs in symmetric product orbifolds. We start by noting that both conjugacy classes and irreducible representations of $S_N$ are labeled by partitions of $N$. Furthermore, the conjugacy classes of $S_N$ can be labeled in terms of their cycle decompositions. 

When we neglect the global symmetries from seed CFTs, all TDLs in the symmetric product orbifold are described by ${\rm Rep}(S_N)$. Simple TDLs are thus labeled by representations of $S_N$. Given an operator in the twisted sector labeled by a conjugacy class represented by $g$, as reviewed in Sec.~\ref{subsec:A-bimodules}, the action is given by 
\ie
\cL_R \cdot O_g = \chi_R(g) \ O_g \,.
\fe
Hence, an operator $O_g$ commutes with $\cL_R$ if and only if $\chi_R(g) = {\rm dim}(R)$. Because we are interested in more general settings (i.e., non-universal TDLs), we will study representations and conjugacy classes that satisfy $|\chi_R(g)| = {\rm dim}(R)$.

We start by recalling that the characters of $S_N$ are integers, and $\abs{\chi_R(g)}\leq \dim(R)$. The simplest representations to consider are thus one-dimensional irreducible representations. The one-dimensional irreducible representations of a finite group $G$ are characterized by its abelianization $G/[G,G]$. The abelianization of $S_N$ is $\mZ_2$, which is the generic normal subgroup of $S_N$ that labels the sign of permutations. Hence, besides the trivial representation, there is one other one-dimensional irrep such that $\chi$ is $-1$ for odd permutations and $1$ for even permutations (called the alternating representation); these representations correspond to the two cosets of $A_N$. 
All operators satisfy $|\chi_R(g)| = \dim (R)$ for these irreps. 

Next, we prove that, except for the exotic case $S_4$, TDLs labeled by higher-dimensional irreps are explicitly broken by twisted operators, namely $|\chi(g)| < \dim (R)$ for arbitrary $g\neq e$ and $\dim(R)>1$. We prove this by contradiction. 
The crucial fact we will need is that for $N\geq 5$, the only nontrivial normal subgroup of $S_N$ is $A_N$.

Assume that given an irrep $R$ with $\dim R>1$, there is a conjugacy class $g$ such that $|\chi_R(g)| = \dim (R)$. Taking the representation to be unitary, this is equivalent to the statement that $\rho_R(g) = c I$, where $c$ is a phase, and $\chi_R(g) = c \dim (R) $. Moreover, we have $\chi_R(g) = \chi_R(g^{-1})$, and it follows that $c=\pm1$. Therefore, $\rho_R(g) \rho_R(h) = \rho_R(h) \rho_R(g)$ for all $h\in S_N$. By Schur's lemma, this means that for all the elements in the minimal normal subgroup that contains the subgroup that is generated by all elements in $[g]$, we have $\rho_R(g) = \pm I$. Since the only nontrivial normal subgroup is $A_N$, this representation must be a direct sum of one-dimensional irreps. We have thus found a contradiction!

The proof holds for all $N\geq 5$. However, for $N=4$, $S_4$ has another normal subgroup $\mZ_2^2$, which consists of the trivial element $e$ and all other elements in the conjugacy class with cycle structure (2,2). As a result, there is a two-dimensional representation (denoted here with the label $(2,2)$) that trivializes the $\mZ_2^2$ subgroup originating from the nontrivial irreducible representation of the coset $S_3 \cong S_4/\mZ_2^2$. One can explicitly check that we have $\chi_{(2,2)}([(2,2)])=\dim((2,2))=2$. This implies that in the $N=4$ symmetric product orbifold, operators in the twisted sector characterized by (2,2) commute with the universal non-invertible TDL labeled by the irrep (2,2). 

It can be explicitly checked that for low values of $N$ there are no other nontrivial normal subgroups, and hence, we have found all instances of $\abs{\chi_R(g)}=\dim(R)$. We summarize our results in the following theorem.

\begin{theorem}
    For all symmetric groups $S_N$, the characters of irreducible representations $R$ satisfying $\dim (R)>1$ always satisfy $|\chi_R(g)| < \dim (R)$, with one exception at $N=4$, characterized by the choice $[g] = (2,2)$ and $R = (2,2)$. 
    \label{thm:symmetric_character}
\end{theorem}

\subsubsection{Non-universal defects}

In this section, we generalize our results to also include non-universal TDLs. First, we argue that only totally symmetric TDLs can commute with local operators in the twisted sector. Then we will show that a necessary condition for a TDL to commute with a twisted sector local operator is that the seed TDL used to construct the TDL is invertible.

Recall the action of a TDL on a local operator~\eqref{action_generic_symmetric_orbifold}
\ie
\cL_{X,R} \cdot O_{[g]} = \sum_{ \{h^i \in H | {}^{h_i}X \in \cO_g(X)\} } \la {}^{h_i}XO_g \ra \chi_R(h_i g h_i^{-1}) O_{[g]} \,,
\fe
Comparing to the quantum dimension~\eqref{eqn:qdim_L}, we note that for $\cL_{X,R}$ to commute with $O_{[g]}$, it must satisfy the following conditions. First, we recall that $\la {}^{h_i}XO_g \ra\leq\la X \ra$. In order for the defect to commute, this bound should be saturated. Second, the character should satisfy $\chi_R(g) = \dim R$. If this is the case, then all terms in the sum are identical, and we can replace the sum by a factor $|\cO_g(X)|$. The TDL then commutes if and only if $|\cO_g(X)| = |\cO(X)|$.  
From this argument, it follows that if not all of the elements in the $S_N$ orbit are stabilized by $g$, its action can only be a portion of its quantum dimension, i.e., the action is strictly smaller than the dimension. As a result, $\cL_{X,R}$ can commute with $O_{[g]}$ only if all elements of the $S_N$ orbit of $X$ are stabilized by $g$. Therefore, it must be totally symmetric. Note that this conclusion does not change if we substitute $S_N$ with a subgroup that does not fix any argument of $X$.

Now suppose that $(X,R)$ is a totally symmetric TDL that commutes with $O_{[g]}$. This means that $X$ can be written in the form $(\cN, \cN, \dots)$, and commutes with $O_{[g]}$. Note that the dual of this simple line, which is just the orientation reversal, denoted as $(\bar\cN, \bar\cN, \dots)$, also commutes with $O_{[g]}$. Recall that $[g]$ is characterized by a partition of $N$. We explicitly give the argument for the case that $[g]$ contains a cycle of length two. The argument generalizes straightforwardly to $[g]$'s with longer cycles. Assume without loss of generality that $g$ permutes the first and the second copy of the seed theory, that is, we pick a representative $g\in [g]$ such that (one of the) length two cycles swaps copy one and two.  Due to Theorem~\ref{thm_commute}, all TDLs in the fusion channel of $\cN$ and $\bar\cN$  also commute with $O_{[g]}$. Now, if we assume that $\cN$ is non-invertible, we have\footnote{Here we use the assumption that the theory is compact and has a unique identity line.}
\ie
\cN \otimes \bar\cN = \id \oplus \cL \oplus \cdots \,.
\fe
Therefore, in the fusion channel of $(\cN, \cN, \dots)\otimes (\bar\cN, \bar\cN, \dots)$, there must be an object labeled by $Y = (\id, \cL, \dots)$. This object $Y$ is not stabilized by $g$ and thus does not commute with $O_{[g]}$. Hence, we found a contradiction with Theorem~\ref{thm_commute}. The generalization of the argument to $[g]$'s containing only longer cycles is straightforward. 
We have thus proven that any totally symmetric TDLs constructed from a non-invertible TDL in the seed theory cannot commute with local operators in twisted sectors. We summarize our result in the following theorem.

\begin{theorem}  \label{thm:4}  Consider a unitary compact symmetric product orbifold $\cT^{\otimes N}/S_N$, whose global symmetries are described by a semisimple tensor category. TDLs that commute with any twisted sector operator must be totally symmetric TDLs, constructed by invertible TDLs in the seed theory, and their $S_N$ representation label $R$ must satisfy Theorem~\ref{thm:symmetric_character}. In other words, they are of the form $((\eta, \dots, \eta); R)$, with $\eta$ an invertible TDL of $\cT$.
\end{theorem}

We can illustrate this criterion using the example of the actions of non-universal TDLs on a single cycle twist-$w$ operator $\phi_{[w]}^l$ constructed in Sec.\,\ref{subsubsec:non-universal_defects}. The condition that ${\mathcal L}_{X,R}$ with $X$ given in (\ref{prod_statesbb}) and $R$ given in (\ref{rep-nonun}) commutes with a single twist  $w$ field  $\phi^l_w$ is
\begin{align}\label{non-un-com}
\frac{\langle \; \mathcal{L}_{X,R} \;\phi^l_{[w]} \rangle }{  \langle \mathcal{L}_{X,R} \rangle}&=\sum_{i=1}^r\delta(n_i\geq w) \frac{(N-w)! n_i!}{N! (n_i-w)!}  \frac{\chi_{R_i}([w] )}{\chi_{R_i} (e)}\frac{\langle \mathcal{L}_i \phi^l\rangle}{\;\langle {\mathcal L}_i\rangle^w}\overset{!}{=} 1\,.
\end{align}
Here, the denominator and numerator on the left-hand side are calculated in (\ref{qdim-bdrystate}) and (\ref{phiL}), respectively.
Both the ratio of the characters as well as $ \langle \mathcal{L}_i \phi^l\rangle$ and $\;\langle {\mathcal L}_i\rangle^w$ are bounded by one. It is easy to see that for any generic partition of $N$, the ratios of the factorials in  (\ref{non-un-com}) are such that the sum is always smaller than one (unless $r=1$). Consequently, only a totally symmetric TDL can commute with a twist  $w$ field $\phi^l_w$. The only possibility is thus a TDL for which a single $n_i=N$ and all the other $n_j=0$. The condition (\ref{non-un-com}) turns into
\begin{align}
\frac{\langle \; \mathcal{L}_{X,R} \;\phi^l_{[w]} \rangle }{  \langle \mathcal{L}_{X,R} \rangle}&= \frac{\chi_{R}([w]) }{\chi_{R} (e)} \frac{\langle \mathcal{L}_i \phi^l\rangle}{\;\langle {\mathcal L}_i\rangle^w}\overset{!}{=} 1\,,
\end{align}
which is possible only if the seed defect is invertible and the representation $R$ is one-dimensional.\footnote{There is also the exotic possibility when $N=4$, and $R$ is the two-dimensional representation corresponding to the partition $(2,2)$. In this case, the operators that commute with this line have two cycles of length two, and thus do not fit within the context of the example discussed here.}

\subsection[Examples: \texorpdfstring{$\mathcal{N}=2$}{N=2} minimal models]{Examples: \texorpdfstring{$\boldsymbol{\mathcal{N}=2$}}{N=2} minimal models}\label{sec:examples}

In this section, we discuss the preservation of TDLs under twisted sector deformations of the symmetric product orbifold of the diagonal $\mathcal{N}=2$ minimal models. The $\mathcal{N}=2$ minimal models are compact, unitary SCFTs with $\mathcal{N}=(2,2)$ supersymmetry, and central charge $c<3$ (see \cite{DiVecchia:1986fwg,Boucher:1986bh,Gray:2008je} for more on the $\mathcal{N}=(2,2)$ minimal models). The minimal models have central charge
\eq{
c=\frac{3k}{k+2}\,,
}
where $k$ is a positive integer.

Furthermore, their operator content consists of a finite number of irreducible representations of the $\mathcal{N}=2$ superconformal algebra. 
The bosonic part of the algebra is realized by the coset (see e.g. \cite{Brunner:2007qu})
\eq{
\frac{\widehat{\mathfrak{su}}(2)_k\oplus\hat{\mathfrak{u}}(1)_{4}}{\hat{\mathfrak{u}}(1)_{2k+4}}\,.
}
The highest weight representations are labeled by
\begin{align}\label{eq:mmreps}
    [l,m,s]\in\mathcal{I}_k\coloneqq\{(l,m,s)\vert 0\leq l\leq k,\ m\in\mathbb{Z}_{2k+4},\ s\in\mathbb{Z}_4,\, l+m+s\in 2\mathbb{Z}\}/\sim\,,
\end{align}
where $[l,m,s]\sim[k-l,m+k+2,s+2]$. These labels determine the conformal dimension and charge of the $\mathcal{N}=2$ superconformal primaries as follows. First, $s\in\{-1,0,1,2\}$ determines the spin structure: operators with $s=\pm 1$ are in the Ramond sector, while in the NS sector $s\in\{0,2\}$. Furthermore, $s$ determines the fermion parity $(-1)^F$ of the highest weight state. The conformal weight and $U(1)$ R-charge of a superconformal primary is given by \cite{DiVecchia:1986fwg}\footnote{Compared to \cite{DiVecchia:1986fwg} the labels used here are related to the ones used there via $k_{{\text{here}}}=m_{{\text{there}}}+2,l_{{\text{here}}}=p_{{\text{there}}}-1$, $m_{{\text{here}}}=s_{{\text{there}}}-r_{{\text{there}}}$, and $s_{{\text{here}}}=r_{{\text{there}}}$. Moreover, in our conventions (see App.\,\ref{app:N=2}), the charges are multiplied by a factor of two compared to \cite{DiVecchia:1986fwg}. This can be seen by examining the OPE of the supercurrents with the $U(1)$ in the stress tensor multiplet.}
\eq{\label{eq:mmhq}
h_{l,m}^s=\frac{l^2+2l-m^2}{4(k+2)}+\frac{\abs{s}}{8}\,,\qquad\text{and}\qquad Q_m^s=\frac{m}{k+2}+\frac{s}{2}\,. 
}
Finally, the modular $S$-matrix elements for these coset representations are given by
\begin{align}\label{eq:S-mat_N=2}
    S_{[L,M,S][l,m,s]} = 
    \frac{1}{k+2} e^{-i\pi \frac{Ss}{2}} e^{i\pi \frac{Mm}{k+2}}
    \sin\left(\pi \frac{(L+1)(l+1)}{k+2}\right)\,.
\end{align}

For different values of $k$, there are different ways to combine the left and right moving characters into a modular invariant that exhibits spectral flow isomorphism between NS-NS and R-R operator spectra. The different possibilities are in one-to-one correspondence to the ADE classification of simply laced Dynkin diagrams. Only the $A$-series minimal models are diagonal and exist for any positive value of $k$; they are denoted by $A_{k+1}$. For simplicity, we restrict to the diagonal theories, but we believe that our results can be generalized to the non-diagonal $D$- and $E$-series as well \cite{Petkova:2001zn}.

The TDLs that preserve $\cN=2$ supersymmetry in the $A$-series minimal models have been described in \cite{Brunner:2007qu}. Since we consider TDLs in theories with $\mathcal{N}=2$ supersymmetry, we have to specify the gluing conditions on the entire stress tensor multiplet. Different choices for the gluing conditions lead to different types of supersymmetry preserved by the defects. We will focus on topological defects preserving B-type supersymmetry on the real line $(z=z^*),$ as in \cite{Brunner:2007qu}\footnote{We expect that our construction will also extend to defects preserving A-type supersymmetry.}
\begin{align}\label{B-type}
    \begin{rcases}
    T(z) - T(z^*) \\
    \overline{T}(\overline{z}) - \overline{T}(\overline{z}^*)\\
    G^\pm(z) - \zeta G^\pm(z^*) \\
    \overline{G}^\pm(\overline{z}) - \overline{\zeta} \;\overline{G}^\pm(\overline{z}^*) \\
    J(z)-J(z^*) \\
    \overline{J}(\overline{z}) - \overline{J}(\overline{z}^*)
    \end{rcases} \xrightarrow{} 0
\end{align}
where $\zeta, \overline{\zeta} \in \{\pm 1\}$, and we refer to App.~\ref{app:N=2} for the $\cN = 2$ algebra. The topological defect operators $\cL : \cH^k \rightarrow \cH^k$ then satisfy the following commutation relations
\begin{align}
    [L_n, \mathcal{L}]  &= [\overline{L}_n, \mathcal{L}]=0\,, \nonumber\\
    \quad G^\pm_r\mathcal{L} - \zeta\;\mathcal{L}\;G^\pm_r &=  
    \overline{G}^\pm_r\mathcal{L} - 
\overline{\zeta}\;\mathcal{L}\;\overline{G}^\pm_r=0\,, \nonumber\\
J_r \mathcal{L} - \mathcal{L} J_r &= \bar J_r \mathcal{L} - \mathcal{L} \bar J_r =0\,,
\end{align}
for all $n \in \mathbb{Z}$ and all $r \in \mathbb{Z} + 
\frac{1}{2}$ ($r \in \mathbb{Z}$) in the NS- (R-) sectors. The defect operators can be written in terms of projectors denoted with $P$ over the bosonic subalgebra
\begin{align}\label{eq:B-type_defect}
     \mathcal{L}_{[L,M,S,\bar{S}]} = \sum_{[l,m,s],\bar{s}} e^{-i\pi \frac{\bar{S}(s+\bar{s})}{2}} 
    \frac{S_{[L,M,S-\bar{S}][l,m,s]}}{S_{[0,0,0][l,m,s]}} P_{[l,m,s,\bar{s}]}\,,
\end{align}
where $s-\bar{s} \in 2 \mathbb{Z}.$ These defect operators are in one-to-one correspondence with the chiral primaries, and hence, we use the highest-weight representations \eqref{eq:mmreps} to label them. They satisfy the following fusion rule
\begin{align}
    \cL_{[L_1,M_1,S_1,\bar{S}_1]}\;\cL_{[L_2,M_2,S_2,\bar{S}_2]} = 
    \sum_{L} N^L_{L_1,L_2} \;
    \cL_{[L,M_1+M_2,S_1+S_2,\bar{S}_1+\bar{S}_2]}\, 
\end{align}
with non-negative integers $N^L_{L_1,L_2}$ that are related to the $S$-matrix elements \eqref{eq:S-mat_N=2} via the Verlinde formula. The quantum dimension of the defect can be computed using its overlap with the vacuum corresponding to $l=m=s=0$, and is given by
\begin{align}
    \langle \cL_{[L,M,S,\bar{S}]} \rangle = \frac{S_{[L,M,S-\bar{S}][0,0,0]}}{S_{[0,0,0][0,0,0]}} = \frac{\sin\left( \frac{\pi (L+1)}{k+2} \right)}{\sin\left( \frac{\pi}{k+2} \right)}\,.
\end{align}
We can compute the action of the TDLs by acting \eqref{eq:B-type_defect} on a state corresponding to a primary operator $\phi^{[l',m',s']}$
\begin{align}
    \cL_{[L,M,S,\bar{S}]} \cdot \phi^{[l',m',s']} = e^{-i\pi \frac{\bar{S}(s'+\bar{s}')}{2}} 
    \frac{S_{[L,M,S-\bar{S}][l',m',s']}}{S_{[0,0,0][l',m',s']}} \;\phi^{[l',m',s']} \,.
\end{align}
Let us now look at the TDLs in the symmetric product of the $\cN = 2$ minimal model. In light of Theorem \ref{thm:4}, we will consider the totally symmetric TDLs as described in Appendix~\ref{app:C.4}, which are constructed by picking the same TDL in every copy of the product theory
\begin{align}
    X= \Big(\underbrace{
    {\mathcal L_{[L,M,S,\bar{S}]}}, {\mathcal L_{[L,M,S,\bar{S}]}},\cdots,{\mathcal L_{[L,M,S,\bar{S}]}} }_{N}\Big) \,.
\end{align}
Given a representation $R$ of $S_N,$ we can make use of \eqref{eq:totally_sym_qdim} to calculate the quantum dimension 
\begin{align}
    \langle \cL_{[L,M,S,\bar{S}],R} \rangle = \text{dim}(R) \; \langle \cL_{[L,M,S,\bar{S}]} \rangle^N\,,
\end{align}
and similarly, \eqref{phiL-max} tells us the action of totally symmetric TDL $\cL_{[L,M,S,\bar{S}],R}$ on a twist-$w$ primary operator $\phi^{[l,m,s]}_{[w]}$
\begin{align}
    \langle \cL_{[L,M,S,\bar{S}],R} \cdot \phi^{[l,m,s]}_{[w]} \rangle = \chi_R([w]) \; e^{-i\pi \frac{\bar{S}(s+\bar{s})}{2}} 
    \frac{S_{[L,M,S-\bar{S}][l,m,s]}}{S_{[0,0,0][l,m,s]}} \; \langle \cL_{[L,M,S,\bar{S}]} \rangle^{N-w} \,.
\end{align}
For the TDL to commute with the twist-$w$ operator, it thus has to solve the following constraint
\begin{align}\label{eq:mmcond}
    \frac{\langle \cL_{[L,M,S,\bar{S}],R} \cdot \phi^{[l,m,s]}_{[w]} \rangle}{\langle \cL_{[L,M,S,\bar{S}]} \rangle} = \frac{\chi_R([w])}{\text{dim}(R)} \; e^{-i\pi \frac{\bar{S}(s+\bar{s})}{2}} 
    \frac{S_{[L,M,S-\bar{S}][l,m,s]}}{S_{[0,0,0][l,m,s]}} \left( \frac{    \sin\left( \frac{\pi}{k+2} \right)}{\sin\left( \frac{\pi (L+1)}{k+2} \right)} \right)^{w} \overset{!}{=} 1\,.
\end{align}

\subsubsection{Exactly marginal operators}

Since we are interested in the preservation of non-universal defects, we need to understand precisely how the relevant TDLs act on the exactly marginal deformations. In other words, we need detailed information on the construction of the twisted sector moduli in symmetric product orbifolds of $\mathcal{N}=2$ minimal models. The existence of such moduli can be read off from the $\frac{1}{2}$-BPS partition function, which can be written in terms of a DMVV-type formula. In the $\frac{1}{2}$-BPS partition function, one then just reads off the chiral primaries of the correct weight and charge. In this way, a complete classification of the moduli was found in \cite{Belin:2020nmp} (see in particular App.\,A therein). There, it is shown that there are many exactly marginal deformations, both in the twisted and untwisted sectors, and of the single- and multi-trace variety. As mentioned before, we focus on deformations coming from twisted sectors. Furthermore, we restrict to single-trace deformations for now, since they are the ones relevant for holography. At the end of this subsection, we will also discuss the marginal deformations with cycle structure $(2)(2)$ because these may lead to non-invertible preserved TDLs when we take $N=4$.

\begin{table}[ht!]
	\centering
	\begin{tabular}{|ccc|}
		\hline
		Seed theory & $k$  & moduli\\
		\hline
		$A_2$ & 1&1 twist 5, 1 twist 7\\
		$A_3$ & 2 &1 twist 3, 1 twist 4, 1 twist 5\\
		$A_5$ & 4 &1 twist 2, 1 twist 3, 1 twist 4\\
		$A_{k+1}$ & odd, $\geq3$ & 1 twist 3\\
		$A_{k+1}$ & even, $\geq6$ &1 twist 2, 1 twist 3\\
		\hline
	\end{tabular}
	\caption{Number of twisted sector single-trace moduli for symmetric product orbifolds of the $A$-series minimal models. The central charge of the minimal models is related to the parameter $k$ by $c=\frac{3k}{k+2}$.}
	\label{t:moduli}
\end{table}

An overview of the single-trace twisted exactly marginal operators in the symmetric product orbifolds of the $\mathcal{N}=2$ $A$-series minimal models is given in Table\,\ref{t:moduli}. 
Note that the exactly marginal operators are not simply constructed using the twisted sector ground state. Their explicit construction was laid out in \cite{Apolo:2022fya}, which we refer to for details. The relevant chiral primaries come from excited states in the twisted sectors. The weight of (a set of) operators in the single-trace twisted sector of length $n$ is given by\footnote{There is a subtlety regarding the boundary conditions for fermionic operators in even and odd twisted sectors. For even values of $n$, the periodicity of the fermions changes from NS to R, and hence the ground state has positive weight $\tilde h=\frac{c}{24}$. This ground state energy that is added for even values of $n$ is usually attributed to a spin field. As we will see below in Table~\ref{tab:tabl2e}, the moduli with odd twist are in the Neveu-Schwarz sector, while those of even twist are in the Ramond sector.}
\eq{
h=\frac{\tilde{h}}{n}+\frac{c}{24}\left(n-\frac{1}{n}\right)\,.
}
Here $\tilde h$ is the weight of a Virasoro primary of the seed theory. There exist additional twisted sector primaries that are constructed using descendants in the seed theory. Their weight can be derived using the cover map procedure. In Table~\ref{tab:tabl2e}
we give, for each modulus, the seed operator that, when inserted in the relevant twisted sector, leads to a chiral primary of weight $h=\frac{1}{2}$ and $U(1)$ charge $Q=1$. 

\begin{table}[ht!]
  \begin{center}\def\arraystretch{1.65}
  \begin{tabular}{|c|c|p{1cm}<{\centering}|p{3.80cm}<{\centering}|}
\hline
  $k$ & $c = \frac{3k}{k+2} $ & twist & seed operator \\ 
  \hline 
  \multirow{2}{*}{1} &  \multirow{2}{*}{1} & 5 &    $G^+_{-3/2}\phi_{[0,0,0]}$  \\
  \cline{3-4}
  &  & 7 &  $G^+_{-3/2}\phi_{[0,0,0]}$ \\
 \hline
   \multirow{3}{*}{2} &  \multirow{3}{*}{$\frac{3}{2}$} & 3 & $G^+_{-1/2}\phi_{[2,0,0]}$ \\
  \cline{3-4}
  &  & 4 & $G^+_{-1}\phi_{[1,-2,1]}$   \\
    \cline{3-4}
  &  & 5 & $G^+_{-1/2}\phi_{[2,0,0]}$  \\
   \hline
 3 & $\frac{9}{5}$ & 3 & $G^+_{-1/2}\phi_{[2,0,0]}$\\
 \hline
   \multirow{3}{*}{4} &  \multirow{3}{*}{2} & 2 &  $\phi_{[4,3,1]}$  \\
  \cline{3-4}
   &  & 3 & $G^+_{-1/2}\phi_{[2,0,0]}$  \\
   \cline{3-4}
  &  & 4 &  $\phi_{[4,3,1]}$  \\
  \hline
  $5,6, \dots$ & $2 < c < 3$ & 3 & $G^+_{-1/2}\phi_{[2,0,0]}$ \\
 \hline
  $6,8,\dots$ & $2 < c < 3$ & 2 & $\phi_{[\frac{k}{2}+2,\frac{k}{2}+1,1]}$\\
 \hline
  \end{tabular}
  \caption{Seed operator used to construct the twisted marginal deformations in the $A$-series minimal models. The states are such that when inserted in the appropriate twisted sector, they lead to chiral primaries of weight $h=\frac{1}{2}$ and $U(1)$ charge $Q=1$ in the symmetric product orbifold theory. In the last column, we denote with $\phi_{[l,m,s]}$ the Virasoro primaries of the minimal models with weight and charge as in \eqref{eq:mmhq}.}
  \label{tab:tabl2e}
  \end{center}
  \end{table}

Next, we return to the exactly marginal operators of cycle type $(2)(2)$. In the $A$-series minimal models, there are two such operators. One is constructed by inserting the seed operator $\phi_{[1,0,1]}$ in both length-2 cycles, and it exists for every value of $k$. For $k>1$, there is an additional modulus constructed by inserting the seed operator $\phi_{[2,1,1]}$ on one of the cycles, and $\phi_{[0,-1,1]}$ on the other cycle.

\subsubsection{Preserved TDLs}\label{sec:prestdlsa}

Now that we have specified the TDLs and exactly marginal operators in the symmetric product orbifolds of $\mathcal{N}=2$ $A$-series minimal models, we can study the preservation of the TDLs under deformations by the marginal operators. From the considerations in Sec.\,\ref{sec:preserveddefects}, it follows that for general values of $N$, and more interestingly in the context of holography, for large values of $N$, only totally symmetric defects made using the trivial or alternating representation of $S_N$ have a chance of being preserved.

Since the minimal models are rational CFTs, we know the quantum dimension and action on local operators in terms of the modular $S$-matrix. In order to study the preservation of the TDLs, we compare the two and find instances where the quantum dimension agrees with the action on the exactly marginal operators. 

First, we focus on the single-trace marginal operators, since these are the ones relevant for holography. The information necessary to study the preservation of TDLs under such deformations is summarized in Table \ref{tab:tabl2e}. The theory is specified by the integer $k$, we need the size of the twist $w$, and finally, we need the seed labels describing the operator used to construct the deformation $l$, $m$, and $s=\bar{s}$. Given these labels, and the data specifying the TDLs, it is simple to work out the condition for the totally symmetric defects to be preserved using \eqref{eq:mmcond}
\eq{\label{eq:mmprescond}
e^{-i \pi \frac{S s + \bar{S}\bar{s}}{2}} e^{i \pi \frac{M m}{k+2} } \left(\frac{\sin\left(\pi \frac{(L+1)(l+1)}{k+2}\right)}{\sin\left(\pi \frac{(l+1)}{k+2}\right)} \right) \left(\frac{\sin\left(\frac{\pi}{k+2}\right)}{\sin\left(\pi\frac{L+1}{k+2}\right)}\right)^w \overset{!}{=} \frac{\dim (R)}{\chi_R([w])}\,.
}
In this equation, we have collected the factors regarding the seed theory data on the left, and the quantities concerning the $S_N$ representation on the right. It is a simple fact from group theory that the absolute value of the right-hand side of \eqref{eq:mmprescond} is at least one. On the other hand, the absolute value of the left-hand side is upper bounded by one.\footnote{This can be shown using general properties of modular $S$-matrices in RCFT, or more explicitly using the specific $S$-matrices of the minimal models and properties of Chebyshev polynomials.} Hence, for a defect to be preserved, both sides of \eqref{eq:mmprescond} have to be equal to $\pm1$. In terms of the $S_N$ group data, this boils down to the fact that the representation has to be the trivial or alternating one.

When $R$ is the trivial representation, the right-hand side of \eqref{eq:mmprescond} is equal to one, and one can completely classify the TDLs that are preserved by the deformations appearing in Table \ref{tab:tabl2e}. The results of this classification are presented in Table \ref{tab:3}, where we provide the conditions on the TDL labels $L$, $M$, $S$, and $\bar{S}$ that solve \eqref{eq:mmprescond}.

\begin{table}[ht!]
  \begin{center}\def\arraystretch{1.65}
  \begin{tabular}{|c|c|p{2cm}<{\centering}|p{4cm}<{\centering}|}
\hline
  twist & $k$ & $L$ condition & $M,S,\bar{S}$ condition \\ 
  \hline 
  \multirow{2}{*}{2} &  0 mod 4 & $0,k$ &    $M-S-\bar{S}=2\text{ mod }4$  \\
  \cline{2-4}
  & 2 mod 4 & $0,k$ &  $M-S-\bar{S}=0\text{ mod }4$ \\
 \hline
   3 &  $\mathbb{Z}_{\ge 2}$ & $0,k$ & $L+M+S-\bar{S}=0\text{ mod }2$ \\
   \hline
 4 &  2 & $0,2$ &  $M+S+\bar{S}=2\text{ mod }4$   \\
   \hline
   4 &  4 & $0,4$ & $M-S-\bar{S}=0\text{ mod }4$ \\
 \hline
 5 &  1 & $0,1$ & $L+M+S-\bar{S}=0\text{ mod }2$ \\
 \hline
 5 &  2 & $0,2$ & $L+M+S-\bar{S}=0\text{ mod }2$ \\
 \hline
  7 &  1 & $0,1$ & $L+M+S-\bar{S}=0\text{ mod }2$  \\
 \hline
  \end{tabular}
  \caption{Conditions on the seed data for the maximally symmetric TDLs in the trivial representation that are preserved under the various single-trace twisted sector marginal deformations of the $A$-series minimal models.}
  \label{tab:3}
  \end{center}
  \end{table}

It is not hard to repeat this analysis for the case when $R$ is the alternating representation. In this case, the sign on the right-hand side of \eqref{eq:mmprescond} depends on the twist of the marginal deformation; when the twist is even, we get a minus sign compared to the previous analysis, affecting the labels of the TDLs that solve the condition \eqref{eq:mmprescond}. The results are summarized in Table \ref{tab:4}.

\begin{table}[ht!]
  \begin{center}\def\arraystretch{1.65}
  \begin{tabular}{|c|c|p{2cm}<{\centering}|p{4cm}<{\centering}|}
\hline
  twist & $k$ & $L$ condition & $M,S,\bar{S}$ condition \\ 
  \hline 
  \multirow{2}{*}{2} &  0 mod 4 & $0,k$ &    $M-S-\bar{S}=0\text{ mod }4$  \\
  \cline{2-4}
  & 2 mod 4 & $0,k$ &  $M-S-\bar{S}=2\text{ mod }4$ \\
 \hline
   3 &  $\mathbb{Z}_{\ge 2}$ & $0,k$ & $L+M+S-\bar{S}=0\text{ mod }2$ \\
   \hline
 4 &  2 & $0,2$ &  $M+S+\bar{S}=0\text{ mod }4$   \\
   \hline
   4 &  4 & $0,4$ & $M-S-\bar{S}=2\text{ mod }4$ \\
 \hline
 5 &  1 & $0,1$ & $L+M+S-\bar{S}=0\text{ mod }2$ \\
 \hline
 5 &  2 & $0,2$ & $L+M+S-\bar{S}=0\text{ mod }2$ \\
 \hline
  7 &  1 & $0,1$ & $L+M+S-\bar{S}=0\text{ mod }2$  \\
 \hline
  \end{tabular}
  \caption{Conditions on the seed data for the maximally symmetric TDLs in the alternating representation that are preserved under the various single-trace twisted sector marginal deformations of the $A$-series minimal models.}
  \label{tab:4}
  \end{center}
  \end{table}

Before we move on to investigate the preservation of non-invertible symmetries in the $N=4$ symmetric product orbifolds deformed by a double-trace twist $(2,2)$ deformation, let us briefly comment on the results so far. The seed TDLs used to construct the preserved totally symmetric TDLs are all invertible, in accordance with Theorem \ref{thm:4}. Furthermore, at least within the context of the $\mathcal{N}=2$ minimal models, the conditions on the seed labels are not particularly strong. For each deformation, there are multiple TDLs preserved by it. This is in strong contrast with the analysis of \cite{Gutperle:2024vyp}, where the preservation of universal defects was considered in the context of symmetric product orbifolds. None of the universal TDLs are preserved by marginal deformations.\footnote{Ref. \cite{Gutperle:2024vyp} considers twist two deformations. However, their arguments generalize straightforwardly to (single-trace) deformations of other twists.}

Finally, let us discuss the exotic case of $N=4$ and double-trace deformations with cycle structure $(2,2)$. As mentioned at the end of the last subsection, when the seed theory is an $A$-series $\mathcal{N}=2$ minimal model with $k>1$, there are two deformations of this cycle type. For $k=1$, there exists a unique exactly marginal deformation of cycle type $(2,2)$. For simplicity, let us focus on the deformation that exists for all $k$. This deformation is constructed by inserting the seed operator $\phi_{[1,0,1]}$ on both of the cycles. As before, only totally symmetric defects have a chance to be preserved. The condition for TDLs to be preserved reduces to 
\eq{\label{eq:mmprescondn4}
 \left(\frac{\sin\left(\pi \frac{2(L+1))}{k+2}\right)}{\sin\left(\pi \frac{2}{k+2}\right)} \right)^2 \left(\frac{\sin\left(\frac{\pi}{k+2}\right)}{\sin\left(\pi\frac{L+1}{k+2}\right)}\right)^4 \overset{!}{=} \frac{\dim (R)}{\chi_R([2,2])}\,.
}
As before, the absolute value of the left-hand side is upper bounded by one, while the absolute value of the right-hand side is at least one, with equality when $R$ is the representation labeled by the partition $(2,2)$. For this representation, the right side of \eqref{eq:mmprescondn4} is equal to one, and as before, we can find conditions on the TDL labels $L$, $M$, $S$, and $\bar{S}$. The left hand side of \eqref{eq:mmprescondn4} is equal to one whenever $L\in\{0,k\}$. Hence, any TDL with the specified value of $L$ is preserved. The conditions on the other labels are merely that they correspond to an actual defect, i.e., $L+M+S-\bar{S}=0\text{ mod }2$. Thus, we have found a set of non-invertible defects that are preserved by an exactly marginal deformation! These non-invertible TDLs exist everywhere along the deformation of the $N=4$ symmetric product orbifold of any $\mathcal{N}=2$ $A$-series minimal model.

\subsection{Applications to holography}\label{sec:hol}

The strongest form of the AdS/CFT correspondence states that any consistent theory of quantum gravity in AdS$_{d+1}$ can equivalently be described by a CFT in $d$ dimensions. An important question that arises is that of the fate of symmetries on either side of the correspondence. The commonly accepted view is that a global symmetry in a holographic CFT is dual to a gauge symmetry in the gravitational dual with the same symmetry structure.

Additionally, any global symmetry on the gravity side leads to an inconsistency in the dual  CFT, in accordance with the no-global symmetries conjecture \cite{Harlow:2018tng,Banks:2010zn,Heckman:2024oot}.
The exact duality of type IIB on $AdS_3\times S^3\times \mathbb{T}^4(\; {\rm or}\; \text{K}3)$ 
with one unit of NS-NS three-form flux \cite{Gaberdiel:2017oqg,Gaberdiel:2018rqv,Eberhardt:2018ouy,Giribet:2018ada,Eberhardt:2019ywk}, gives some indication of the realization of TDLs on the worldsheet, where the cover spaces in the calculation of correlations functions of the symmetric product orbifold can be identified with the worldsheet of the AdS$_{3}$ string theory as conjectured in \cite{Pakman:2009zz} and proven in \cite{Eberhardt:2019ywk}.
This suggests that correlation functions computed in the presence of TDLs can be thought of as topological defects inserted on the string worldsheet.  From the spacetime perspective, the defects are interpreted as some brane-like extension of the worldsheet defect into the bulk \cite{Knighton:2024noc}. For recent discussions on the worldsheet and spacetime perspective of the bulk description of topological defects in the AdS$_3$ context, see e.g. \cite{Gutperle:2024vyp, Knighton:2024noc,Belleri:2025eun,Heckman:2024obe,Harris:2025wak}.

As discussed in the introduction,  symmetric product orbifold CFTs do not have an AdS dual description in terms of (local) supergravity at low energies.  It is believed that such a description may be obtained for certain seed CFTs by deforming the CFT  to a strongly coupled point on its moduli space.
However, whether there is a holographic point on their moduli space is generally a hard question to answer. In the well-known cases where the seed CFT is given by a nonlinear sigma model on K3 or $\mathbb{T}^4$ there is a holographic point on moduli space, and a big step towards establishing this fact is the existence of the supergravity theory on AdS$_{3}\times S^3\times M_4$, with $M_4\in\{K3,\mathbb{T}^4\}$. In particular, the agreement of the elliptic genus between the symmetric product orbifold of K3 and supergravity on AdS$_{3}\times S^3\times$ K3 was seen as strong evidence \cite{deBoer:1998kjm}. See \cite{Apolo:2022fya,Benjamin:2022jin} for some evidence that deformations of the symmetric product orbifolds with $\mathcal{N}=2$ SCFT seed theories can lead to semiclassical gravity duals; however, what form (if any) such a dual takes is at present not known. We should also note that seed theories that can lead to semiclassical gravity duals are very special in the space of $\mathcal{N}=2$ SCFTs \cite{Belin:2020nmp,Benjamin:2022jin}. 

The results of Sec.\,\ref{sec:4} may be useful to pin down a putative supergravity background dual to the deformation of more general symmetric product orbifolds. One of the upshots of our analysis in this section is that any supergravity background dual to a deformation of a symmetric product orbifold of an $A$-series $\mathcal{N}=2$ minimal model has to be consistent with the preserved TDLs found in Sec.\,\ref{sec:prestdlsa}. One worry might be that even though the TDLs are preserved by the deformation, all the operators they act on might lift and get removed from the low-lying spectrum. In other words, it could happen that the preserved TDLs become unfaithful under the deformation. It is easy to check that this is not the case by considering the action of the preserved TDLs on operators that are also preserved, i.e., chiral primaries. Generically, the TDLs act nontrivially on such operators and hence the TDLs correspond to genuine symmetries on the entire conformal manifold. At minimum, the massless sector of any potential bulk supergravity should thus be consistent with these preserved symmetries.

We end this section with some speculative comments on the nature of the preserved defects and their interpretations in holography. One of our main results is that only invertible symmetries can be preserved (for $N>4$, i.e., the cases relevant for holography). This result is compatible with arguments coming from string theory, where it is argued that non-invertible symmetries on the worldsheet 
 are generically broken by string loop effects \cite{Heckman:2024obe}. The string theory argument boils down to the fact that selection rules on genus-zero correlation functions coming from non-invertible symmetries are not valid at higher genus. Our results provide evidence for this general rule in all setups of string theory where the worldsheet CFT is related to a symmetric product orbifold (whenever $N>4$). Moreover, it is consistent with more general symmetric product orbifold theories having a place in string theory.

A consequence of the fact that only invertible symmetries can be preserved is that almost all of the $S_N$ symmetry is broken as one moves away from the symmetric product orbifold point in moduli space (using a twisted sector deformation). However, we have found that TDLs constructed not just with the trivial representation of $S_N$ can be preserved. There are also preserved TDLs made using the alternating representation (that are nevertheless invertible). This suggests that there can be some remnant of the $\mathbb{Z}_2$ symmetry associated with the sign of $S_N$ permutations that remains after deforming.\footnote{From the expressions in \cite{Knighton:2024noc} it is not hard to see that this also occurs in Sym$^N(\mathbb{T}^4)$. It would be interesting to study how this remnant of $\mathbb{Z}_2$ gets manifested in the supergravity dual.} It would be interesting to obtain a deeper understanding of this nontrivial $S_N$-remnant that is present on the entire conformal manifold.

\section{Discussion}\label{sec:discussion}

In this work, we construct generalized symmetries represented by TDLs in symmetric product orbifolds, generalizing the universal and maximally symmetric TDL constructed in  \cite{Gutperle:2024vyp}.  We determine which of the TDLs are preserved by exactly marginal deformations in the twisted sector away from the symmetric product orbifold point on the moduli space of the CFT.   The main result is contained in Theorem \ref{thm:4}, stating that (except for a single isolated case) a commuting TDL is maximally symmetric, constructed from an invertible seed TDL, labelled by the alternating or identity representation, and is an invertible TDL. We derive these results using the mathematical language of $G$-equivariantization of fusion categories. In addition, we provide an explicit construction of such TDLs (and an illustration of the general theorem) using the methods of Petkova and Zuber for the symmetric product orbifold of the $A$-series ${\cal N}=(2,2)$ minimal models. This explicit example allows us to give a classification of all preserved TDLs in these models. We should emphasize that the mathematical argument is more general and not limited to rational seed CFTs. 

Our results may be slightly disappointing in the sense that no non-invertible TDL commutes with deformations in the twisted sector for $N>4$. However, if a deformation leads to a theory with a semiclassical gravity dual, then the defect we found exists in this theory and can be used to constrain possible duals.
The only non-invertible commuting TDL we find is an isolated case for $N=4$ (see Theorem \ref{thm:symmetric_character}), hence this example is not interesting for holography, but for certain seed CFTs it may be relevant in mathematics or condensed matter theory.

There are several directions in which the results of the present work can be extended. As mentioned in \cite{Chang:2018iay}, if a TDL anticommutes with a deformation operator, the symmetry is not preserved, but instead it follows that an interface exists between two CFTs where the marginal operator is turned on with opposite sign. It would be very interesting to investigate these interfaces for symmetric product orbifolds and, in particular, the interpretation in their possible holographic duals. The analysis on the CFT side that one would have to do is very similar to the one contained in this work. All one has to do is add a minus sign to \eqref{eq:comm}, i.e., we have to solve the equation $\cL\cdot\Phi=-\langle\cL\rangle \,\Phi$. This minus sign can then be traced throughout the analysis. In particular, in the examples of seed theories given by the $A$-series $\cN=2$ minimal models there will be a very similar looking classification of anticommuting TDLs as the ones presented in Table \ref{tab:3} and Table \ref{tab:4}, in the sense that only the condition on the labels $M$, $S$, and $\bar{S}$ will change.

We have investigated the preservation of TDL symmetries under marginal deformations; however, the general theorems derived in Sec.\,\ref{sec:4} are also valid for relevant deformations (in the twisted sector), and, as first discussed in \cite{Chang:2018iay}, these can be used to restrict possible endpoints of renormalization group flows in symmetric product orbifold CFTs. As discussed in the previous section on holographic applications, it would be interesting to determine whether deformations to a semiclassical gravity dual for the $\mathcal{N}=(2,2)$ seed theories exist and how the preserved TDL can be useful in constraining these duals. We plan to return to some of these questions in future work.

\subsection*{Acknowledgments}
We thank Yichul Choi, Sarah M. Harrison, Seolhwa Kim, Stathis Vitouladitis, and Yifan Wang for helpful discussions. CL is especially grateful to Sarah M. Harrison for suggesting looking at generalized symmetries in symmetric orbifolds to him. SB would like to thank CERN for its hospitality during the completion of this work. MG is grateful to the Department of Physics and Astronomy, University of Washington for hospitality when this paper was finalized. 

The research of SB and MG is supported in part by NSF grant PHY-22-09700.  SB, MG, and DR are grateful to the  Bhaumik Institute for Theoretical Physics for support. The research of NB, YJC, and CL is supported by the US Department of Energy, Office of Science, Office of High Energy Physics, under Award Number DE-SC0025704.

\appendix

\section{More on \texorpdfstring{$\boldsymbol{G}$}{G}-equivariantizations}
\label{app:equivariantization}

Consider a fusion category $\cC$ along with a group action $G$ on it. That is to say, there is a homomorphism from a finite group $G$ to the automorphism group of the fusion category $\cC$ that preserves its fusion structure\footnote{In mathematical terms, this is a monoidal functor from the skeletal $G$ category to the category of tensor auto-equivalences of $\cC$.  Its monoidal structure is given by group multiplication laws. Such a $G$ action naturally gives rise to an extension to $\cC$, known as $G$-crossed categories. More precisely, it can be viewed as a generalization of group extensions, where we instead extend a group by a category $\cC$.}
\begin{equation}
\alpha: G \to {\rm Aut}_{\otimes}(\cC) \,.
\label{action}
\end{equation}
Under this homomorphism, a group element $g$ is sent to the automorphism $\alpha_g$ that encodes how $g$ acts on objects in the fusion category $\cal C$.

One can then extend the symmetry by including $G$. This is done by the $G$-extension of fusion categories\cite{Etingof:2009yvg}. In the following, we focus on trivial extensions, which are directly applicable to global symmetries in symmetric product orbifolds. We denote the category after the trivial extension by $\cC \rtimes G$. In this case, we can describe the simple objects in $\cC \rtimes G$ in terms of Deligne's tensor products of simple objects in $\cC$ and $G$, while the fusion rules can be described in terms of the semi-direct product
\begin{equation}
(X\boxtimes g ) \otimes (Y \boxtimes h) := ( X \otimes \alpha_g(Y)) \boxtimes gh, \quad X, Y \in \cC, \; g,h \in G \,,
\end{equation}
where $\alpha_g$ is given by~\eqref{action}. In what follows, we denote $X\boxtimes g$ as $X^g$ and $\alpha_g(X)$ as ${}^gX$ for short. The associator is naturally induced from $\cC$ via
\ie
\alpha_{\cC \rtimes G; X^g, Y^h, Z^l} = \alpha_{\cC; X, {}^gY, {}^{gh}Z^{ghl}} = \alpha_{\cC; X, {}^gY, {}^{gh}Z} \,.
\fe
Here in the second equality, we make use of the fact that we can freely add lines from $G$ to $Z$ without changing the associator.

In this construction, the subsymmetry $G$ is always non-anomalous, and we can gauge it. The corresponding dual fusion category is known as the $G$-equivariantization of $\cC$, denoted as $\cC^G$. 
In the following, we give a derivation of its categorical structures in terms of symmetry gauging. 

Before doing so, we comment on nontrivial $G$-extensions in the context of two-dimensional QFT, and why they are not expected in the context of symmetric product orbifolds. One can define a $G$-extension by writing down a fusion category as
\ie
\cD = \bigoplus_{g\in G} \cC_g \,. 
\fe
such that $\cC_e = \cC$ and $\cL^g \otimes \cL^h \in \cC_{gh}$, for any $\cL^g \in \cC_g$ and $\cL^h \in \cC_h$. 
Roughly speaking, there are two different origins of nontriviality. First, there is the possibility of a nontrivial extension in the same sense as a nontrivial group extension. Namely, the fusion rules do not split, and generically there does not exist $\cL^g \in \cC_g$, $\cL^{g^{-1}} \in \cC_{g^{-1}}$ such that $\cL^g \otimes \cL^{g^{-1}}$ contains $\id$. The simplest example in this direction is to consider the extension of a $\mZ_2$ symmetry by a $G = \mZ_2$ action that acts trivially on the $\mZ_2$. A non-split extension leads to a $\mZ_4$ symmetry. A nontrivial extension like this drastically modifies the fusion rules of permutation TDLs of $S_N$ and obviously cannot be the right description for the permutation symmetry of a tensor product QFT $\cT^{\otimes N}$. 

The other origin of nontriviality is more subtle. It comes from the possibility that the three-way junctions for TDLs in $\cC$, as a linear vector space, carry nontrivial charges of $G$. This is the two-dimensional version of the well-known symmetry fractionalization\cite{Barkeshli:2014cna}, and it leads to the modification that dual TDLs now may carry the labels of projective representations of the stabilizer subgroups. In fact, in two-dimensional QFT, because both topological orders and symmetry operators are described by TDLs, it is also possible that the three-way junctions for TDLs in $G$ carry nontrivial charges of $\cC$.
And both the symmetry fractionalization of $\cC$ in $G$, and $G$ in $\cC$, are captured by the mixed 't Hooft anomaly between $\cC$ and $G$. We do not expect a tensor product QFT $\cT^{\otimes N}$ to exhibit such a mixed 't Hooft anomaly, because the different copies of the seed theory are not coupled to one another in any way.

\subsection[Simple objects in \texorpdfstring{$\cC^{G}$}{CG} as \texorpdfstring{$A$}{A}-bimodules]{Simple objects in \texorpdfstring{$\cC^{\boldsymbol{G}}$}{CG} as \texorpdfstring{$\boldsymbol{A}$}{A}-bimodules}
\label{app:bimodule_equivariantization}

In this section, we give a more systematic derivation of the set of simple objects in $\cC^G$ in the language of gauging. One reason we do that is to make the description in Sec.~\ref{subsec:equivariantization} precise, and the second reason is that the bimodule structures capture the information of TDL actions on (defect) local operators, as discussed in Sec.~\ref{subsec:A-bimodules}. Recall from Sec.~\ref{subsec:gauging} that a gauging procedure can be described by an algebra object. For our purpose here, the algebra object is simply $A = \sum_{g\in G} \id^g$. In general, there is a choice of discrete torsion characterized by $H^2(G, U(1))$, represented by the multiplication maps $m_{g,h}$. In this section, we assume that the torsion is trivial up to a normalization factor
\ie
m_{g,h} = \frac{1}{\sqrt{|G|}} \,.
\fe
It is not hard to check that the simple objects and fusion rules of $\cC^G$ do not depend on this assumption. Within a unitary gauge, the normalization by $\sqrt{\abs{G}}$ is chosen such that one can remove a bubble of $A$ without any cost. 

Recall that discrete gauging is a topological operation that corresponds to stacking a fine enough network of $A$ on the manifold where the QFT is defined. Furthermore, the dual TDLs are described by $A$-bimodules. In the same flavor used to solve the bimodules of the gauging $G \to {\rm Rep}(G)$, one can solve the bimodule structures of $\cC^G$ from the perspective of the gauging by $A = \sum_{g \in G} \cL_g$ in $\cC \rtimes G$. The results are as follows. 

We first solve the left $A$-module structure. A simple left $A$-module, in terms of the induced module argument, is given as an object in $\cC\rtimes G$ as
\ie
M = \sum_{g\in G} {}^g X^g, \quad X \in \cC \,.
\label{bimodule_object}
\fe
Its left action can be gauge fixed to be $\rho_{e, X^g} = \mathbb{1}/\sqrt{|G|}$ for all $X, g$. Note that the outgoing objects of the left and right actions are completely fixed by the ingoing ones, so we can denote these maps only by their ingoing legs. The left associativity reads
\ie
\rho_{h,X^l} \rho_{g,{}^hX^{hl}} = \rho_{gh,X^l} / \sqrt{|G|} \,.
\label{left_associativity}
\fe
It follows that we can use the residual gauge freedom of the left $A$-module to fix the left action to be proportional to the identity matrix, with proportionality $1/\sqrt{|G|}$. 

Due to the semi-direct product structure of $\cC\rtimes G$, the right $A$-action mixes different simple left $A$-modules, and the general form of a simple $A$-bimodule, as an object in $\cC$, is given by
\ie
M = n \sum_{Y \in \cO(X), g\in G} Y^g \,,
\fe
where we denote the $G$ orbit in ${\rm Irr}(\cC)/G$ that $X$ belongs to as $\cO(X)$, and $n$ is a positive integer to be fixed by the end of this subsection. In other words, we assume that the left and right actions are described by $n \times n$ matrices. From now on, we use $X$ to denote a fixed representative $X \in \cO(X)$, and use $Y$ to denote arbitrary elements in $\cO(X)$. Sometimes, we will also denote elements in $\cO(X)$ with ${}^hX$ for some $h \in G/G_X$. It is important to note that there are $|G|/|G_X|-1$ gauge degrees of freedom unfixed in the $A$-bimodule structures, which correspond to rotating the simple left $A$-modules within the simple $A$-bimodule. To completely fix this gauge freedom, we take a set of representatives $h_i \in G_X\backslash G$, $i=1,2,\dots, |G|/|G_X|$, and assume
\ie
\tilde\rho_{X,h_i} = \frac{\mathbb{1}}{\sqrt{|G|}} \,.
\label{gauge_fixing}
\fe 
Henceforth, we denote this set as $H$. Note that in the trivial coset, we pick the identity as representative $h_i=e$. 

The right associativity is given by
\ie
\tilde\rho_{Y^g,h} \tilde\rho_{Y^{gh},l} = \tilde\rho_{Y^g,hl} /\sqrt{|G|} \,.
\fe
For $g=e$, this implies that all right action maps can be represented by $\tilde\rho_{Y,h}$
\ie
\tilde\rho_{Y^h,l} = \tilde\rho_{Y,h}^{-1} \tilde\rho_{Y,hl} /\sqrt{|G|} \,.
\label{right_associativity}
\fe
The bimodule structures are completely solved once we find $\tilde\rho_{{}^hX,g}$ for arbitrary $h \in G/G_X$, $g\in G$. 

The bimodule compatibility condition says that
\ie
\tilde\rho_{{}^hY^{hg},l} = \tilde\rho_{Y^g,l} \,.
\label{compatibility}
\fe
Combining this with~\eqref{right_associativity}, by taking $h= g^{-1}$, we have
\ie
\tilde\rho_{Y,g} \tilde\rho_{{}^{g^{-1}}Y,l} = \tilde\rho_{Y,gl} / \sqrt{|G|}, \quad g\in G, \; l\in G \,.
\label{eqn:right_asso_2}
\fe
If we focus on $X$ and further require $g\in G_X$ in~\eqref{eqn:right_asso_2}, we obtain the following relation
\ie
\tilde\rho_{X,g} \tilde\rho_{X,l} = \tilde\rho_{X,gl} / \sqrt{|G|}, \quad g\in G_X, \; l\in G \,.
\label{eqn:right_rep}
\fe
This result has three important consequences.
First, it means that $\tilde\rho_{X,g}$ forms a linear representation $R$ of $G_X$: $\tilde\rho_{X,g} = R(g)/\sqrt{|G|}$, and $n = {\rm dim(R)}$. Second, with our convention of gauge fixing~\eqref{gauge_fixing}, it implies that $\tilde\rho_{X,gh_i} = \tilde\rho_{X,g}$ for any $h_i$. Finally, we can take $Y=X$, $h=h_i$, $l = h^{-1}_i g h_j$ in~\eqref{compatibility}, which leads to
\ie
\tilde\rho_{{}^{h_i^{-1}}X,h_i^{-1}gh_j} = \tilde\rho_{X,gh_j} \,, \quad i,j = 1,2,\dots,|G|/|G_X|, \; g\in G_X \,.
\label{eqn:right_action}
\fe
Because $h_i^{-1}gh_j$ covers every element in $G$, we have completely described the bimodule structures. Thus, we have completed our proof that simple $A$-bimodules are labeled by $G$ orbits $\cO(X) \in {\rm Irr}(\cC)/G$ and the corresponding irreducible representations of $G_X$. Note that the stabilizer subgroups $G_Y$ of different elements $Y \in \cO(X)$ are generally different; however, they are related by conjugation and hence are isomorphic. 

\subsection{Actions on local operators}
\label{app:action}

In this section, we give a detailed derivation of~\eqref{action_generic_symmetric_orbifold}. This is based on the interpretation of the action in terms of the actions of bimodules on the states of $S^1$, as illustrated in Figure~\ref{fig:bimodule_insertion}. 

We consider a local operator $O_{[g]}$ in the twisted sector labeled by $[g]$. By this, we mean to choose the basis of $g$ in the algebra object $A$ that goes through the temporal circle, where $g$ is a representative in $[g]$. The extended operator
\ie
\tikz[baseline=-0.75]{\draw[thick] (-2,0) -- (2,0); \draw[black, -{Stealth[round, length=5pt, width=5pt, bend]}] (-2,0)--(0,0); \draw[black, -{Stealth[round, length=5pt, width=5pt, bend]}] (-2,0)--(-1.1,0); \draw (-2,-1) -- (-2,1); \draw (2,-1) -- (2,1); \draw (-2,-1) -- (2,-1); \draw[thick] (-0.5,-1) -- (0.5,0); \draw[black, -{Stealth[round, length=5pt, width=5pt, bend]}] (-0.5,-1)--(0,-0.5); \draw[thick] (-0.5,0) -- (0.5,1); \draw[black, -{Stealth[round, length=5pt, width=5pt, bend]}] (-0.5,0)--(0,0.5); \node[below] at (0.2,-1) {$|O_{[g]}\ra$}; \node[left] at (-0.20,-0.5) {$\cL_g\subset A$}; \node[above] at (-1.1,0) {$A$}; \node[left] at (0,0.7) {$\cL_g\subset A$}; \node[right] at (0,0.25) {$A$}; }
\fe
is a projector that kills $G$ charged states. So we will assume $O_{[g]}$ is neutral from now on and focus on the action from the bimodule $M$. 

Given a $g$, and a dual TDL labeled by $(X,R)$, we have
\ie
M = \dim(R) \sum_{Y\in \cO(X), h\in G} Y^h \,.
\fe
The bases in $M$ are compatible with the fusion channel on the cylinder only when ${}^gY = Y$ and $h\in C_g$, where $C_g = \{h\in G| gh=hg, \forall g\in G\}$ is the centralizer of $g$. We also denote $\cO_g(X) = \{Y \in \cO(X)| {}^gY = Y\}$. 

Now we look at the bimodule actions of these channels. For compactness, we use a different normalization in~\eqref{eqn:equality_1} -~\eqref{eqn:equality_3} such that the overall $1/\sqrt{|G|}$ factor is thrown away. We will recover it right after. Firstly, we look at the specific representative $X$ from $\cO(X)$. Using~\eqref{right_associativity} we have
\ie
\tilde\rho_{X^h,g} = \tilde\rho^{-1}_{X,h} \tilde\rho_{X,hg} = \tilde\rho^{-1}_{X,h} \tilde\rho_{X,gh} = \tilde\rho^{-1}_{X,h'h_i} \tilde\rho_{X,gh'h_i} = \tilde\rho^{-1}_{X,h'} \tilde\rho_{X,gh'} = \tilde\rho_{X,g} \,,
\label{eqn:equality_1}
\fe
where we have uniquely decomposed $h = h'h_i$ for $h'\in G_X$ and $h_i \in H$. This shows that the right action is independent of $h \in C_g$. 

Then we consider a ${}^{h_j^{-1}}X \in \cO_g(X)$. Using~\eqref{compatibility} and~\eqref{right_associativity} we have
\ie
\tilde\rho_{{}^{h_j^{-1}}X,g} = \tilde\rho_{X^{h_j},g} = \tilde\rho_{X,h_j}^{-1} \tilde\rho_{X,h_j g} = \tilde\rho_{X,h_j g} = \tilde\rho_{X,h_i g h_i^{-1} h_i} = \tilde\rho_{X,h_i g h_i^{-1}} \,.
\fe
We note that the choice $h_j \in G_X \backslash G$ such that ${}^{g h_j^{-1}}X = {}^{h_j^{-1}}X$ satisfies $h_j g h_j^{-1} \in G_X$ because ${}^{h_j g h_j^{-1}}X = {}^{h_j h_j^{-1}}X = X$. Finally, we discuss the most general case, ${}^{h_j^{-1}}X^h$, and show that the right action is independent of $h \in C_g$. Using~\eqref{compatibility},~\eqref{right_associativity} and~\eqref{eqn:right_rep}, we have
\ie
\tilde\rho_{{}^{h_j^{-1}}X^h,g} = \tilde\rho_{X^{h_j h},g} = \tilde\rho_{X,h_j h}^{-1} \tilde\rho_{X,h_j hg} = \tilde\rho_{X,h_j h}^{-1} \tilde\rho_{X,h_j g h} = \tilde\rho_{X,h_j h}^{-1} \tilde\rho_{X,h_j g h_j^{-1}} \tilde\rho_{X,h_j h} \,.
\label{eqn:equality_3}
\fe
By implementing the equation in Fig.~\ref{fig:bimodule_insertion}, we thus find the following relation
\ie
\cL_{X,R} \cdot O_{[g]} = \sum_{g\in [g]} \sum_{ \{h^i \in H | {}^{h_i}X \in \cO_g(X)\} }\la {}^{h_i}XO_g \ra |C_g| \frac{\chi_R(h_i g h_i^{-1})}{|G|} O_{[g]} \,.
\label{eqn:action_1}
\fe
Here, $\la {}^{h_i}XO_g \ra = \la O_g| {}^{h_i}X | O_g \ra$ denotes the action in the ungauged theory of ${}^{h_i}X$ on $O_g$, which is a local operator attached to a TDL labelled by $g$. In the context of symmetric product orbifolds, it is completely captured by the actions of seed TDLs on factorized cycles of Hilbert spaces of $[g]$, as discussed in Sec.~\ref{subsubsec:non-universal_defects}. Finally, we point out that the contributions from different $g\in [g]$ are all identical because they are related by an overall conjugation, and we can simplify~\eqref{eqn:action_1} to
\ie
\cL_{X,R} \cdot O_{[g]} = \sum_{ \{h^i \in H | {}^{h_i}X \in \cO_g(X)\} } \la {}^{h_i}XO_g \ra \chi_R(h_i g h_i^{-1}) O_{[g]} \,,
\fe
where we make use of the identity $|C_g||[g]| = |G|$. 

The quantum dimension of $\cL_{X,R}$ is identical to the action on the identity operator. In this case $g = e$, so $O_g=1_e$, and hence $\langle{}^{h_i}XO_g\rangle = \la X \ra$. Furthermore, $\chi_R(h_i h_i^{-1}) = \dim R$, and $\cO_g(X) = \cO(X)$. We thus find
\ie
\la \cL_{X,R} \ra = |\cO(X)| \dim(R) \la X \ra \,.
\label{eqn:qdim_L}
\fe
Using the fact that
\ie
|\la {}^{h_i}XO_g \ra| \leq \la X \ra \,,
\fe
which can be proven by extending the arguments for \eqref{eqn:lemma_qdim} to $g$-twisted Hilbert spaces, we can verify that\footnote{This is possible because in this case one can blow an $X$ bubble on the $g$ TDL without cost. } 
\ie
|\la O_{[g]} |\hat\cL_{X,R} | O_{[g]} \ra| \leq \la \cL_{X,R} \ra\,.
\fe

\subsection{Some comments on F symbols}
\label{app:F_symbols}

We conclude this appendix with some comments on F symbols. We will not work out a general expression for F symbols of $\cC^G$, but leave that for future work.  
Instead, we will argue what parts of the categorical information contribute to F symbols, and in the process, shed some light on the possible complexity. The key point is that the F symbols are not fully determined by the $6j$ symbols of the representation category of the common stabilizer subgroup that participates in the three-to-one fusion channel. In addition, the information of the representation label of the full stabilizer subgroup for each simple object that is relevant to the F symbols contributes. 

To illustrate this point, we go back to the toy example $\mZ_2^2 \rtimes \mZ_2$ and work out its F symbols by explicitly solving the three-way and four-way $A$-bimodule vertices. We focus, in particular, on one set of F symbols $F_{x\cN y}^\cN$, where $x,y \in \{(\id,\id) \oplus (\id, \theta') \oplus (\eta_1 \eta_2, \id) \oplus (\eta_1 \eta_2, \theta')\}$. In this case, the common stabilizer subgroups are always trivial, but we also know the F symbols must be nontrivial. 

\begin{figure}[htbp]
     \centering
     \begin{subfigure}[b]{0.45\textwidth}
         \centering
         \includegraphics[width=\textwidth]{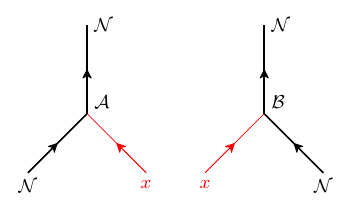}
         \caption{Three-way bimodule vertices}
         \label{fig:3way_bimodule}
     \end{subfigure}
     \hfill
     \begin{subfigure}[b]{0.45\textwidth}
         \centering
         \includegraphics[width=\textwidth]{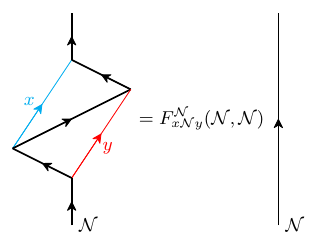}
         \caption{F symbols of bimodules}
         \label{fig:bimodule_fsymbol}
     \end{subfigure}
     \caption{F symbols in the language of $A$-bimodule}
\end{figure}

First, we rewrite the dual TDLs in terms of $A$-bimodules. The non-invertible TDL, as an object in ${\rm Vec}_{\mZ_2^2} \rtimes \mZ_2$ is given by $M = \eta_1 \oplus \eta_1^{\theta} \oplus \eta_2 \oplus \eta_2^{\theta}$, and both left and right actions are trivial. Four invertible lines are given by $M = \id \oplus \id^{\theta}$ or $\eta_1\eta_2 \oplus \eta_1\eta_2^{\theta}$ with trivial left actions. Their right actions are $\tilde\rho_{-,\theta} = \tilde\rho_{-^{\theta},\theta} = \pm 1$ depending on the label of ${\rm Rep}(\mZ_2)$. One can fill in the hyphen with $\id$ or $\eta_1\eta_2$ and similarly in the following.  

We then solve two sets of three-way bimodule vertices in Figure~\ref{fig:3way_bimodule} using the compatibility conditions of the $A$-bimodule tensor product. The results are as follows:
\ie
\cA_{\eta_1,-} = \cA_{\eta_2^{\theta},-} = \cA_{\eta_2,-^{\theta}} = \cA_{\eta_1^\theta,-^\theta} = \frac{1}{\sqrt{2}} \,, \\
\cA_{\eta_2,-} = \cA_{\eta_1^{\theta},-} = \cA_{\eta_1,-^{\theta}} = \cA_{\eta_2^\theta,-^\theta} = \pm \frac{1}{\sqrt{2}} \,, \\
\cB_{-,\eta_1} = \cB_{-^\theta,\eta_1} = \cB_{-^\theta,\eta_1^\theta} = \cB_{-,\eta_1^\theta} = \frac{1}{\sqrt{2}} \,, \\
\cB_{-,\eta_2} = \cB_{-^\theta,\eta_2} = \cB_{-^\theta,\eta_2^\theta} = \cB_{-,\eta_2^\theta} = \pm \frac{1}{\sqrt{2}} \,.
\label{eqn:3-way_junction}
\fe
Note that one needs to fix a gauge for each fusion channel of dual TDLs. The three-way vertices are normalized such that the quantum dimensions of dual TDLs are properly normalized. Then, we can express the F symbols of $A$-bimodules in terms of three-way vertices, as shown in Figure~\ref{fig:bimodule_fsymbol}, and find the following result:
\ie
F_{x\cN y}^{\cN} = \begin{pmatrix}
    1 & 1 & 1 & 1 \\
    1 & 1 & -1 & -1 \\
    1 & -1 & 1 & -1 \\
    1 & -1 & -1 & 1 
\end{pmatrix} \,.
\fe
These are exactly the F symbols for ${\rm Rep}(D_8)$. Note that it does not depend on the choice of signs in~\eqref{eqn:3-way_junction}.  

\section{TDLs and folded boundary states}
\label{app:c}

In this appendix, we describe TDLs as boundary states using the folding trick. We illustrate the construction for Verlinde lines in RCFTs and then apply them to symmetric product orbifold CFTs and present the details of calculations of the quantum dimension and the action on local operators, which we use in the main body of the paper. 
While we use RCFTs as seed CFTs in this appendix, we expect the calculations to generalize straightforwardly to commutative fusion categories.

\subsection{Topological defect lines in rational CFTs}

We consider RCFTs with diagonal partition functions
\begin{align}
    Z(\tau,\bar{\tau}) = \sum_{i,j} \delta_{ij}\;  \chi_i(\tau) \bar{\chi}_j(\bar{\tau})~,
\end{align}
where $i,j\in\{1,2,\dots,b\}$ label the primary fields.
Consider two copies of the rational CFT, which we call CFT$_I$ and CFT$_{II}$, that are separated by a TDL. Following Zuber and Petkova, the TDL operators can be constructed using the following projection operators \cite{Petkova:2000ip}
\begin{align}\label{projecta}
    P_{i\bar{i}} = \sum_{n,m} \ket{i,n}^{(I)} \otimes {\overline{\ket{i,m} }}^{(I)}  \;\;{}^{(II)}\bra{i,n} \otimes  {}^{(II)}\overline{\bra{i,m}} \,.
\end{align}
Here, the sum over $n,m$ is over all Virasoro descendants (or more generally descendants of an extended chiral algebra)  of the primary $i$ in analogy with the construction of Ishibashi states in BCFTs \cite{Ishibashi:1988kg}. 
The TDLs are then constructed using the modular S-matrix of the RCFT
\begin{align}\label{rcft defect}
    {\mathcal L}_a = \sum_i \frac{S_{ai}}{S_{0i}} P_{i \bar{i}}~.
\end{align}
The defects commute with the holomorphic and antiholomorphic Virasoro modes separately and are hence topological
\begin{align}
    L^{(I)}_{n} \; {\mathcal L} _a  &= {\mathcal L} _a \; L^{(II)}_n~, \qquad  \text{and}\qquad \bar L_{n}^{(I)} \;  {\mathcal L} _a  = {\mathcal L} _a  \;\bar L^{(II)}_n~.
 \end{align}
 The quantum  dimension $\langle {\cal L}_a\rangle $ of the TDL is given by
 \begin{align}
     \langle {\mathcal L}_a\rangle  = \langle 0 |\hat{\mathcal L}_a | 0 \rangle = \frac{S_{a0}}{S_{00}}\,.
 \end{align}
The action of a TDL on a primary  operator $\phi^l$ is defined as
\begin{align}\label{actionphi}
    {\mathcal L}_a \cdot \phi^l =  \langle {\mathcal L}_a \phi^l\rangle \; \phi^l = \frac{S_{al}}{S_{0l}} \; \phi^l \,.
\end{align}

\begin{figure}[htbp]
  \centering
  \begin{tikzpicture}[scale=1.7]
    \draw[very thick] (0,0) rectangle (3,2);
    \draw[red, very thick] (1.5,0) -- (1.5,2);
    \node at (0.75,1) {$\mathrm{CFT}_{(I)}$};
    \node at (2.25,1) {$\mathrm{CFT}_{(II)}$};
    \node[above] at (1.5,2) {\color{red}$\mathcal{L}_a$};

    \draw[->, thick] (3.6,1) -- (4.5,1);

    \coordinate (A) at (5,0);
    \coordinate (B) at (5,2);
    \coordinate (C) at (6.5,2);
    \coordinate (D) at (6.5,0);
    \draw[very thick] (C) -- ++(220:1.4) coordinate (F);
    \draw[very thick] (D) -- ++(220:1.4) coordinate (G);
    \draw[very thick] (F) -- (G);
    \draw[very thick] (A) -- (B) -- (C);
    \draw[red, very thick] (C) -- (D);

    \path[name path=bottom] (A) -- (D);
    \path[name path=foldline] (F) -- (G);
    \path[name intersections={of=bottom and foldline, by=I}];
    \draw[very thick] (A) -- (I);

    \node at (5.45,1.75) {$\mathrm{CFT}_{(I)}$};
    \node at (6,0.45) {$\overline{\mathrm{CFT}}_{(II)}$};
    \node at (6.8,1) {\color{red}{$|B_a\rangle\!\rangle$}};
  \end{tikzpicture}
  \caption{Folding trick: CFT$_{(II)}$ is folded along the defect line, turning it into a boundary state of CFT$_{(I)} \otimes \overline{\text{CFT}}_{(II)}$. We take time to run orthogonal to the defect.}
  \label{fig:folding-trick}
\end{figure}
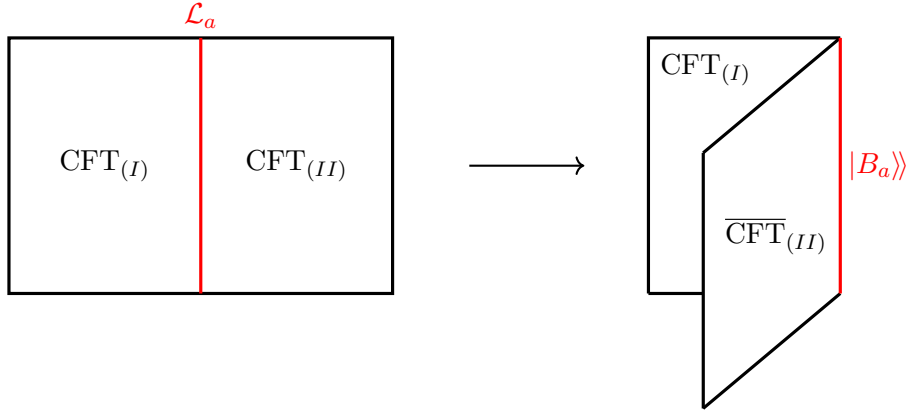

Using the folding trick, illustrated in Figure \ref{fig:folding-trick}, the TDL can alternatively be described as a boundary state in CFT$_{(I)} \otimes \overbar{\text{CFT}}_{(II)}$ \cite{Oshikawa:1996dj,Bachas:2001vj}. Here $\overbar{\text{CFT}}_{(II)}$ is the folded CFT where the folding is performed by a worldsheet parity transformation that exchanges the holomorphic and antiholomorphic fields. The projector (\ref{projecta}) turns into
\begin{align}
    \mid B_i \rangle\rangle = \sum_{n,m}  \ket{i,n}^{(I)} \otimes   \overbar{\ket{i,n}}^{(II)}  \otimes \overbar{\ket{i,m}}^{(I)} \otimes {\ket{i,m}}^{(II)}  \,,
\end{align}
and satisfies \begin{align}
    \Big(L^{(I)} _n-\bar L^{(II)}_{-n}\Big) \mid B_i\rangle\rangle\  =  \Big(L^{(II)} _n-\bar L^{(I)}_{-n}\Big) \mid B_i \rangle\rangle\  =0 \,.
\end{align}
The TDL then  corresponds to the superposition
\begin{align}
    \mid B_a\rangle\rangle = \sum_i \frac{S_{ai}}{S_{0i}} \; \mid B_i\rangle\rangle\,.
\end{align}
The quantum dimension of the TDL is identified with the vacuum overlap
\begin{align}\label{qdbstate}
    \langle {\mathcal L}_a\rangle  = {}^{(I)}\langle 0 \mid  \otimes  {}^{(II)}\langle 0 \mid \mid B_a\rangle\rangle\,,
\end{align}
and the action of a primary field on the TDL is determined by the overlap
\begin{align}
    {}^{(I)}\langle \phi^l  \mid B_a\rangle\rangle = \langle {\mathcal L}_a \phi^l\rangle \mid \phi^l\rangle^{(II)}\,.
\end{align}

\subsection{Boundary states for non-universal TDLs}
In this section, we construct the folded boundary states for non-universal TDLs in symmetric product orbifold CFTs. These lines depend on the details of the TDLs of the seed theory. For a RCFT the seed theory has $r$ simple TDL denoted by $\cL_i, i \in\{ 1, \dots, r\}$. The non-universal TDLs of the symmetric product orbifold CFT are characterized by two sets of data.  First, an ordered list $X$ of TDLs in the tensor product ${\mathcal T}^{\otimes N}$
\begin{align}\label{prod_states}
 X= \Big(\underbrace{
    {\mathcal L_1}, {\mathcal L_1},\cdots,{\mathcal L_1} }_{n_1}, \underbrace{
    {\mathcal L_2}, {\mathcal L_2},\cdots,{\mathcal L_2} }_{n_2}, \cdots, \underbrace{
    {\mathcal L_r}, {\mathcal L_r},\cdots,{\mathcal L_r} }_{n_r}\Big)\,.
\end{align}
The set $X$ can alternatively be characterized by a set of $r$ integers 
\begin{align}
 \vec{n}=(n_1,n_2,\cdots, n_r)\,,   
\end{align}
 which is a partition of $N$, that is,  $\sum_i n_i =N$. This partition indicates how many copies of ${\mathcal L_i}$ are present in the tensor product.  The stabilizer of $X$ is 
\begin{align}
    G_X= S_{n_1}\times S_{n_2}\times \cdots \times S_{n_r}\,.
\end{align}
The second set of data is a representation of $G_X$, i.e. a set of representations $R_i$  of $S_{n_i}$
\begin{align}
    R= (R_1,R_2, \cdots, R_r)\,.
\end{align}
In the following, we construct the folded boundary state corresponding to the TDL ${\mathcal L}_{X,R}$.
 For a given $\vec{n}$ we partition the ordered set of $N$ integers into $r$ sets $N_1 = \{1,\cdots, n_1\}, \ N_2=\{n_1+1,\cdots, n_1+n_2\},\ \cdots,\  N_r=\{N-n_r+1, N\}$.  
The symmetric group $S_{n_i}$ then acts on the set $N_i$.  For each $i\in\{1,\dots,r\}$, an element  $g_i\in S_{n_i}$ can be decomposed into a product of cycles 
\begin{align}\label{gielement}
    g_i = \sigma_1^{(i)}\cdots  \sigma_2^{(i)} \cdots \sigma_{m_i}^{(i)} \in S_{n_i}(N_i)\,.
\end{align}
Here, we denote with $S_{n_i}(N_i)$ the symmetric group $n$ acting on the set $N_i$, and the single cycle $\sigma_k^{(i)}$ is associated with  $\sigma_k^{(i)}$ twisted sector boundary state
\begin{align}\label{ishibashi}
  \mid a_i, \sigma_k^{(i)} \rangle\rangle^{(I)} \mid a_i, \sigma_k^{(i)} \rangle\rangle^{(II)}=  \sum_{j=1}^r \frac{S_{a_i j}}{S_{j0}} \sum_{n,m} \ket{j,n}^{(I)}_{\sigma_k^{(i)}} \otimes   \overbar{\ket{j,n}}^{(II)}_{\sigma_k^{(i)}}  \otimes \overbar{\ket{j,m}}^{(I)}_{\sigma_k^{(i)}}  \otimes {\ket{j,m}}^{(II)} _{\sigma_k^{(i)}}  \,,
\end{align}
where the sum over $m,n$ is over the (fractional) Virasoro descendents in the $\sigma_k^{(i)}$ twisted sector, and the labels $a_i$ label the boundary states of the seed theory.
The boundary state in the $g_i$ twisted sector is then the product of the single cycle twists (\ref{gielement})
\begin{align}
    \mid (a_i)_{g_i} \rangle\rangle^{(I)} \mid (a_i)_{g_i}  \rangle\rangle^{(II)}= \prod_{k=1}^{m_i}  \mid a_i, \sigma_k^{(i)} \rangle\rangle^{(I)} \mid a_i, \sigma_k^{(i)} \rangle\rangle^{(II)}.
    \end{align}
\noindent
The  TDL is labeled by the data $X$ and $R$ is then given by
\eqsp{\label{foldedbst}
    &\mid X,R\;\rangle\rangle\coloneqq \frac{1}{N!} \sum_{\substack{h^{(I)}\in  S_N \\ h^{(II)}\in S_N}} \prod_{i=1}^r \\ 
    &\hspace{40pt}\left( \sum_{g_i \in S_{n_i}(N_i)} \frac{\chi_{R_i}(g_i)}{n_i!} \; \mid (a_i)_{h^{(I)}\;g_i\;(h^{(I)})^{-1}}\rangle\rangle^{(I)} \otimes \mid(a_i)_{h^{(II)}\;g_i\;(h^{(II)})^{-1}}\rangle\rangle^{(II)} \right)\,.
}
The summation over $h^{(I)} \in S_N$ and $h^{(II)} \in S_N$ is to ensure   invariance under the full $S_N$ permutation group for both $\text{CFT}_{(I)}$ and $\text{CFT}_{(II)}$. The action is by conjugation acting on the group element $g= \prod_{i=1}^r g_i$ in a slight abuse of notation.

\subsection{Quantum dimension and action on local operators}

The quantum dimension of the TDL ${\cal L}_{X,R}$ corresponding to the doubled boundary state $\mid X,R\;\rangle\rangle$ is determined by the overlap of the vacuum in the product CFT. Since the vacuum  is in the untwisted sector, it follows that only the terms with $g_i$ equal to the identity permutation acting on the set $N_i$ contribute to the quantum dimension
 \begin{align}\label{qdim-bdrystate}
  \langle \mathcal{L}_{X,R} \rangle&=  {}^{(I)}\bra{0} \otimes{}^{(II)}\bra{0}  X,R\rangle\rangle\nonumber\\
  &= \frac{1}{N!} \sum_{\substack{h^{(I)}\in S_N \\ h^{(II)}\in S_N}}  \prod_{i=1}^r \left( \frac{\chi_{R_i}(e)}{n_i!} {}^{(I)}\bra{0} \otimes{}^{(II)}\bra{0} \mid (a_i)_{\rm{id}} \rangle\rangle^{(I)} \otimes \mid (a_i)_{\rm{id}} \rangle\rangle^{(II)}  \right)\nonumber \\
    &= N! \prod_{i=1}^r \frac{\chi_{R_i}(e)}{n_i!} \left(\frac{S_{a_i 0 }}{S_{00} }\right)^{n_i} \nonumber \\
    &=
    \frac{|S_N|}{|G_X|} \prod_{i=1}^r  {\rm dim}(R_i) \langle\mathcal L_i\rangle^{n_i}\,.
\end{align}
Let us now compute the action of the TDL ${\mathcal L}_{X,R}$ on a single cycle  twist-$w$ state of the $l$-th primary  $\phi^l_{[w]}$ of the seed CFT
\begin{align}
     {\mathcal L}_{X,R} \cdot \phi^l_{[w]} = \langle  \mathcal{L}_{X,R}   \phi^l_{[w]}\rangle \;   \phi^l_{[w]}\,.
\end{align}
Using the boundary state formalism, this quantity can be calculated from the overlap of the state corresponding to $\phi^l_{[w]}$ in CFT${}^{(I)}$, which is 
normalized as follows
\begin{align}
    \ket{\phi^l_{[w]}} ^{(I)}= \frac{1}{\sqrt{wN!(N-w)!}} \sum_{h\in S_N} \ket{\phi^l_{h(1\dots w)h^{-1}}}^{(I)}.
\end{align}
The overlap with the folded boundary state (\ref{foldedbst}) is then
   \begin{align} \label{nr_woverlap}
{}^{(I)}\langle \phi^l_{[w]}\mid X,R\rangle\rangle& = \frac{1}{N!} \frac{1}{\sqrt{w\;N!(N-w)!}}   \sum_{h\in S_N}  {}^{(I)}\bra{\phi^l_{h(1\dots w)h^{-1}}}\sum_{\substack{h^{(I)}\in S_N \\ h^{(II)}\in S_N}} \nonumber\\
& \times \prod_{i=1}^r \left( \sum_{g_i \in S_{n_i}(N_i)} \frac{\chi_{R_i}(g_i)}{n_i!} \; \mid(a_i)_{h^{(I)}\;g_i\;(h^{(I)})^{-1}}\rangle\rangle^{(I)} \otimes \mid (a_i)_{h^{(II)}\;g_i\;(h^{(II)})^{-1}}\rangle\rangle^{(II)} \right)
\nonumber\\
&= \frac{1}{\sqrt{w\;N!(N-w)!}} \;{}^{(I)}\bra{\phi^l_{(1\dots w)}} \sum_{\substack{k\in S_N \\ h^{(II)}\in S_N}}\prod_{i=1}^r  \Big(\sum_{g_i \in S_{n_i}(N_i)} \frac{\chi_{R_i}(g_i)}{n_i!} \nonumber\\
&\quad \mid(a_i)_{k \;g_i\;k^{-1}}\rangle\rangle^{(I)} \otimes \mid(a_i)_{h^{(II)}\;g_i\;(h^{(II)})^{-1}}\rangle\rangle^{(II)} \Big)\,.
\end{align}
In going to the second line, we have made use of $S_N$ invariance and the  following identity
\begin{align}
  {}^{(I)} \langle \phi^l_{h(1\dots w)h^{-1}} \mid(a_i)_{h^{(I)}\;g_i\;(h^{(I)})^{-1}}\rangle\rangle^{(I)} &= {}^{(I)} \bra{ \phi^l_{(1\dots w)^{-1}} }U^\dagger(h) U(h^{(I)})  (a_i)_{g_i}\rangle\rangle^{(I)} \nonumber\\
 &= {}^{(I)} \bra{ \phi^l_{(1\dots w)^{-1}} }  (a_i)_{k g_i k^{-1}}\rangle\rangle^{(I)}\,,
\end{align}
where $k= h^{-1} h^{(I)}$ and $U(h)$ denotes the conjugation by $h\in S_N$.  One factor of $N!$ in the first line of (\ref{nr_woverlap}) is cancelled by one sum over $S_N$ on which the overlap does not depend. The resulting inner product is non-zero only if $g = \prod_{i=1}^r g_i$ contains exactly one cycle of length $w$ with the other cycles being trivial and $k$ is such that $kgk^{-1} = (1 \cdots w)^{-1} = (w \cdots 1).$ Let us count in how many ways that is possible. Without loss of generality, suppose the $w$-cycle lies in $S_{n_i}$ and that $g_{j\neq i} = e$. First, we require $\delta(n_i \geq w)$. Now, from the $n_i$ labels, there are $\binom{n_i}{w}$ ways to uniquely pick $w$ labels. Up to cyclic permutations, there are then $(w-1)!$ many ways to permute the labels within each $w$-cycle. Thus, there are $\sum_{i=1}^b \delta(n_i \geq w) \frac{n_i!}{w(n_i - w)!}$ choices for $g$ that lead to a single $w$-cycle. For each such choice of $g$, there are then $w(N-w)!$ choices for $k$ that map the given $w$-cycle to $(w \dots 1).$ Putting all of this together, we find
\begin{align} {}^{(I)}\langle{\phi^l_{[w]}}\vert{X,R}\rangle\rangle& =w(N-w)! \Bigg( \sum_{i=1}^r\delta(n_i\ge w)\frac{\chi_{R_i}(w)}{w(n_i-w)!}\nonumber \\
& \qquad\times {}^{(I)}\langle \phi^l_{(1\cdots w)}\mid (a_i)_{(w\cdots 1)}\rangle\rangle^{(I)}\big(\langle 0\mid a_i\rangle\rangle^{(I)}\big)^{n_i-w}\prod_{\substack{j=1\\j\neq i}}^r\frac{\chi_{R_j}(e)}{n_j!}\big(\langle 0\mid a_j\rangle\rangle^{(I)}\big)^{n_j} \Bigg) \nonumber \\
& \qquad\times \frac{1}{\sqrt{wN!(N-w)!}} \sum_{h^{(II)} \in S_N}  \ket{\phi^l_{h^{(II)}(w\dots 1)(h^{(II)})^{-1}}}^{(II)} \nonumber \\
&\eqqcolon \langle  \; \mathcal{L}_{X,R} \;\phi^l_{[w]}\rangle \ket{(\phi^l)_w}^{(II)}.
\end{align}
Note that in the overlap of ${}^{(I)}\bra{(\phi^l)_{h(1\dots w)h^{-1}}}$ with $\ket{(a_i)_{h^{(I)}\;g_i\;(h^{(I)})^{-1}}}^{(I)},$ the sum over primaries in the definition of the Ishibashi states \eqref{ishibashi} will collapse to just the primary corresponding to the twisted primary field  $(\phi^l)_w.$ Furthermore, since we sum over the same primary in copies $(I)$ and $(II)$, this overlap restricts $\ket{(a_i)_{h^{(I)}\;g_i\;(h^{(I)})^{-1}}}^{(II)}$ to be such that $(a_i) = \phi^l$. As a result, we end up with $\ket{(\phi^l)_w}^{(II)}$. Evaluating the various inner products in RCFT boundary states gives
\begin{align}\label{phiL}
 \langle \; \mathcal{L}_{X,R} \;\phi^l_{[w]} \rangle &=   w(N-w)! \sum_{i=1}^r\delta(n_i\ge w)\frac{\chi_{R_i}(w)}{w(n_i-w)!} \frac{S_{il}}{S_{l0}} \left( \frac{S_{i 0}}{S_{00}}\right)^{n_i-w} \prod_{\substack{j=1\\j\neq i}}^r\frac{\chi_{R_j}(e)}{n_j!} \left( \frac{S_{j 0}}{S_{00}}\right)^{n_j}\nonumber \\
 &\hspace{-10pt}= w(N-w)! \sum_{i=1}^r\delta(n_i\ge w)\frac{\chi_{R_i}(w)}{w(n_i-w)!}  \langle {\mathcal L}_i \phi^l \rangle \langle {\mathcal L}_i\rangle^{n_i-w}\prod_{\substack{j=1\\j\neq i}}^r\frac{{\rm dim}(R_j)}{n_j!} \langle {\mathcal L}_j\rangle^{n_j}\,.
\end{align}

We can compare this expression to \eqref{action_generic_symmetric_orbifold}. To do so, we fix $g = (w\dots 1)$ in \eqref{action_generic_symmetric_orbifold}, and we need to enumerate all the elements in $\cO(X)$ that are stabilized by $g$. This stabilization is possible for each $n_i \geq w$ when there are $w$ $\cL_i$'s in the first $w$ copies. We can freely arrange all the remaining seed TDLs in the remaining $N-w$ slots (in this way of thinking about the defects, the list in \eqref{prod_statesbb} should be thought of as an unordered list, so that any $\cL_i$ can be in any slot). There is a combinatorial factor of $\frac{(N-w)!}{(n_i-w)!\prod_{j\neq i}n_j!}$ coming from the inequivalent $h_i$'s that contribute equally to the action. Finally, we note the correspondence
\ie
\la {}^{h_i}XO_g \ra &= \langle {\mathcal L}_i \phi^l \rangle \langle {\mathcal L}_i\rangle^{n_i-w} \prod_{j\neq i} \la \cL_j \ra^{n_j} \,, \\
\chi_R(h_i g h_i^{-1}) &= \chi_{R_i}([w]) \prod_{j\neq i} \dim(R_j) \,.
\fe
With these ingredients, we can evaluate \eqref{action_generic_symmetric_orbifold} and find complete agreement with \eqref{phiL}.

\subsection*{Totally symmetric defects}\label{app:C.4}
Special cases of non-universal TDLs are the so-called totally symmetric TDLs. This special case corresponds to a partition of $N$ where one single $n_i=N$ and all other $n_j$'s are zero. Hence, $X$ and $G_X$ are given by
\begin{align}\label{prod_statesb}
 X= \Big(\underbrace{
    {\mathcal L_a}, {\mathcal L_a},\cdots,{\mathcal L_a} }_{N}\Big), \qquad \text{and}\qquad G_X=S_N\,.
\end{align}
The totally symmetric defect is thus characterized by the choice of seed TDL ${\mathcal L}_a$ and representation $R$ of $S_N$. The quantum dimension and action of a local operator on the defect can be obtained by specializing the non-universal case to (\ref{prod_statesb}). The quantum dimension is given by
\begin{align}\label{eq:totally_sym_qdim}
    \langle {\mathcal L}_{a,R}\rangle= \text{dim}(R) \langle \mathcal L_a\rangle^N\,,
\end{align}
and the action of the TDL on a twist-$w$ state of the $l$-th primary  $\phi^l_{[w]}$  is given by
\begin{align}\label{phiL-max}
 \langle  \; \mathcal{L}_{a,R}  \; \phi^l_{[w]}\rangle &= \chi_R([w]) \langle {\mathcal L}_a \phi^l \rangle \langle {\mathcal L}_a\rangle^{N-w}\,.
 \end{align}

\section{\texorpdfstring{$\boldsymbol{\mathcal{N}=2}$}{N=2} superconformal algebra}\label{app:N=2}

The $\mathcal{N}=2$ superconformal algebra contains the following field content: a weight 2 stress tensor $T$, a weight 1 $U(1)$ current $J$, and 2 fermionic weight $3/2$ currents $G^\pm$. The fermionic currents have $U(1)$ charge $\pm1$ respectively, as is reflected in the commutator of the generators below. In terms of the generators of the currents, the $\mathcal{N}=2$ algebra is given by
\begin{equation}
\begin{aligned}\label{superVir comms}
\bigl[L_m,L_n\bigr]
 ={}&
  (m-n)L_{m+n} +\frac{c}{12}m(m^2-1)\delta_{m+n,0}\,,
   \\
\bigl[J_m,J_n\bigr]
={}&
\frac{c}{3}m\,\delta_{m+n,0}\,,
  \\
\bigl\{G^+_r,G^-_s\bigr\}
 ={}&
2L_{r+s}+(r-s)J_{r+s} +\frac{c}{3}\Big(r^2-\frac{1}{4}\Big)\delta_{r+s,0}\,,
 \\
\bigl[L_m,J_n\bigr]
={}&
 -n\, J_{m+n}\,, \\ 
 \bigl[J_m,G^\pm_r\bigr]
  ={}&
  \pm G^\pm_{m+r}\,,
 \\
  \bigl[L_m,G^\pm_r\bigr]
  ={}&
   \Big(\frac{m}{2}-r\Big) G^\pm_{m+r}\,,
   \\
\bigl\{G^\pm_r,G^\pm_s\bigr\}
={}&
0\ .
\end{aligned}
\end{equation}
The central charge is denoted as $c$, and the $U(1)_R$ level is $\hat t=c/6$. The weight and $U(1)$ charge of a state are given by the eigenvalues of $L_0$ and $J_0$, respectively.

The $\mathcal{N}=2$ superconformal algebra is invariant under the spectral flow automorphism, which acts on the generators as
\cite{Schwimmer:1986mf}:
\begin{equation}
\begin{aligned}\label{spectral flow 1}
L_n &\quad \to \quad\,\,\,\, L'_{n} ={} L_n + \gamma J_{n} + \frac{\gamma^{2}}{6}c \,\delta_{n,0}\,,
\\
J_n &\quad \to  \quad \,\,\,\, J'_{n}  ={} J_n + \frac{c}{3}\gamma\,  \delta_{n,0}\,,
\\
G^{\pm}_{r} &\quad \to \quad  G_{r}^{\pm '}  ={} G^{\pm}_{r\pm \gamma}\,.
\end{aligned}
\end{equation}
Here $\gamma\in\mathbb{Z}+\frac{1}{2}$  is a spectral flow parameter that maps the NS to R sectors and vice versa. Values of $\gamma\in\mathbb{Z}$, on the other hand, map the respective sectors to themselves. In the NS sector, fermionic operators are anti-periodic. This is reflected in the fact that the modes of the supercurrent $G_r$ are half-integer, that is, $r\in\mathbb{Z}+\frac{1}{2}$. In the Ramond sector, on the other hand, the fermions are periodic, reflected in integer-modded supercurrents ($r\in\mathbb{Z}$).

\bibliographystyle{ytphys}
\bibliography{ref}
\end{document}